\documentclass[12pt,a4paper]{article_x}
\usepackage{epsfig}
\usepackage{amssymb}
\usepackage{latexsym}
\usepackage[british]{babel}
\usepackage{chicago}
\usepackage{supertabular}
\usepackage{float}

\newcommand{\bbox}[1]{\mbox{\boldmath ${{#1}}$}}

\renewcommand{\cite}[1]{\shortcite{#1}}
\renewcommand{\citeNP}[1]{\shortciteNP{#1}}
\newcommand{\tcite}[1]{\shortciteANP{#1} \citeyear{#1}}

\newcommand{\subsubsubsection}[1]{\bigskip {\bf #1}}

\renewcommand{\theequation}{\arabic{section}.\arabic{equation}}

\newcounter{saveequation}

\newenvironment{mathletters}%
{\addtocounter{equation}{1}%
\setcounter{saveequation}{\value{equation}}%
\setcounter{equation}{0}%
\renewcommand{\theequation}%
{\mbox{\arabic{section}.\arabic{saveequation}\alph{equation}}}}%
{\renewcommand{\theequation}{\mbox{\arabic{section}.\arabic{equation}}}%
\setcounter{equation}{\value{saveequation}}}

\newcommand{\mlabel}[1]{\immediate\write1{\string\newlabel{#1}%
{{\arabic{section}.\arabic{saveequation}}{\thepage}}}}

\newcommand{\ba}{{\bf a}}
\newcommand{\bA}{{\bf A}}
\newcommand{\bb}{{\bf b}}
\newcommand{\bB}{{\bf B}}
\newcommand{\be}{{\bf e}}
\newcommand{\bE}{{\bf E}}
\newcommand{\bff}{{\bf f}}
\newcommand{\bF}{{\bf F}}

\newcommand{\bH}{{\bf H}}
\newcommand{\bj}{{\bf j}}
\newcommand{\bk}{{\bf k}}
\newcommand{\bm}{{\bf m}}
\newcommand{\bp}{{\bf p}}
\newcommand{\bP}{{\bf P}}
\newcommand{\bq}{{\bf q}}
\newcommand{\bQ}{{\bf Q}}
\newcommand{\br}{{\bf r}}

\newcommand{\bu}{{\bf u}}
\newcommand{\bv}{{\bf v}}
\newcommand{\bx}{{\bf x}}
\newcommand{\bX}{{\bf X}}

\newcommand{\bz}{{\bf z}}

\newcommand{\bN}{{\bf 0}}

\newcommand{\bzeta}{{\bbox \zeta}}

\newcommand{\grad}{{\bbox \nabla}}

\newcommand{\tc}{\widetilde{c}}

\newcommand{\tV}{\widetilde{V}} 
\newcommand{\tw}{\widetilde{w}}
\newcommand{\tz}{\widetilde{z}} 

\newcommand{\tD}{\widetilde \Delta}
\newcommand{\tDii}{\widetilde \Delta^{(2)}}

\newcommand{\tcQ}{\widetilde \cQ}

\newcommand{\tbu}{\widetilde{\bf u}}

\newcommand{\hS}{\widehat{S}}

\newcommand{\hD}{\widehat \Delta}

\newcommand{\cE}{{\cal E}}
\newcommand{\cF}{{\cal F}}
\newcommand{\cH}{{\cal H}}
\newcommand{\cP}{{\cal P}}
\newcommand{\cQ}{{\cal Q}}
\newcommand{\cS}{{\cal S}}

\newcommand{\atr}{a_\triangle}
\newcommand{\bg}{b_{\rm g}}
\newcommand{\cg}{c_{\rm g}}
\newcommand{\cli}{c_{\rm Li}}
\newcommand{\co}{c}
\newcommand{\cs}{c_{\rm s}}
\newcommand{\dE}{\Delta \cE}
\newcommand{\dF}{\Delta \cF}
\newcommand{\Dii}{\Delta^{(2)}}
\newcommand{\Do}{\Delta_0}
\newcommand{\dS}{\Delta \cS}

\newcommand{\el}{\varepsilon_{\rm l}}
\newcommand{\eo}{\varepsilon_0}

\newcommand{\Fm}{F_{\rm m}}
\newcommand{\gE}{{\gamma_{\rm Eu}}}
\newcommand{\Gi}{{\rm Gi}}
\newcommand{\Hc}{H_{\rm c}}
\newcommand{\Hci}{H_{\rm c1}}
\newcommand{\Hcii}{H_{\rm c2}}
\newcommand{\Hmi}{H_{\rm m1}}
\newcommand{\Hmii}{H_{\rm m2}}

\newcommand{\Ldis}{L_{\rm disloc}}
\newcommand{\Ls}{L_{\rm s}}
\newcommand{\muo}{\mu^{*}}

\newcommand{\ns}{n_{\rm s}}
\newcommand{\Qtr}{Q_\triangle}
\newcommand{\rhon}{\rho_{\rm n}}
\newcommand{\st}{\varepsilon_\parallel}
\newcommand{\txig}{t_{\xig}}
\newcommand{\Tc}{T_{\rm c}}
\newcommand{\Tco}{T_{\rm c0}}
\newcommand{\Teff}{T_{\rm eff}}
\newcommand{\Tg}{T_{\rm g}}
\newcommand{\Th}{T_{\rm h}}
\newcommand{\Tm}{T_{\rm m}}
\newcommand{\Tsh}{T^{\rm sh}}
\newcommand{\uo}{u^{*}}
\newcommand{\vm}{v_{\rm m}}
\newcommand{\xig}{\xi_{\rm g}}

\newcommand{\tauc}{\tau_{\rm c}}
\newcommand{\taug}{\tau_{\rm g}}
\newcommand{\tauco}{\tau_{\rm c0}}

\newcommand{\vphi}{\varphi}
\newcommand{\aphi}{\alpha}

\newcommand{\fcaption}[2]{
\parbox{12cm}{\caption{\small {#1}} \label{#2} }}

\begin{document}
\selectlanguage{british}



\begin{title}
  {Vortex-glass phases in type-II superconductors\footnote{\tt Accepted
      for publication in {\em Advances in Physics}}}
\end{title}

\begin{author}
  {T. Nattermann\footnote{tel.: +49-[0]221-470-2907; fax:
      +49-[0]221-470-5159; email: natter@thp.uni-koeln.de} 
    \enspace
    and S.  Scheidl\footnote{tel.: +49-[0]221-470-4213; fax:
      +49-[0]221-470-5159; email: sts@thp.uni-koeln.de}
    \\
    Institut f\"ur Theoretische Physik, Universit\"at zu K\"oln
    \\
    Z\"ulpicher Stra{\ss}e 77, D-50937 K\"oln, Germany}
\end{author}

\maketitle

\begin{abstract}
  A review is given on the theory of vortex-glass phases in impure
  type-II superconductors in an external field.  We begin with a brief
  discussion of the effects of thermal fluctuations on the
  spontaneously broken $U(1)$ and translation symmetries, on the
  global phase diagram and on the critical behaviour.  Introducing
  disorder we restrict ourselves to the experimentally most relevant
  case of weak uncorrelated randomness which is known to destroy the
  long-ranged translational order of the Abrikosov lattice in three
  dimensions.  Elucidating possible residual glassy ordered phases, we
  distinguish between {\em positional} and {\em phase-coherent vortex
    glasses}.  The study of the behaviour of isolated vortex lines and
  their generalization -- directed elastic manifolds -- in a random
  potential introduces further important concepts for the
  characterization of glasses.  The discussion of {\em elastic} vortex
  glasses, i.e., topologically ordered dislocation-free positional
  glasses in two and three dimensions occupy the main part of our
  review.  In particular, in three dimensions there exists an elastic
  vortex-glass phase which still shows quasi-long-range translational
  order: the `Bragg glass'.  It is shown that this phase is stable
  with respect to the formation of dislocations for intermediate
  fields. Preliminary results suggest that the Bragg-glass phase may
  not show phase-coherent vortex-glass order.  The latter is expected
  to occur in systems with weak disorder only in higher dimensions (or
  for strong disorder, as the example of unscreened gauge glasses
  shows).  We further demonstrate that the linear resistivity vanishes
  in the vortex-glass phase.  The vortex-glass transition is studied
  in detail for a superconducting film in a parallel field.  Finally,
  we review some recent developments concerning driven vortex-line
  lattices moving in a random environment.
\end{abstract}

\newpage
\tableofcontents



  \section{Introduction}

Since its discovery by Kammerlingh Onnes in 1911, superconductivity
has attracted generations of physicists. In 1935 Fritz and Heinz
London (see e.g. \tcite{London50}) developed a very successful
phenomenological theory which describes both the perfect conductivity
as well as the perfect diamagnetism of superconductors.  As discussed
later by \tcite{London50} this theory can be motivated by considering
superconductivity as a phenomenon characterized by long-range order of
momentum $\bp$.  Bohr-Sommerfeld quantization $\oint\bp_sd{\bf s}=nh$
on a torus gives fluxoid quantization \cite{London50}.
\tcite{Ginzburg+50} combined London's electrodynamics of a
superconductor with Landau's theory of phase transitions, creating a
powerful phenomenological description of superconductivity. The
transition to the superconducting phase corresponds here to the
breaking of the U(1) symmetry of the complex order parameter $\Psi$
and the appearance of off-diagonal long-range order (ODLRO).

In a pioneering work \tcite{Abrikosov57:t} showed the existence of a
second type of superconductors which (for sufficiently strong external
magnetic field) allows for a penetration of quantized magnetic flux in
the form of vortex lines, which form a triangular lattice, reducing
the perfect diamagnetism and creating a source for dissipation due to
the motion of the vortex-line core driven by the transport current.
In this case the continuous translational symmetry of the system is
broken in addition to the $U(1)$ symmetry.  Effects from thermal
fluctuations, although studied already since 1960, were considered to
be extremely small because of the large correlation length and low
transition temperatures of conventional superconductors
\cite{Ginzburg60:t}.

To keep the superconducting properties, vortices have to be prevented
from moving by pinning centers. An early theory of pinning of isolated
vortex lines \cite{Anderson+64} shows the absence of dissipation only
at zero temperature. Thermally activated hopping leads to a small but
finite dissipation at low temperatures (as compared to the height of
energy barriers).  \tcite{Larkin70:t} extended this theory to the
Abrikosov vortex-line lattice, showing the destruction of its
translational long-range order.  Although -- in principle -- the
Abrikosov phase could thence be considered not to really differ from a
pinned vortex-line liquid (and hence from the normal phase), the
generic phase diagram of conventional superconductors was assumed to
practically be that of \tcite{Abrikosov57:t} with a now finite
correlation length of the vortex-line array.

In the 1980s this picture was changed by two initially unrelated
developments: the discovery of high-$T_c$ superconductors by
\tcite{Bednorz+86} and the much better understanding of random system,
in particular of spin glasses and of random-field systems (for a
recent review see \tcite{Young98}). In the high-$T_c$ superconductors
with their elevated transition temperatures and pronounced anisotropy,
fluctuation effects became now very important as can be seen for
instance from the observed melting of the vortex-line lattice
\cite{Cubitt+93,Zeldov+95}.  Moreover, for pure systems it was
demonstrated \cite{Moore89,Moore92,Ikeda+92} that thermal fluctuations
-- which prohibit true long-range translational order of the
vortex-line lattice (VLL) only in $d \leq 2$ dimensions -- destroy the
ODLRO of the gauge invariant order parameter in the Abrikosov phase
even in higher dimensions.  Thus, in three dimensions thermal
fluctuations restore the $U(1)$ symmetry of the Ginzburg-Landau
Hamiltonian but nevertheless allow for the existence of a vortex
lattice.  This finding is paralleled by an earlier observation of
\tcite{Schafroth55} that an external magnetic field above a critical
strength destroys Bose-Einstein condensation of an ideal Bose gas
which still shows some remanent diamagnetic moment.

For systems with disorder the idea emerged that despite of the
destruction of true translational long-range order the system could
show a phase with some kind of glassy long-range order, the ``vortex
glass'' \cite{FisherMPA89}.  Because of the residual rigidity in the
vortex-line array, arbitrarily large energy barriers now exist,
leading to a highly non-linear resistivity
\cite{Feigelman+90,FisherMPA89,Feigelman+89,Nattermann90,FisherDS+91:tf}
\begin{equation}
\rho (j)\sim  e^{-(j_{\rm t}/j)^{\mu}}
\label{nonlin-res}
\end{equation}
where $\mu$ denotes an exponent $0\le\mu\le 1$ and $j_{\rm t}$ a
threshold current.  Since the {\em linear} resistivity vanishes, the
system is truly superconducting.

In the following years vortex pinning and depinning as well as flux
creep under the action of an external current was investigated by many
researchers to a great extent.  A brilliant summary of the results of
these efforts till 1994 is given in the extensive review article by
\tcite{Blatter+94} (see also \tcite{Brandt95}, \tcite{Gammel+98:rev},
\tcite{Giamarchi+98:rev}).  It is {\it not} the intention of the
present article to provide an updated version of these reviews by
discussing the results obtained since then.  Instead, we want to focus
here mainly on one particular aspect of the theory, namely on the
discussion of the equilibrium phase diagram of weakly disordered
type-II superconductors in an external magnetic field. We want to
demonstrate that the notion of a `vortex glass' is not a blurred
expression for the hopelessly intricate situation in a disordered
system (as some physicists may still claim), but that it has a well
defined meaning.  The knowledge of the equilibrium properties is also
important for the proper understanding of situations close to
equilibrium, e.g. for the discussion of flux creep under the influence
of a small external current.

Despite of the conclusion common to all authors mentioned above of
expecting a non-analytic current-density dependence of the resistivity
in the glassy phase it seems to be indicated to refer here also to the
differences between these approaches.  \tcite{FisherMPA89} and
\tcite{FisherDS+91:tf} in their definition of the glassy phase started
from correlation functions measuring ODLRO and focused on the
long-range (glassy) order of phases (see chapter
\ref{subsec.infl.of.disorder}).  On the other hand,
\tcite{Feigelman+89}, \tcite{Nattermann90} and subsequently
\tcite{Korshunov93:vg} and \tcite{Giamarchi+94} considered primarily
the glassy order of the vortex-line array, i.e., they focused on the
positions of the vortex lines.  In both cases, the expression `vortex
glass' was used.  The above mentioned differences in the lower
critical dimensions for the breaking of the $U(1)$ and of the
translational symmetry in pure systems however suggests, that there
may be also different lower critical dimensions for a {\em
  phase-coherent} and a {\em positional vortex glass}.  In this review
we concentrate mainly on the positional glass. If, in particular, the
positional vortex glass is free of large dislocation loops such that
an elastic description of the vortex array is possible, the positional
vortex glass will be called {\em elastic vortex glass}.  The most
prominent example of an elastic vortex glass is its three-dimensional
version: the so-called `Bragg-glass' (for a recent brief review see
\tcite{Giamarchi+98:rev}).  Whether non-elastic vortex-glass phases
exist is still unclear.

Systems with columnar disorder which leads to the formation of the
so-called {\em `Bose-glass' phases} (see e.g. \tcite{Tauber97}) will
be completely neglected in this review since their physics is
substantially different from that for systems with point disorder, to
which we restrict ourselves here.  However, we also include the
discussion of various phases found for driven systems far from
equilibrium, which share some features with the equilibrium phase
diagram.

The article is organized as follows. In chapter
\ref{sec.GLdescription} we present a brief summary of the
Ginzburg-Landau theory of type-II superconductors and a short
discussion of the influence of thermal fluctuations and of the effect
of point disorder in the critical region.  We also define the
different types of vortex-glass order and give a brief account of
results obtained for models with strong disorder -- the so-called
gauge glasses.  In chapter \ref{sec.ran.man} we review the behaviour
of a {\it single} vortex line and its generalizations --
$D$-dimensional directed manifolds -- in a random potential. This
simple, but not at all trivial system allows for a discussion of
different aspects of glassiness of a system.  Chapter
\ref{sec.film.perp} is devoted to the superconducting film in a
parallel field, a geometry which allows for a very detailed
description of the vortex glass phase as well as of the transition to
the normal phase both for the static and dynamic quantities.  In
chapter \ref{sec.SF.perp.filed} we discuss an impure superconducting
film in a field perpendicular to the film plane. It turns out that
dislocations destroy the positional vortex-glass phase in this
geometry.  The `Bragg-glass' phase of a bulk superconductor as well as
its stability with respect to dislocations is considered in chapter
\ref{sec.bulk}.  A short account of recent activities on driven vortex
lattices in impure superconductors is presented in chapter
\ref{sec.dyn.phases}.  We close the paper with a brief summary of the
results of this article (chapter \ref{sec.conclusions}).  The appendix
contains some technicalities and a list of recurrent symbols.

  \setcounter{equation}{0}
  \section{Ginzburg-Landau description}
\label{sec.GLdescription}

In this chapter we give a very brief introduction into the mean-field
theory and the effects arising from thermal and disorder fluctuations
in type-II superconductors in the framework of the Ginzburg-Landau
theory.  Since there is extensive (and partially contradicting)
literature on thermal effects it is impossible to include all related
references.  However, we attempt to include the most recent articles
on the subject which may serve as more comprehensive guides to further
references.

\subsection{The Ginzburg-Landau model}
\label{subsec.GLmodel}

In 1950 Ginzburg and Landau proposed a phenomenological description of
superconductors by introducing a two-component order parameter
$\Psi(\br)=|\Psi(\br)| e^{i \phi(\br)}$ which couples in a
gauge-invariant form to the magnetic field described by the vector
potential $\bA(\br)$ \cite{Ginzburg+50}. The density $\ns(\br)$ of
superconducting charge carriers (i.e., of the Cooper pairs), which is
a central quantity of the earlier London theory \cite{London+35}, is
related to $\Psi(\br)$ by $\ns(\br)=|\Psi(\br)|^2$.  The
Ginzburg-Landau (GL) free energy is given by
\begin{eqnarray}
{\cal H}_{\rm GL}&=&\frac{1}{2}\int d^dr \ \Big\{
\beta \Big( |\Psi |^2 + \frac\alpha \beta \Big)^2 
 \nonumber \\ &&
+\frac{\hbar^2}{m} 
\Big| \Big(i\grad -\frac{2\pi}{\Phi_0}{\bf A}\Big)\Psi\Big|^2 
+\frac{1}{4\pi}(\grad\wedge {\bf A} -{\bf H })^2\Big\},
\label{eq:GLHamiltonian}
\end{eqnarray}
where $\Phi_0=h\co/2e$ denotes the flux quantum, $\alpha(T) \propto
(T-\Tco)$, $\Tco$\label{intro.Tco} is the mean-field transition
temperature, ${\bf H}$ is the external field, and $m$ denotes the mass
of a Cooper pair.  The GL free energy is characterized by two basic
length scales, the coherence length $\xi$ and the penetration depth
$\lambda$, which are related to the parameters of ${\cal H}_{\rm GL}$
by
\begin{mathletters}
\begin{eqnarray}
\xi^2(T) & = & \frac{\hbar^2}{2m |\alpha(T)|},
\label{intro.xi}
\\
\lambda^2(T) &=&\frac{m\co^2}{4\pi|\Psi_0|^2(2e)^2}.
\label{intro.lambda}
\end{eqnarray}
\end{mathletters}
Here $|\Psi_0|^2=|\alpha|/\beta$ denotes the saturation value of
$|\Psi|^2$ in a homogeneous current free state for $T<\Tco$ and
$\bH=\bN$.  For our further discussion it is convenient to use the
following rescaling to introduce dimensionless quantities $\Psi'$,
$\bA'$ and $\br'$
\begin{eqnarray}
\Psi & = & \Psi' |\alpha/\beta|^{1/2},\quad 
\bA =\frac{\Phi_0}{2\pi\xi}\bA ',\quad
\br =\br '\xi.
\label{eq:rescaling}
\end{eqnarray}
This leads  to
\begin{eqnarray}
\frac 1T {\cal H}_{\rm GL}&=&\frac{1}{4\pi}\left(
\frac{|\tauco|^{4-d}}{2 \Gi}\right)^{1/2}
\int d^dr'\Big\{ \frac 12 ( |\Psi '|^2 + \alpha/|\alpha|)^2
\nonumber\\ && 
+ | (i\grad ' - \bA ')\Psi'|^2
+   \kappa^2
\Big( \grad '\wedge\bA '-{\bf H}/{\Hcii^{\rm MF}}\Big)^2\Big\}.
\label{eq:rescaledHamiltonian}
\end{eqnarray}
Here we introduced the Ginzburg number 
\begin{equation}
\label{intro.Gi}
\Gi \equiv \left(\frac{T}{4\pi\xi^d(0) \beta |\Psi_0(0)|^4}\right)^2
= \left(\frac{T}{\xi^d(0) \Hc^2(0) }\right)^2,
\end{equation}
the Ginzburg-Landau parameter $\kappa \equiv \lambda/\xi$ and the
reduced temperature $\tauco \equiv (\Tco-T)/\Tco$.  $\Hcii^{\rm MF}
\equiv\Phi_0/2\pi\xi^2 \label{intro.Hcii}$ is the mean-field upper
critical field and $\Hc^2=4 \pi \beta |\Psi_0|^4 \label{intro.Hc}$ is
the thermodynamic critical field.

\subsection{Mean-field theory}\label{subsec:MFtheory}
\label{sec.MF.mod}

Within {\em mean-field (MF) theory}, the GL free energy has to be
minimized with respect to the fields $\Psi(\br)$ and $\bA(\br)$. The
resulting GL-equations
\begin{equation}
(i\grad^{\prime}-\bA^{\prime})^2\Psi^{\prime}+\frac{\alpha}{|\alpha|}
\Psi^{\prime}
+|\Psi^{\prime}|^2\Psi^{\prime}=0
\label{eq.GL1}
\end{equation}
and (for $\alpha<0$)
\begin{equation}
\kappa^2\frac{1}{|\Psi^{\prime}|^2}\grad\wedge(\grad\wedge\bA^{\prime})
+\bA^{\prime}=-\grad^{\prime}\phi
\label{eq.GL2}
\end{equation}
then have to be solved with the appropriate boundary conditions. As is
clear from equation (\ref{eq:rescaledHamiltonian}), the only two
parameters which will enter the solution in the bulk, are the
GL-parameter $\kappa$, which plays the role of an inverse effective
charge of the $\Psi$ field, and the strength of the external magnetic
field $\bH'=\bH/\Hcii^{\rm MF}$.

For $\kappa<1/\sqrt 2$ (type-I superconductors), mean-field theory
yields for $T<\Tco$ and $H<\Hc(T)$ a phase with vanishing resistance
and perfect diamagnetism. The transition to the normal phase at
$\Hc(T)$ is first order.

For $\kappa>1/\sqrt 2$ (type-II superconductors), on the other hand,
perfect diamagnetism exists only up to the field $\Hci^{\rm MF}
\approx (\Hcii^{\rm MF}/{2\kappa^2})(\ln \kappa +
0.08)$.\label{intro.Hci} For larger fields, magnetic flux penetrates
the sample in the form of quantized vortex lines, each carrying a flux
quantum $\Phi_0$.  The energy per unit length of the vortex line is
therefore given by $\el = (\Phi_0/4\pi) \Hci \simeq \eo \ln \kappa$
with $\eo=(\Phi_0/4 \pi \lambda)^2$ as the important energy scale per
unit length.

The vortex lines form a triangular `Abrikosov' lattice
\cite{Abrikosov57:t,Kleiner+64} of spacing $\atr=(2/\sqrt 3)^{1/2} a$,
where $a \equiv (\Phi_0/B)^{1/2}$.  The Abrikosov lattice (or `mixed')
phase shows both broken translational symmetry and off-diagonal
long-range order (ODLRO), i.e., broken U(1) symmetry of the order
parameter.  Both broken symmetries vanish simultaneously if $H$
reaches $\Hcii^{\rm MF}$.  One should however take into account that
the correlation function for ODLRO (see chapter
\ref{subsubsec.finite.ext.field})
$\langle\Psi^{\ast}(\br)\,\Psi(\br^{\prime})\rangle$ -- even if
calculated in MF approximation -- shows strong spatial variations due
to the rapid change of the phase. Indeed, for a system of radius $R$
the tangential phase gradient on the boundary is of the order $R/2a^2$
which corresponds to a phase change of $2\pi$ on a distance
$l_{\phi}\approx 4\pi a^2/R$ (with $a\approx 100{\rm nm}$ and
$R\approx 1{\rm cm}$, $l_{\phi}\approx 10^{-2}{\rm nm}$ which is
smaller than an atom, see \tcite{Brandt74}).  Thus
$\langle\Psi^{\ast}(\br)\,\Psi(\br^{\prime})\rangle$ cannot be very
meaningful as a physical observable.  Quantities with a physical
significance should be in particular gauge invariant. In the
GL-description these are the amplitude $|\Psi|$ of the order
parameter, the `super-velocity' $\grad\phi
-\frac{2\pi}{\Phi_0}\bA\equiv\grad\tilde\phi$ and the magnetic
induction $\bB=\grad\wedge\bA$.  All other physical quantities can be
expressed in these fields, e.g. the current density can be written as
\begin{equation}
\bj=-\frac{2e\hbar}{m}|\Psi|^2\grad\tilde\phi\,.
\label{eq.curr.den}
\end{equation}

In treating the vortex system, two main approximations have been used:
{\it the lowest Landau level} (LLL) {\it approximation}, which is
valid sufficiently close to $\Hcii^{\rm MF}$ ($H \gtrsim \frac 13
\Hcii^{\rm MF}$), and the {\it London approximation}, which is valid
at intermediate and small fields $B \lesssim 0.2\Hcii^{\rm MF}$, where
$\xi\ll a$. The precise range of applicability of the LLL is still
under debate (see e.g. \tcite{Oneill+93}, \tcite{Li+99} and references
therein).

In the London approximation one neglects amplitude inhomogeneities,
$|\Psi|=|\Psi_0|$, which leads to a diverging energy density at the
vortex cores.  The position $\br_i(s)$ of these cores is parameterized
by the label $i$ of a vortex and the variable $s$ along the contour of
the vortex lines. The Ginzburg-Landau Hamiltonian takes then the form
\begin{equation}
{\cH}_{\rm London}=\frac{1}{2}
\int d^dr\left\{\frac{\hbar^2}{m}|\Psi_0|^2
(\grad\phi+\frac{2\pi}{\Phi_0}\bA)^2 +\frac{1}{4\pi}
(\grad\wedge\bA -\bH)^2\right\}\,.
\label{eq.HLondon}
\end{equation}
This functional has to be regularized near the vortex cores, e.g. by
excluding tubes of radius $\xi$ around the vortex cores from the
volume integration.  The phase $\phi(\br)$ of the complex order
parameter is a multivalued function since $\phi$ changes by $2\pi$
along a path surrounding a vortex line.  We decompose $\phi$ now into
a vortex part $\phi_{\rm v}$ and a ``spin-wave'' part $\phi_{\rm sw}$.
The vortex part is assumed to fulfill the saddle point equation apart
from the position of the vortices $\br_i(s)$.  With the London gauge
$\grad \cdot \bA =0$ this yields
\begin{equation}
\grad^2 \phi_{\rm v}(\br)=0\,,\quad
\br\not=\br_i(s)
\label{eq.saddle.point1}
\end{equation}
and
\begin{equation}
\grad \wedge (\grad\phi_{\rm v})=2\pi\bm(\br)\,.
\label{eq.saddle.point2}
\end{equation}
Here $\bm(\br)$ denotes the vortex-density field
\begin{equation}
\bm(\br)=\sum\limits_im_i \int ds \frac{d \br_i(s)}{ds}
\delta^{(3)}(\br_i(s)-\br)\,,
\label{eq.vortex.density.field}
\end{equation}
where the integration is along the vortex line $i$ which carries the
vorticity $m_i=\pm 1$. If the spin-wave part $\phi_{\rm sw}(\br)$
vanishes on the surface of the sample, $\phi_{\rm sw}$ and $\phi_{\rm
  v}$ decouple.  Since the vector potential $\bA$ appears only
quadratically in $H_{\rm GL}$ it can be integrated out by using the
saddle-point equation which is the second GL-equation in the
phase-only approximation
\begin{equation}
\bA +\lambda^2\grad\wedge(\grad\wedge\bA)=
-\frac{\Phi_0}{2\pi}\grad\phi_{\rm v}.
\label{eq.sec.GL.eq}
\end{equation} 
Taking the curl of (\ref{eq.sec.GL.eq}) gives the modified London
equation
\begin{equation}
\lambda^2 \grad^2 \bB(\br)-\bB(\br)=\Phi_0\bm(\br),
\label{eq.mod.London.eq}
\end{equation}
where $\bB=\grad \wedge \bA$ can now completely be expressed in terms
of the vortex degrees of freedom given by the vortex density field
$\bm(\br)$, equation (\ref{eq.vortex.density.field}).  The London
Hamiltonian then takes the form
\begin{equation}
\cH_{\rm London}=\frac{1}{2}\int d^dr\left\{\frac{\hbar^2}{m}|\Psi_0|^2
(\grad\phi_{\rm sw})^2+\frac{\lambda^2}{4\pi}(\grad\wedge\bB)^2+
\frac{1}{4\pi}(\bB-\bH)^2\right\}\,.
\label{eq.HLondon1}
\end{equation}
In most parts of this review we will use the London picture,
since it remains valid in the vortex phases we will describe, in
particular in the elastic glass phases.

The {\it elasticity theory} of the Abrikosov lattice for an isotropic
superconductor was worked out by Brandt
\citeyear{Brandt77:I,Brandt77:II}.  The distortion of a vortex line is
described by a two-component displacement field $\bu(\bX,z)$, where
the lattice vector $\bX\equiv\bX_{n,m}=((2n+m)\atr/2,m\sqrt{3}\atr/2)$
denotes the rest position of the vortex line in the plane
perpendicular to $\bH=H\hat{\bz}$.  In many cases one can go over to
the continuum description: $\bu(\bX,z)\to\bu(\bx,z)\equiv\bu(\br)$.
On large scales $L\gg\lambda$ the elastic energy of the vortex-line
lattice is then given by
\begin{eqnarray}
\cH_{\rm el} = \frac 12 \int d^2x d^{d-2} z 
\left\{ c_{11} (\grad_\perp \cdot \bu)^2 
+ c_{66}  (\grad_\perp \wedge \bu)^2 
+ c_{44} (\grad_\parallel \bu)^2  \right\},
\label{H.el.3D}
\end{eqnarray}
where $c_{11} \approx c_{44} \approx \bB^2/4\pi$.  The Abrikosov phase
is characterized in particular by a non-zero shear modulus $c_{66}
\approx \frac{B \Phi_0}{(4 \pi \lambda)^2}\left(1-\frac B{\Hcii^{\rm
      MF}} \right)^2$, which vanishes both at $\Hcii$ {\em and}
$\Hci$, reaching a maximum of $c_{66} \approx \Hci \Hcii/56 \pi$ in
between.  One should however take into account that the elastic free
energy is in general {\it non-local}, which is expressed in a strong
dispersion of $c_{11}$ and $c_{44}$ on scales smaller than $\lambda$
(see, e.g., \tcite{Brandt91}).

In the absence of pinning centres, the system in the Abrikosov phase
behaves superconducting only for currents parallel to the magnetic
field. For currents with components perpendicular to the field the
Lorentz force drives the vortex-line array, which leads to metallic
behaviour with resistivity $\rho \approx \rho_{\rm n} B/\Hcii$
\cite{Bardeen+65}.  Here $\rho_{\rm n}$ is the resistivity of the
normal phase.

Before we come to the discussion of fluctuation effects, we want to
consider a possible extension of the model (\ref{eq:GLHamiltonian})
which describes an isotropic superconductor.  High-$\Tc$
superconductors, however, are characterized by a pronounced layer
structure, which results in an inhomogeneity and in a strong spatial
{\em anisotropy} of the effective mass of the Cooper pair such that
$m$ is replaced by $M=m/\epsilon^2 \label{intro.eps.aniso}$ for
electrons moving perpendicular to the layers.  Typical values for
$\epsilon$ are $\epsilon_{\rm YBCO} \approx 0.16$ and $\epsilon_{\rm
  BSCCO} \approx 10^{-2}$ for the two high-$\Tc$ materials YBCO and
BSCCO.  It was shown by \tcite{Blatter+92} (for a more detailed
discussion, see \tcite{Blatter+94}), that in the case of $\kappa \gg
1$ the result for a thermodynamic quantity
$\cQ(\vartheta,H,T,\xi,\lambda,\epsilon,\Delta)$ of an anisotropic
superconductor (where $\vartheta$ denotes the angle between the
magnetic field direction and the $xy$ plane, and $\Delta$ the strength
of disorder) can be obtained from the corresponding result $\tcQ$ for
the isotropic system by the relation
\begin{eqnarray}
  \cQ(\vartheta,H,T,\xi,\lambda,\epsilon,\Delta) = s_\cQ \ 
  \tcQ(\epsilon_\vartheta H, T/\epsilon, \xi, \lambda,\Delta/\epsilon),
\label{scal.aniso}
\end{eqnarray}
where $\epsilon_\vartheta^2=\epsilon^2 \cos^2 \vartheta + \sin^2
\vartheta$, $s_V=s_E=s_T=\epsilon$ for volume, energy and temperature,
and $s_B=s_H=1/\epsilon_\vartheta$ for magnetic fields.  Since $T$ and
$\Delta$ are increased by a factor $1/\epsilon$ with respect to the
isotropic system, it is clear that fluctuation effects, which will be
considered in the following sections, are drastically enlarged (by a
factor up to 100) in these materials.

If the spatial anisotropy is so large that the coherence length
$\xi_z=\xi \epsilon$ in $z$ direction becomes of the order of the
layer spacing $s$, the discreteness of the layer structure becomes
relevant. In this case, new effects such as a decoupling of the layers
[for $B>B_{\rm 2D} \approx (\epsilon/s)^2 \ln (s/\xi)$] may occur. The
appropriate description is then the Lawrence-Doniach model
\cite{Lawrence+71}.  We will not attempt to cover in this review also
these particular features of strongly layered materials, instead we
restrict ourselves in the following to the discussion of the isotropic
superconductor, knowing that the results for the anisotropic case can
be found from the relation (\ref{scal.aniso}).

The neglect of layer-effects is also supported by the following
argument: Our theoretical analysis will be mainly based on the elastic
description of the vortex lattice and our main interest concerns
features on large length scales.  Sufficiently weak disorder will
indeed effectively couple to the vortex lattice only on very large
length scales, where the elasticity of the lattice may be described by
{\em local} elasticity theory.  Since within the London approach all
information about anisotropy and even about the layered structure is
encoded in the dispersion of the elastic constants, we expect that the
large-scale properties are independent of these details (provided that
one is in the parameter regime where the London approach is valid and
that disorder is sufficiently weak).

\subsection{Thermal fluctuations}
\label{subsec.th.fluct}

So far we have ignored the influence of fluctuations, i.e., of
configurations which do not fulfill the GL equations.  These can be
taken into account if we interpret the GL free energy as an effective
Hamiltonian from which the true free energy ${\cal F}(T,\bH)$ has to
be calculated as
\begin{eqnarray}
  {\cal F}(T,\bH) = - T \ln \left( \int {\cal D}\Psi {\cal D} \bA
    e^{-{\cal H}_{\rm GL}/T} \right).
\label{GL.free}
\end{eqnarray}
Since the only material independent common feature of type-II
superconductors is flux quantization it is natural to build from
$\Phi_0$ and $T$ a characteristic length scale \cite{FisherDS+91:tf}
\begin{eqnarray}
  \Lambda_T \equiv \frac{\Phi_0^2}{16 \pi^2 T} \approx 2 \cdot 10^8
  \frac {\textrm \AA} {T[{\textrm K}]},
\label{intro.Lambda_T}
\end{eqnarray}
which is the same for all materials.  Since the energy per unit length
of a vortex line (and hence also its stiffness constant $\st$, see
below) is of the order $\eo=(\Phi_0/4 \pi \lambda)^2$, $\Lambda_T$
denotes the length scale on which the mean squared displacement of a
vortex line is of order $\lambda$. Since $\Lambda_T$ is so large,
thermal fluctuation effects are expected to be small (however, see our
remark about strongly anisotropic systems in the previous chapter
\ref{sec.MF.mod}).  In $d=3$ dimensions the Ginzburg number can be
expressed as $\Gi \approx (\kappa \lambda(0)/\Lambda_T)^2 $.  As can
be seen directly from (\ref{eq:rescaledHamiltonian}) and
(\ref{GL.free}) the contribution from fluctuations in $\Psi(\br)$ and
$\bA(\br)$ will indeed be small, if both $\Gi \ll |\tauco|^{4-d}$ and
$\kappa \gg1$.

\subsubsection{Zero external field}
\label{subsubsec.0ext.field}

In {\em zero external field}, $\bH=\bN$, to begin with, it was shown
by \tcite{Halperin+74} that for type-I superconductors, where
$\kappa<1/\sqrt2$ and typically $\Gi \ll 1$ (e.g., $\Gi \approx
10^{-13}$ for aluminium), fluctuations in the vector potential
$\bA(\br)$ render the transition first order.

For type-II superconductors, on the other hand, the situation is less
clear.  It has been argued that the transition remains second order
\cite{Helfrich+80,Dasgupta+81}.  In the high-$\Tc$ compounds with
large values for $\kappa$ ($\kappa_{\rm YBCO} \approx 100$,
$\kappa_{\rm BSCCO} \approx 60$) and large Ginzburg numbers ($\Gi_{\rm
  YBCO} \approx 10^{-2}$, $\Gi_{\rm BSCCO} \approx 1$) fluctuations in
the vector potential are weak compared to those of the order
parameter.  Then there exist two critical regions. In the outer
critical region
\begin{eqnarray}
  |\tauc| \ll \Gi^{1/(4-d)}, \quad |\tauc| \gg (\Gi/\kappa^4)^{1/(4-d)},
\end{eqnarray}
where $\tauc \equiv (\Tc-T)/\Tc$ now denotes the reduced temperature
with respect to the {\em true} transition temperature $\Tc$,
fluctuations of the order parameter lead to an $XY$-like critical
behaviour.  Fluctuations in the vector potential can be neglected in
this regime.  Since the coherence length $\xi(T) \approx \xi(0)
|\tauc|^{-\nu}$ with $\nu=\nu_{XY}\approx \frac 23$ in $d=3$
dimensions increases now more strongly than the penetration depth
$\lambda \approx \lambda_0 |\tauc|^{-\beta+ \eta \nu/2}$, the
effective value of $\kappa \sim \xi^{-(4-d)/2}$ decreases until both
lengths are of the same order.  This signals a cross-over to a second
critical regime with (probably) inverted $XY$ behaviour
\cite{Helfrich+80,Dasgupta+81}. In this asymptotic regime $\lambda$
and $\xi$ scale in the same way with the correlation exponent
$\nu_{XY}$ \cite{Olsson+99}.  It should be mentioned, however, that
other scenarios have been proposed (for a recent discussion of earlier
results see \tcite{Kiometzis+95}, \tcite{Radzihovsky95:epl},
\tcite{Herbut+96}, \tcite{Herbut97:jpa}, \tcite{Folk+99},
\tcite{Nguyen+99}).

In $d=2$ dimensions fluctuations prevent the formation of a long-range
ordered phase.  As was shown by Pearl \citeyear{Pearl64,Pearl65}, the
effective London penetration depth $\Ls=2 \lambda^2/s$ diverges with
decreasing $s$, where $s$ denotes the film thickness. Therefore
fluctuations in the vector potential can be neglected and the system
in zero external field shows a Kosterlitz-Thouless transition to a
quasi-long-range ordered phase \cite{Doniach+79,Halperin+79}.

\subsubsection{Finite external field}
\label{subsubsec.finite.ext.field}

Next we consider the case of {\em finite external field}.  The most
obvious effect of thermal fluctuations on the vortex-line lattice is
{\it melting} \cite{Eilenberger67,Nelson88}.  Melting has been seen
experimentally in YBCO
\cite{Safar+92,Kwok+92,Charalambous+93,Kwok+94:a,Liang+96,%
  Schilling+96,Welp+96} and BSCCO
\cite{Pastoriza+93,Zeldov+95,Hanaguri+96}.  To estimate the melting
temperature one may use the phenomenological {\it Lindemann criterion}
\begin{eqnarray}
  \langle \bu^2 \rangle^{1/2} = \cli \atr,
\label{intro.Linde}
\end{eqnarray}
where $\bu$ denotes the displacement of a vortex line from its rest
position, $\langle \dots \rangle$ the thermal average and $\cli\approx
0.1 \dots 0.2$ is the Lindemann number. Since the shear modulus
$c_{66}$ vanishes at $\Hci^{\rm MF}$ and $\Hcii^{\rm MF}$, melting
will occur by approaching {\em both} critical fields. Close to $\Hci$
the melting line $\Hmi(T) \label{intro.Hm1}$ is roughly given by
\cite{FisherDS+91:tf}
\begin{eqnarray}
\frac{\Hmi-\Hci}{\Hci} \approx \left( \frac \lambda {\Lambda_T} \right)^2
\label{intro.Hmi}.
\end{eqnarray}
The region $\Hci < H <\Hmi$, where the vortex lines form a liquid, is
extremely small, except for the vicinity of $\Tc(H=0)$ where $\lambda$
diverges. $\Hci$ is reduced with respect to $\Hci^{\rm MF}$ due to
fluctuations \cite{Nelson88,Nelson+89}.  We note, however, that for a
proper calculation of the melting curves the dispersion of the elastic
constants has to be taken into account.  In anisotropic and layered
superconductors, \tcite{Blatter+96:prl}, following an earlier
suggestion by \tcite{Brandt+96}, found an additional
fluctuation-induced attractive van-der-Waals interaction between
vortex lines, which may lead at very low temperatures to a first order
transition between the Meissner and the Abrikosov lattice phase.

At large fields the melting line $\Hmii(T)$ is reached if
\cite{Brandt89,Houghton+89}
\begin{eqnarray}
\atr(\Hmii) \approx  \frac{\lambda^2}{4 \cli^2 \Lambda_T} 
\label{intro.Hmii}.
\end{eqnarray}
For $\Hmi < H < \Hmii$ the vortex lines form a solid (cf. figure
\ref{fig.phadi_th.reg}).

\begin{figure}
\centering
\epsfig{file=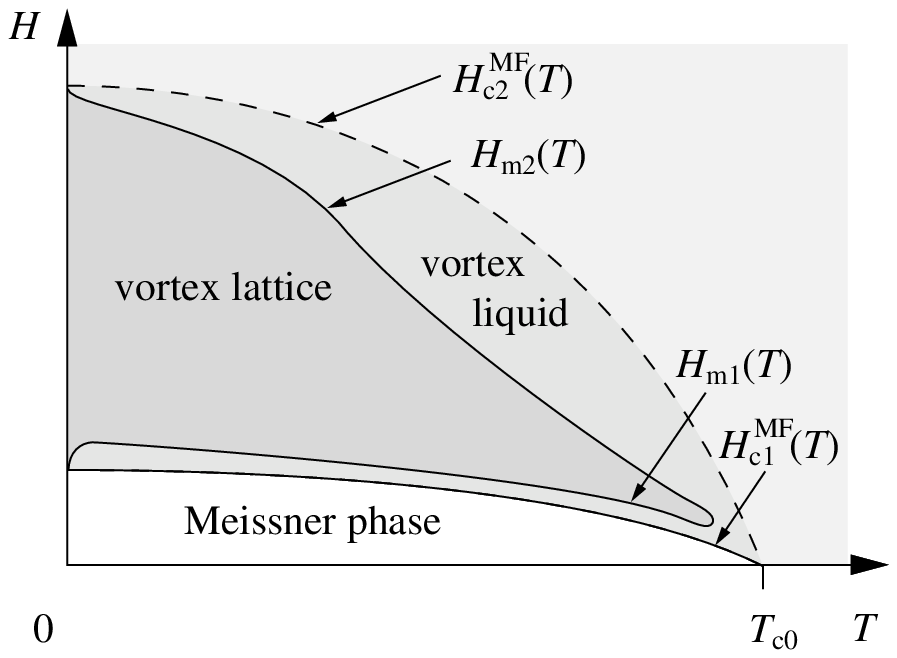,height=6cm}
\fcaption{Schematic illustration of the phase diagram of type-II
  superconductors in the $(T,H)$ plane.  In mean-field theory,
  vortices form the Abrikosov vortex lattice between the critical
  fields $\Hci^{\rm MF}$ and $\Hcii^{\rm MF}$.  Due to thermal
  fluctuations the vortex lattice is solid only between the melting
  fields $\Hmi$ and $\Hmii$ (dark shaded area) along which it melts
  into a vortex liquid.}{fig.phadi_th.reg}
\end{figure}

Alternatively, one could start directly from the GL-Hamiltonian and
study the fluctuation corrections in the vicinity of $\Hcii^{\rm MF}$
(see, e.g., \tcite{Lee+72}, \tcite{Bray74}, \tcite{Thouless75},
\tcite{Ruggeri+76}, \tcite{Ruggeri79}, \tcite{Brezin+85},
\tcite{Affleck+85}, \tcite{Brezin+90}, \tcite{Radzihovsky95:NS}).  As
first observed by \tcite{Lee+72}, the fluctuations in a
$d$-dimensional superconductor in an external field are like those of
a $(d-2)$-dimensional system in zero field, suggesting that the
mean-field phase transition to an Abrikosov phase {\em with} ODLRO is
destroyed by fluctuations in dimensions $d<d_{\rm cl}^{\rm ODLRO}=4$
(the upper critical dimension of the system is now $d_{\rm cu}^{\rm
  ODLRO}=6$).  This somewhat surprising conclusion is in agreement
with calculations of Moore \citeyear{Moore89,Moore92} and others
(\citeANP{Glazman+91:pc} \citeyear{Glazman+91:pc,Glazman+91:prb},
\tcite{Ikeda+92}), who found the {\it destruction} of ODLRO by {\it
  strong phase} fluctuations below a lower critical dimension, $d \leq
d_{\rm cl}^{\rm ODLRO}$, starting from the existence of a periodic
solution, i.e.  of a vortex lattice within the GL-theory.

Detailed considerations show that the lower critical dimension $d_{\rm
  cl}^{\rm ODLRO}=4$ for systems {\em with screening} and $d_{\rm
  cl}^{\rm ODLRO}=3$ for systems {\em without screening} (Moore
\citeyearNP{Moore89,Moore92}).  Since the aforementioned calculations
on fluctuation effects close to $\Hcii^{\rm MF}$ use the LLL
approximation which usually {\em neglects} screening, it is clear that
a simple dimensionality shift by 2 does not work here. Qualitatively,
it is plausible that screening as an additional source of fluctuations
increases $d_{\rm cl}^{\rm ODLRO}$.  A similar effect is also observed
for gauge glasses (see section \ref{subsec.infl.of.disorder}).
Quantitatively, the shift of $d_{\rm cl}^{\rm ODLRO}$ can be traced
back to the strong dispersion of the tilt modulus for $k>\lambda^{-1}$
\cite{Moore92}. (Note that the compression modulus does {\em not}
affect the formation of ODLRO.)  Breakdown of dimensionality reduction
by 2 was also shown by \tcite{Radzihovsky+96} in a layered
superconductor in a parallel field, where $d_{\rm cl}^{\rm ODLRO}=2.5$
but the upper critical dimension $d_{\rm cu}^{\rm ODLRO}=5$, and the
vortex lattice is still stable in $d \geq 2$ dimensions.

ODLRO is conventionally defined by a non-vanishing limit of the gauge
invariant pair correlation function
\begin{eqnarray}
  C_2(\br_1,\br_2;\Gamma)= \langle \Psi(\br_1) \Psi^*(\br_2) e^{i (2
    \pi/\Phi_0) \int_\Gamma d\br \cdot \bA} \rangle
\label{intro.corr.C2}
\end{eqnarray}
for $|\br_1-\br_2|\to\infty$. ($\langle \cdot \rangle$ denotes the
average over thermal fluctuations.) Note that
$C_2(\br_1,\br_2;\Gamma)$ itself depends on the path $\Gamma$ between
$\br_1$ and $\br_2$ along which the vector potential is integrated.
The proposal of \tcite{Moore92} to keep only the longitudinal
component of $\bA$ to make $C_2$ path {\em independent} but preserve
its gauge invariance corresponds in the London gauge to the complete
neglection of the phase factor $e^{i (2 \pi/\Phi_0) \int_\Gamma d\br
  \cdot \bA} $ in equation (\ref{intro.corr.C2}).

In fact, a non-vanishing asymptotic expression for
$C_2(\br_1,\br_2;\Gamma)$ if $|\br_1-\br_2|\to\infty$ may not be the
appropriate definition for the existence of long-range order in cases
in which topological defects are forced into the system by external
boundary conditions or (as in type-II superconductors) external
fields.  The most simple counter example is an Ising magnet below
$T_c$ with anti-periodic boundary conditions, which force a domain
wall into the system. Wall fluctuations will then suppress magnetic
correlations.  The loss of ODLRO due to thermal fluctuations in
type-II superconductors -- if concluded from the asymptotic behaviour
of $C_2$ -- is related to the fact that phase fluctuations
$\delta\phi(\br)$ of the order parameter are related to (shear)
distortions $\bu(\br)$ of the vortex-line lattice via
\cite{Moore92,Ikeda+92,Oneill+93,Baym95}
\begin{equation}
\delta\phi(\bk)=
\frac{2\pi i}{a^2}\frac{k_x u_y(\bk)-k_y u_x(\bk)}{k^2}.
\label{delta.varphi(bq)}
\end{equation}
Here we used the Fourier transforms $\delta\phi(\bk)$ and $\bu(\bk)$
of the fields $\delta\phi(\br)=\phi(\br)-\phi_0(\br)$ and $\bu(\br)$
and $\bk=(k_x,k_y,\bk_\parallel)$.  $\phi_0(\br)$ denotes the phase in
the ground state.  Note that equation (\ref{delta.varphi(bq)}) is not
in contradiction to equation (\ref{eq.saddle.point1}) since $\grad^2
\phi_{\rm v}(\br) \neq 0$ only at the vortex sites.  Equation
(\ref{delta.varphi(bq)}) implies that on large length scales phase
fluctuations are amplified displacement fluctuations of the
vortex-line lattice.  Destruction of ODLRO below $d=4$ dimensions does
not mean however -- this is the point of view we will take here -- the
absence of any ordered phase. This can be concluded from the London
approximation (\ref{eq.HLondon1}): Expressing $\bB(\br)$ with the help
of (\ref{eq.mod.London.eq}) in terms of the vortex degrees of freedom,
the only phase fluctuations left are spin-wave like which lead to a
destruction of conventional long-range order only in $d\le 2$
dimensions. The energy of the vortex lattice can be approximated at
low $T$ by an elastic Hamiltonian (\ref{H.el.3D}) which may be
supplemented by a contribution from dislocations.  For $d \geq 3$ such
a system shows a phase with {\em translational long-range order}
(TLRO) for $\Hmi(T) < H < \Hmii(T)$.  (In $d=2$ dimensions, elastic
fluctuations reduce the order to quasi-TLRO characterized by a
power-law decay of correlations.)  In the elastic description, which
we will use in most parts of this review, the appropriate order
parameter for TLRO is
\begin{equation}
\psi_{\bQ}(\br)=e^{i\bQ \cdot \bu(\br)}
\label{eq.psi.Q(r)}
\end{equation}
with $\langle\psi_{\bQ}(\br)\rangle\not= 0$ in the Abrikosov phase.
$\bQ$ denotes a reciprocal lattice vector of the Abrikosov lattice.
From this point of view the loss of ODLRO has no physical
significance.  $C_2$ is therefore not the appropriate quantity to
define order.  This conclusion is in agreement with a number of
numerical investigations in which a clear indication for a transition
to an ordered phase was seen in two and three dimensions.
\tcite{Hu+93} and \tcite{Kato+93:b} have considered the pair
correlation function of the superfluid density
\begin{eqnarray}
  C_4(\br_1,\br_2)=\langle |\Psi(\br_1)|^2|\Psi(\br_2)|^2  \rangle.
\end{eqnarray}
Using the LLL approximation both groups found a clear indication for a
first order melting transition in $d=2$.  Their work was extended by
\citeANP{Sasik+93} \citeyear{Sasik+93,Sasik+94:prl} to
three-dimensional systems.  They considered the helicity modulus
\begin{eqnarray}
\gamma_{\alpha \beta}(\br,\br')=
\left. \frac{\partial \langle j_\alpha(\br) \rangle}
{\partial  A'_\beta(\br')} \right|_{\bA'=\bN}=
\frac \co{4 \pi \lambda^2} \delta(\br-\br') \delta_{\alpha\beta} 
- \frac 1{\co T} \langle
j_\alpha(\br) j_\beta(\br') \rangle_c, 
\end{eqnarray}
where $\bA'$ is an additional vector potential and $\bj$ the current
density.  In the LLL (where $\lambda \to \infty$) they found a rapid
increase of $\gamma_{zz}$ at the freezing temperature, indicating the
formation of a vortex-line lattice.  \tcite{Sasik+95} have
subsequently shown that the vortex liquid-to-solid transition is {\em
  not} accompanied by a divergence of the correlation length for phase
coherence.  In their simulation $C_2$ decays exponentially even in the
solid phase in agreement with the predictions of \tcite{Moore92},
\tcite{Ikeda+92}, and \tcite{Baym95}.

It has to be mentioned, however, that the point of view we adopt in
this article -- namely that the loss of ODLRO does not rule out the
existence of a vortex lattice -- is not shared by all authors. In
particular Moore \citeyear{Moore89,Moore92,Moore97} argues that there
is no mixed phase in type-II superconductors at finite temperatures.
The observed effects in the behaviour of resistance and magnetization
are explained as cross-over phenomena which would disappear in the
thermodynamic limit. Moreover, the vortex lattices found in the
Monte-Carlo simulations mentioned before is considered to be an
artifact produced by the quasi-periodic boundary conditions used in
these simulations. Moore and co-workers
\cite{Oneill+92,Oneill+93,Lee+94,Dodgson+97,Kienappel+99,Moore+99}
have tried in their own Monte-Carlo simulations to avoid these effects
(which they consider to be crucial) by placing the two-dimensional
superconductor on the surface of a sphere.  In their studies no
freezing transition to a vortex-lattice state was observed.  It should
however be mentioned that zero-energy modes connected with the rigid
rotations of the film on the sphere and disclinations arising from
topological constraints for a triangular lattice on the sphere may
obscure the transition.

To conclude: Although the absence of ODLRO at finite temperatures in
the mixed phase of type-II superconductors (see, e.g.,
\citeNP{Houghton+90}) and its consequences for the existence of a
vortex lattice are still under debate, the most simple scenario is the
existence of a vortex lattice in the absence of ODLRO below $d=4$
dimensions.  The continuous translational invariance of the system is
reduced to finite translations by lattice vectors $\bX$ of the
Abrikosov lattice.  
In contrast to ODLRO, the lower critical dimension for TLRO in pure
systems is $d_{\rm cl}^{\rm TLRO}=2$.

Since the critical regime is enlarged in a non-zero external field (in
$d=3$) to \cite{Ikeda+89}
\begin{eqnarray}
\frac{\Tc(H)-T}{\Tc(H)} \lesssim
\Gi^{1/3} \left( \frac{\Hcii^{\rm MF}(T)}{\Hcii^{\rm MF}(0)}\right)^{2/3}
\end{eqnarray}
with respect to the zero-field critical region, but is still too small
to describe the difference between ${\Hcii^{\rm MF}(T)}$ and the
melting line, \tcite{Feigelman+93} have argued that between the
Abrikosov lattice and the normal phase there may be an intermediate
liquid phase which still shows longitudinal superconductivity.  We
will not follow this idea here since more recent extensive simulations
show only a single transition between the Abrikosov and the normal
phase \cite{Hu+97,Hu+98,Nguyen+98,Nordborg+97,Nordborg+98,Olsson+99},
although further proposals for an intermediate phase were made
recently \cite{Tesanovic99,Nguyen+99}.

In the rest of this article we will ignore the existence of these
exotic phases in pure superconductors (although their existence cannot
be ruled out completely), but concentrate instead on the influence of
randomly distributed frozen impurities on the vortex-line lattice
phase.  The impurities will be assumed to be completely uncorrelated
as already mentioned in the introduction.  Columnar or planar defects
lead to very different physics (for a recent review on the so-called
Bose glass, see \tcite{Tauber97} and references therein) and will not
be discussed in this paper.

\subsection{The influence of disorder}
\label{subsec.infl.of.disorder}

Disorder can be introduced in the Ginzburg-Landau model
(\ref{eq:GLHamiltonian}) by assuming small random contributions to all
parameters which characterize the system, i.e.  $\alpha,\;\beta,\;m$
and $\bH$. We will restrict ourselves here to the case of systems with
random mean-field transition temperature $\Tco$, i.e. we substitute
\begin{equation}
\Tco \to \Tco + \delta \Tco(\br)
\label{system+rf}
\end{equation}
with
\begin{equation}
\overline{\delta \Tco(\br)\,\delta \Tco(\br^{\prime})}=
\xi^2\,\delta_{\xi}(\br -\br^{\prime})\,\delta \Tco^2
\label{overbar.delta.T}
\end{equation}
where the overbar denotes the disorder average and $\delta_{\xi}(\br)$
is a $\delta$-function of widths $\xi$.

In mean-field theory, a randomness in $\Tco$ would correspond to a
smearing out of the transition. However, as first shown by
\tcite{Khmelnitskii75}, thermal fluctuations permit the occurrence of
a sharp phase transition. According to the so-called Harris criterion
\cite{Harris74} weak randomness does not change the critical behaviour
if the exponent of the specific heat of the pure system $\alpha_{\rm
  pure}$ is negative, or, equivalently, if the correlation length
exponent $\nu_{\rm pure}$ obeys $\nu >2/d$.  Since for the $XY$-model
in three dimensions $\alpha$ is negative \cite{Lipa+96}, the
zero-field critical behaviour for type-II superconductors should be
unchanged in the outer critical region. This would also apply in case
that behaviour of the pure system in the inner critical region is also
of (inverted) $XY$-type. On the contrary, for type-I superconductors
it was shown by \tcite{Boyanovsky+82} that a sufficiently large amount
of disorder may convert the first order transition into a second order
one.

If an external field is applied, the situation is quite different.
Because of the problems connected with ODLRO even in pure systems,
which were discussed in the previous chapter, it seems to be expedient
to first consider the influence of disorder on the structural
properties of the mixed phase.

Inside the Abrikosov phase, disorder leads to a destruction of
translational long-range order as first shown by \tcite{Larkin70:t}.
This follows from the fact that a randomness in the local value of
$T_c(\br)$ leads to a random potential acting on the vortices.  The
specific form of the resulting coupling of the disorder to the vortex
displacements will be given in the following chapters.  The disorder
averaged order parameter $\overline{\langle e^{i\bQ \cdot
    \bu}\rangle}$ for TLRO vanishes since disorder leads to diverging
displacement $\bu$ in the limit of large system sizes.  However, it
will be shown below that the correlation function
\begin{equation}
S(\bQ,\br)=\overline{\langle e^{i\bQ \cdot[\bu(\br)-\bu(\bf 0)]}\rangle}
\label{eq.D2}
\end{equation}
may still obey an algebraic decay, which shows up in Bragg peaks in
the structure factor (Giamarchi and Le Doussal
\citeyearNP{Giamarchi+94,Giamarchi+95}, \citeNP{Emig+99}).  Moreover,
in analogy with spin glass theory (see \tcite{Binder+86}) one may
consider the {\em positional} glass correlation function
\begin{equation}
S_{\rm PG}(\bQ,\br)=\overline{\Big|\langle 
e^{i\bQ \cdot[\bu(\br)-\bu(\bf 0)]}\rangle\Big|^2}
\label{eq.D4}
\end{equation}
as another signature of the existence of some residual, `glassy' order
of the Abrikosov lattice.  Below we will call a system a {\em
  positional vortex glass} if $S_{\rm PG}(\bQ,\br)$ decays not faster
than a power law for $|\br| \to \infty$.  For $T\to 0$, $S_{\rm
  PG}(\bQ,\br)$ approaches one.  At non-zero temperature,
(\ref{eq.D4}) measures the strength of thermal fluctuations of the
vortex lines around the ground state.  Two limiting cases seem to be
conceivable: If the disorder acts effectively as a random force on the
vortex lines, then the thermal fluctuations around the disordered
ground state are identical to those of the pure system. In this case
(\ref{eq.D4}) is non-zero above $d=2$ dimensions. In the opposite case
of very strong pinning the vortex lines can be considered to fluctuate
thermally in a kind of narrow parabolic potential and (\ref{eq.D4})
may be finite even in $d<2$. The relevance of these structural
correlation functions for the glassy behaviour of the mixed phase will
be further discussed in the following chapters.

A complementary discussion of the glassy behaviour was proposed by M.
P. A. Fisher \citeyear{FisherMPA89} and \tcite{FisherDS+91:tf}. These
authors use the correlation function for ODLRO as the starting point.
Clearly, because of the relation (\ref{delta.varphi(bq)}),
$\overline{\langle\Psi(\br)\Psi^{\ast}({\bf 0})\rangle}$ will vanish
for $r\to\infty$ in $d<4$ dimensions.  In analogy to spin glasses they
therefore introduce the correlation function for {\it phase-coherent
  vortex-glass order} (using the London gauge $\grad \cdot \bA=0$):
\begin{equation}
C_{\rm VG}(\br)=
\overline{\big|\big<\Psi^{\ast}(\br)\Psi({\bf 0})\big>\big|^2}\,.
\label{C_VG}
\end{equation}
It is instructive to consider (\ref{C_VG}) in the London limit where
$C_{\rm VG}(\br)\approx|\Psi_0|^2 \overline{|\langle
  e^{i[\delta\phi(\br)- \delta\phi({\bf 0})]}\rangle|^2}$.
$\delta\phi(\br)$ describes the fluctuations around the ground state
pattern $\phi_0(\br)$.  If the disorder is of random-force type,
according to (\ref{delta.varphi(bq)}) thermal fluctuations should
destroy $C_{\rm VG}(\br)$ below $d=4$. On the other hand for thermal
fluctuations in a parabolic potential around the ground state, $C_{\rm
  VG}(\br)$ should be finite.

\tcite{Dorsey+92} and Ikeda \citeyear{Ikeda96,Ikeda96:t} have
calculated the vortex glass susceptibility $\chi_{\rm VG}=\int d^dr \ 
C_{\rm VG}(\br)$ for a GL-Model with random transition temperature in
the LLL approximation.  \tcite{Dorsey+92} found in a mean-field
calculation a second-order transition with a diverging vortex-glass
susceptibility by approaching the vortex-glass transition temperature
$\Tg(H)$, which is slightly below $T_{\rm c2}(H)$.  In mean-field
theory, the vortex-glass correlation length $\xi_{\rm VG}$ diverges
with a mean-field exponent $\nu^{\rm MF}_{\rm VG}=1/2$.  From this one
can expect the existence of a non-zero limit of $C_{\rm VG}(\br)$ for
$|\br|\to\infty$ below $\Tg(H)$. Taking critical fluctuations into
account, which can be done within a $d=6-\epsilon$ expansion, the
model is found to be in the some universality class as the Ising spin
glass.  In particular, the dynamical critical exponent is found to be
$z=2(2-\eta)$, where $\eta\simeq -\epsilon /6$.  It should be observed
that a similarity between the Ising spin glass and vortex glasses was
also mentioned for so-called gauge-glass models (see
\tcite{Huse+90}).  On the other hand, the results of
\tcite{Dorsey+92} and Ikeda \citeyear{Ikeda96,Ikeda96:t} -- if applied
to $d=3$ dimensions -- cannot be so easily reconciled with the
findings from the elastic description of a vortex lattice in a random
potential. As we will discuss in chapter \ref{sec.bulk}, the vortex
lattice exhibits indeed a glassy phase (the Bragg glass) for $2<d\leq
4$, in which the correlation function $C_{\rm VG}(\br)$, if calculated
to lowest order in $\epsilon=4-d$, vanishes for large $|\br|$
exponentially.  The reason for this decay consists in the strong
thermal fluctuations of the phases around the (distorted) ground
state. To order $\epsilon$, these phase fluctuations are only weakly
suppressed by disorder and hence $C_{\rm VG}(\br)$ decays
exponentially. However, higher order terms in $\epsilon$ may still
change this result. In principle, there could be two glassy phases,
which show a non-vanishing $S_{\rm PG}(\bQ,\br)$ and $C_{\rm VG}$,
respectively.  For the moment we consider it to be more likely that
there is only one glassy phase and the discrepancy between the results
follows from the use of different approximations valid in
$d=6-\epsilon$ and $d=4-\epsilon$ dimensions, respectively.

So far we assumed that the disorder is weak. On the other hand
gauge-glass like models were proposed for the description of {\em
  granular superconductors} or systems with {\em strong disorder}
\cite{Ebner+85,John+86}.  Each superconducting grain of centre position
$\br_i$ is described by the phase $\phi_i$ of the order parameter
which is assumed to be constant within a grain. The Hamiltonian then
reads (see e.g. \tcite{Wengel+96})
\begin{equation}
\cH=-J\sum\limits_{<ij>}\cos{(\phi_i-\phi_j-A_{ij}-\lambda_0^{-1}a_{ij})}+
\frac{1}{2}\sum\limits_{\Box}(\grad\wedge\ba)^2\,,
\label{eq.Hwengel}
\end{equation}
where $a_{ij}=\int\limits_{\br_j}^{\br_i}\ba(\br)\,d\br$.
$\sum\limits_{<ij>}$ is the sum of all nearest neighbors of a cubic
lattice and $\sum\limits_{\Box}$ is the sum over all plaquettes.
$\ba$ denotes the fluctuations of the vector potential which are
limited by the bare screening length $\lambda_0$. The influence of the
external field as well as the contribution from randomness are assumed
to be included in the $A_{ij}$ which are taken to be independent
random variables with a distribution between 0 and $2\pi$. A detailed
discussion of the relation between the vortex glass (the expression is
here understood in the sense that it describes the glassy phase of an
impure type-II superconductor in an external field) and gauge glasses
is given in \tcite{Blatter+94}. The model (\ref{eq.Hwengel}) is in
particular isotropic in contrast to the GL-Hamiltonian which shows a
pronounced anisotropy due to the presence of the external field $\bH$.
In addition, the disorder of the gauge glass has a nature which is
completely different from the one in equation (\ref{system+rf}). The
first one couples to the vortices via the phase of the order
parameter, while the latter one couples only via the amplitude.  As a
consequence, the gauge-glass disorder distorts the vortices much more
than local variations of $\Tco$.

For $\lambda_0\to\infty$, i.e. in the absence of screening, the gauge
glass was investigated numerically by a number of authors (see
\tcite{FisherMPA+91}, \tcite{Gingras92}, \tcite{Cieplak+91},
\tcite{Reger+91}, \tcite{Cieplak+92}, \tcite{Hyman+95},
\tcite{Maucourt+98}, \tcite{Kosterlitz+98}, \tcite{Olson+99}).  While
in two dimensions a transition to a glass phase is found to be only at
$T=0$, this transition takes place at finite temperatures in three
dimensions.  \tcite{Huse+90} found at the transition a diverging
gauge-glass susceptibility
\begin{equation}
\chi_{\rm GG}=\sum\limits_jC_{\rm GG}(\br_i-\br_j)\,,
\label{eq.chiGG}
\end{equation}
where $C_{\rm GG}$, in analogy with (\ref{C_VG}), denotes the gauge 
glass correlation function
\begin{equation}
C_{\rm GG}(\br_i-\br_j)=\overline{\big|\langle e^{i(\phi_i-\phi_j)}\rangle
\big|^2}\,.
\label{eq.C.GG}
\end{equation}
The divergence of $\chi_{\rm GG}$ may signal the transition to a phase
with non-zero limit of $C_{\rm GG}(\br)$ for $|\br|\to\infty$.  If one
assumes (\ref{C_VG}) as the definition of vortex glass order, then the
gauge glass would be a vortex glass.

More recently, the case of a finite $\lambda_0$ was considered
\cite{Bokil+95,Wengel+96,Wengel+97,Kisker+98}. It turns out that
screening seems to destroy the gauge glass transition in three
dimensions. More recently \tcite{Kawamura99} and \tcite{Pfeiffer+99}
have attempted to include the effect of anisotropy into the gauge
glass model by assuming an extra contribution to $A_{ij}$ arising from
the external field.  However, also in this case no finite temperature
glass transition was found.

  \setcounter{equation}{0}
  \section{Directed elastic manifolds in a random potential}
\label{sec.ran.man}

In the course of this article we will restrict our consideration of
fluctuations to vortices and we will ignore other independent
fluctuations such as those of the condensate amplitude or of the
magnetic field.  Vortices will be treated within the London picture,
where vortex lines are represented as string-like objects.  For low
temperatures and weak disorder the vortex lines will fluctuate
only weakly around the ground state of the vortex-line lattice (VLL),
the Abrikosov lattice, where all lines are directed along the magnetic
field.

Before we analyze the VLL as an ensemble of vortex lines in its full
complexity it is instructive to study a {\em single} vortex line.  In
general, a vortex can be characterized by two dimensions which depend
on the physical realization under consideration: the `internal'
dimension $D$\label{intro.D} of the vortex considered as a `directed
manifold', and the number $N\label{intro.N}$ of its displacement
components.  For example, a vortex line in a bulk superconductor has
$(D,N)=(1,2)$, a single vortex line in a superconducting film has
$(D,N)=(1,1)$, and a point vortex in a film corresponds to
$(D,N)=(0,2)$.  This concept of elastic manifolds also applies to
other physical systems such as interfaces between magnetic domains,
for which $(D,N)=(2,1)$.  In all these examples the spatial dimension
of the system is $d=D+N$\label{intro.d}, since displacements are
possible only in directions orthogonal to the $\bz$ direction(s) along
which the manifold is spanned.

It is worthwhile to mention at this point that the analysis of random
manifolds applies not only to {\em single} vortex lines, but to a
certain extent (i.e., within a regime of length scales) also to vortex
line {\em lattices}.  For example, the VLL in a weakly disordered bulk
superconductor resembles over a large range of length scales an
elastic manifold with $(D,N)=(3,2)$.  In this case, where the elastic
manifold is spanned in all spatial dimensions including the
displacement directions, $d=D$.  Nevertheless, there is a fundamental
difference between manifolds and VLLs, which is crucial for the
physics on large scales: VLLs have a periodic structure in contrast to
manifolds.

To parameterize vortex conformations, we use the
$D$-dimensional coordinate $\bz$ along the direction of the magnetic
field and a $N$-dimensional coordinate $\bx \equiv \bx(\bz)$ in
transverse directions, see figure \ref{fig.mani.coord}.

\begin{figure}
\centering
\epsfig{file=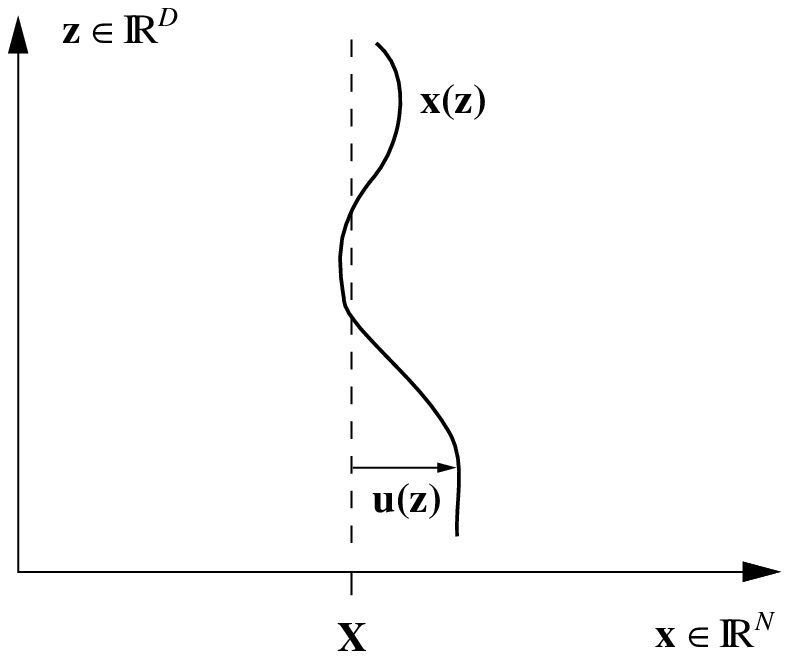,height=6cm}
\fcaption{Illustration of the manifold coordinate system.  The manifold
  conformation (solid line) is described by the $N$-component vector
  $\bx$ as a function of the $D$-dimensional coordinate $\bz$. The
  displacement $\bu(\bz) \equiv \bx(\bz)-\bX$ is defined with respect
  to a flat reference position $\bX$ (dashed line).}
{fig.mani.coord}
\end{figure}

In this chapter we are mainly interested in qualitative aspects of
vortex fluctuations in the presence of disorder.  To this end we
describe the elastic energy of a vortex line, which arises from the
kinetic energy of the supercurrents and the magnetic field energy, in
a harmonic approximation (i.e., we keep only terms of second order in
the displacement) and we ignore non-localities or possible
anisotropies in the elastic stiffness constant $\st$. Then the elastic
energy can be written as
\begin{eqnarray}
{\cal H}_{\rm el}&=& \int d^D z \ \frac \st 2 (\grad_\parallel {\bf x})^2 .
\label{H.el.mani}
\end{eqnarray}
Therein the gradient $\grad_\parallel \equiv (\frac \partial {\partial
  z_1},\dots,\frac \partial {\partial z_D})$ acts along the
longitudinal $\bz$ directions.

Pinning of vortex lines due to the presence of impurities, grain
boundaries etc. can be described by a potential $V(\bx,\bz)$ which is
`quenched', i.e., frozen on the relevant time scale for vortex
fluctuations.  It yields a contribution
\begin{eqnarray}
{\cal H}_{\rm pin}&=& \int d^D z \ V({\bf x}(\bz),{\bf z})
\end{eqnarray}
to the total energy $\cH={\cal H}_{\rm el} + {\cal H}_{\rm pin}$ of a
vortex line.  For simplicity, we will discuss here only point-like
disorder. We will assume that this potential is Gaussian distributed
with zero average and variance
\begin{eqnarray}
\overline{V({\bf x},{\bf z})V({\bf x}',{\bf z}') } &=&
\Delta({\bf x}-{\bf x}') \delta({\bf z}-{\bf z}') .
\label{intro.Delta}
\end{eqnarray}
For simplicity we will restrict ourselves to disorder with
short-ranged and isotropic correlations. Then the correlator
$\Delta(\bx)=\Delta(|\bx|)$ can be characterized by its integral
\begin{eqnarray}
\Do &\equiv& \hD(\bN)= \int d^N x \ \Delta(\bx)
\label{intro.Delta_0}
\end{eqnarray}
and a correlation length.  Such a pinning potential can arise from
point-like impurities that locally suppress the density of the
superconducting condensate.  The correlation length of the disorder
then also coincides with the superconducting coherence length $\xi$.
In principle, one can retain a finite correlation length in the $\bx$
and $\bz$ directions.  However, it turns out a posteriori that -- as
long as the correlation length does not exceed the smallest scale of
vortex conformations, which also is $\xi$ -- correlations in $\bz$
direction are irrelevant.  Subsequently, we will often write
$\Delta({\bf x})= \Do \delta_\xi({\bf x})$, where $\delta_\xi$ denotes
a Dirac delta-function smeared out on a scale $\xi$.  For
semi-quantitative purposes we occasionally use $\delta_\xi({\bf x})=
(2 \pi \xi^2)^{-N/2} \exp(-\bx^2/2\xi^2) $.  If we denote the pinning
force of an individual impurity by $f_{\rm pin}$, then
\begin{eqnarray}
  \Do \approx f_{\rm pin}^2 n_{\rm i}^{(d)} \xi^{2N+2},
\end{eqnarray}
where $n_{\rm i}^{(d)}$ denotes the concentration of the impurities in
the $d$ dimensional space.

Directed manifolds in disorder represent a paradigm for systems which
are dominated by disorder.  Since disorder leads to a frustrating
competition between elastic and pinning energies, the structural order
can be reduced substantially.  In addition, the dynamics of the system
can get extremely slow due to the quenched nature of disorder.
Therefore such a system can be called a `glass'.  Because of the
simplicity of the manifold model, it plays a paradigmatic role for
glassy systems and it allows the identification and the understanding
of many characteristic features of a more complicated `vortex glass'.

\subsection{Equilibrium properties}

As we have seen, vortices in impure superconductors are a realization
of `elastic manifolds' in `random media' which are paradigms for
disordered systems.  We now proceed to summarize briefly some key
properties of the latter model system.  For more detailed
presentations about these random manifolds the reader is referred to
review articles \cite{Kardar94,Halpin+95,Lassig98}, in particular to
\tcite{Blatter+94} for a discussion in the context of vortex systems,
as well as to \tcite{Nattermann+89}, \tcite{Belanger+91} and
\tcite{Young98} in the context of disordered magnets.  Subsequently we
will focus our attention on the question, in what physical quantities
`glassiness' appears in these systems.

From a principal point of view, it is important to note that disorder
breaks symmetries of the Hamiltonian of the pure system: {\em
  translation} invariance ${\bf x}({\bf z}) \to {\bf x}({\bf z}) +
{\bf x}_0$ for a constant shift ${\bf x}_0$ of vortex lines, as well
as an analogous {\em rotation} symmetry ${\bf x}({\bf z}) \to {\bbox
  {\cal R}} \cdot {\bf x}({\bf z})$, for a rotation matrix ${\bbox
  {\cal R}}$.  Therefore it is interesting to examine to what extent
the physical state of the manifold in disorder actually reflects these
broken symmetries, or whether these symmetries are restored due to
thermal fluctuations.

\subsubsection{Structure}
\label{sec.struct}

One quantity of primary interest is the structure of the manifold in
disorder which can be described in terms of displacement correlations.
In the absence of disorder the vortex line has a flat ground state,
$\bx(\bz) \equiv \bX$ for all $\bz$.  Introducing the displacement
${\bf u}({\bf z}) \equiv {\bf x}({\bf z}) - {\bf X}$, shape
fluctuations can be described by the relative displacement at points
separated by a distance ${\bf z}-{\bf z}'$ parallel to the magnetic
field:
\begin{eqnarray}
  W({\bf z}-{\bf z}') &\equiv & 
\overline{\langle [{\bf u}({\bf z})-{\bf u}({\bf z}') ]^2 \rangle} 
\sim |{\bf z}-{\bf z}'|^{2 \zeta}.
\label{intro.W}
\end{eqnarray}
As already introduced above, $\langle \dots \rangle$ and
$\overline{\dots \rule{0pt}{5pt}}$ denote the thermal and disorder
average respectively. For large distances this quantity typically
follows a power law with the roughness exponent $\zeta=\zeta(D,N)
\label{intro.zeta}$.  The manifold is called {\em flat}, if $W$ is
finite for $|{\bf z}-{\bf z}'| \to \infty$ (then the convergence of
$W$ to its asymptotic value can be described by an exponent $\zeta
<0$), whereas it is called {\em rough} if $W$ diverges (i.e., $\zeta
\geq 0$, where $\zeta=0$ typically corresponds to $W(\bz) \sim
\ln^\alpha z$ with some power $\alpha$).

In the absence of disorder thermal fluctuations are described by an
exponent
\begin{equation}
\zeta_{\rm th}=\frac{2-D}2 .
\label{zeta.th}
\end{equation}
The presence of disorder may increase the roughness and lead to a
larger exponent $\zeta>\zeta_{\rm th}$.  In particular, disorder
always induces roughness in dimensions $2<D \leq 4$, as we will discuss
in more detail below.

For $D<2N/(2+N)$ disorder is irrelevant at sufficiently high
temperatures. There is a phase transition (see e.g.  \tcite{Imbrie+88}
for the case $D=1$) between a low-temperature phase, where the
manifold is disorder dominated and essentially shows the structure of
the ground state, and a high-temperature phase, where the manifold is
entropically driven out of the ground state and shows a structure as
in the absence of disorder.

Actually, the physical situation can be more complicated if an upper
critical dimension $N_{\rm cu}$ exists, above which the
low-temperature phase is governed by the Gaussian exponents.  The
existence of an upper critical dimension is still 
controversial. \tcite{Moore+95:kpz} argue for $N_{\rm cu}=4$ and
\citeANP{Lassig+97} \citeyear{Lassig+97,Lassig+98} argue for $N_{\rm
  cu} \leq 4$ in $D=1$.  We will not further discuss this possible
complication here, assuming $N_{\rm cu}$ to be large enough such that
the vortex systems of physical interest are not concerned.

For $D \geq 2N/(2+N)$ one can think of the manifold as having a unique
disorder-dominated ground state.  For all temperatures entropic
effects are too weak to detach the manifold from its ground state.

\subsubsubsection{\ref{sec.struct}a Structural order parameter}

In order to have a tool to quantify the structural order of the
manifold, we introduce
\begin{eqnarray}
\psi_{\bf k}({\bf z})&\equiv& e^{i {\bf k} \cdot {\bf u}({\bf z})}
\label{intro.psi_k}
\end{eqnarray}
as order parameter.  If the manifold performs only weak thermal
fluctuations around its ground state, $\langle \psi_{\bf k}({\bf z})
\rangle \neq 0$ and disorder actually breaks the translation symmetry
of the manifold.  Otherwise, if thermal fluctuations detach the
manifold from its ground state, $\langle \psi_{\bf k}({\bf z})\rangle
= 0$.  The correlation function of this order parameter,
\begin{eqnarray}
S(\bk,\bz -\bz') \equiv
\overline{\langle \psi_{\bf k}^*({\bf z}) \psi_{\bf k}({\bf z}') \rangle}
\approx \exp\left(- \frac 12 {\bf k} \cdot {\bf W}({\bf z}-{\bf z}')
\cdot {\bf k}\right),
\label{intro.Psi_k}
\end{eqnarray}
is related to the displacement correlation
\begin{eqnarray}
W_{\alpha \beta}({\bf z}-{\bf z}') &\equiv & 
\overline{\langle [u_\alpha({\bf z})-u_\alpha({\bf z}')] 
[u_\beta({\bf z})-u_\beta({\bf z}')]\rangle}. 
\label{intro.W.mat}
\end{eqnarray}
The function $W$, previously introduced in (\ref{intro.W}), is simply
the trace of the matrix $W_{\alpha \beta}$.  The approximate relation
in (\ref{intro.Psi_k}) 
neglects higher cumulants of the displacement distribution and
holds in general only for small displacement
fluctuations.  It even holds for large displacement fluctuations and
actually is an identity, if the fluctuations of ${\bf u}$ have
Gaussian distribution.

\subsubsubsection{\ref{sec.struct}b Perturbative analysis}

A first qualitative insight into the relevance of disorder can be
obtained from an elementary perturbative analysis.  Such an analysis
was performed originally by \tcite{Larkin70:t} for vortex lattices and
by \tcite{Efetov+77:t} for the closely related charge-density waves.

In this approach the manifold (at $T=0$) is considered in an
absolutely flat reference state $\bx(\bz)=\bX$, for which the pinning
force ${\bf F}^{\rm pin}({\bf z}) \equiv - \grad_\perp V({\bf
  x}(\bz),{\bf z})$ is calculated. We denote by $\grad_\perp \equiv
(\frac \partial {\partial x_1},\dots,\frac \partial {\partial x_N})$
gradients in $\bx$ directions in contrast to $\grad_\parallel$ for
$\bz$ directions.  From this force the disorder induced manifold
displacement is obtained in linear response theory as $u_\alpha(\bq)
\equiv \int d^Dz \ e^{- i \bq \cdot \bz} u_\alpha (\bz) =G_{\alpha
  \beta}({\bf q}) F^{\rm pin}_\beta(\bq)$ using the response function
\begin{eqnarray}
G_{\alpha \beta}({\bf z}-{\bf z}') \equiv
\frac{\delta u_\alpha({\bf z})}{\delta F^{\rm pin}_{\beta}({\bf z}')},
\quad \quad
G_{\alpha \beta}({\bf q})=
\frac 1 {\st q^2} \delta_{\alpha \beta}
\label{intro.G}
\end{eqnarray}
of the free manifold. Thus the displacement correlations are given in
linear response by
\begin{eqnarray}
\overline{\langle u_\alpha({\bf q}) u_\beta(-{\bf q}) \rangle} 
= G_{\alpha \gamma}({\bf q}) \ \overline{ F^{\rm pin}_\gamma(\bq)
  F^{\rm pin}_\delta(-\bq)} \ G_{\beta \delta}(-{\bf q}) = \frac
{\Delta^{(2)}}{\st^2 q^4} \delta_{\alpha \beta},
\label{C.rf}
\end{eqnarray}
where we introduced the variance of the pinning force $\Delta^{(2)}$
via
\begin{eqnarray}
\Dii \delta_{\alpha \beta} \equiv
-\partial_\alpha \partial_\beta \left. \Delta({\bf x}) \right|_{{\bf
    x}={\bf 0}}.
\label{intro.Delta2}  
\end{eqnarray}
Its value is related to the variance and correlation length of the
pinning potential approximately through
\begin{eqnarray}
\Dii = \frac{\Do}{(2 \pi)^{N/2}\xi^{2+N}}.
\label{intro.Dii.sec4}
\end{eqnarray}
Here we have assumed a Gaussian form for $\Delta(\bx)$.

The perturbative correlation function (\ref{C.rf}) is characterized by
a roughness exponent 
\begin{equation}
\zeta_{\rm rf}=\frac{4-D}2,
\label{zeta.rf}
\end{equation}
which we refer to as the `random force' value. This exponent
characterizes the actual correlation function only on sufficiently
small scales $|\bz| \lesssim L_\xi$ below the Larkin length $ L_\xi$
\cite{Larkin70:t}, since the perturbative treatment is justified only
as long as $W(\bz) \lesssim \xi^2$.  The Larkin length can be
estimated by equating the elastic energy $E_{\rm el} \approx \st
L_\xi^{D-2} \xi^2$ with the pinning energy in the random-force
approximation $E_{\rm pin}\approx (L^D \xi^2 \Dii)^{1/2}$ as
\cite{Larkin70:t,Bruinsma+84}
\begin{eqnarray}
L_\xi \approx [\st^2 \xi^2/\Dii]^{1/\epsilon},
\label{intro.L_xi}
\end{eqnarray}
where $\epsilon \equiv 4-D$ (to be distinguished from $\st$).  From
this rough analysis the manifold is found to be flat in $D>4$ (since
$\zeta_{\rm rf}<0$), logarithmically rough in $D=4$ (with $\zeta_{\rm
  rf}=0$), and rough with an exponent $\zeta_{\rm rf}>0$ in $D<4$.
Although the perturbative analysis can provide a good approximation
only on small length scales $|\bz| \lesssim L_\xi$ (however, in
certain cases intermittency can be relevant on small scales
\cite{Bouchaud+95}), the roughness of the manifold in $D \leq 4$
persists in more sophisticated approaches (such as a self-consistent
or renormalization-group analysis), which go beyond a perturbative
approach.

Thus, a single vortex line ($D=1$) in a bulk superconductor ($d=3$),
which is roughened by pure thermal fluctuations (cf. equation
(\ref{zeta.th})), is also roughened by disorder at $T=0$. In contrast,
a VLL in a bulk superconductor, which can be considered as a manifold
with $D=d$, is not roughened by pure thermal fluctuations but by
pinning.

\subsubsubsection{\ref{sec.struct}c Flory analysis}

As already mentioned, the above perturbative analysis breaks down on
length scales $|\bz| \gtrsim L_\xi$ because perturbation theory does
not adequately treat a system with many minima in the potential energy
(for a pedagogical example see \tcite{Villain+83}).  Although
perturbation theory could be continued to higher orders, it will never
be able to describe the displacements on largest scales $|\bz| \to
\infty$, where the multi-stability (existence of many local minima) of
the potential energy landscape is crucial.

For the further analysis it is convenient to start from the replica
Hamiltonian \cite{Edwards+75}
\begin{eqnarray}
\cH_n &=& \cH_{{\rm el},n} + \cH_{{\rm pin},n} 
\nonumber \\ 
&=&\sum_{a,b =1}^n \int d^Dz \ \left\{
\frac \st 2 \delta^{ab} (\grad_{\bf z} {\bf u}^a)^2
- \frac 1 {2T} \Delta(\bu^a(\bz) - \bu^b(\bz)) \right\},
\label{H.rep.mani}
\end{eqnarray}
which is obtained from replicating the original system $n$ times and
performing a disorder average.  Note that Greek lower indices denote
transverse spatial components, whereas Roman upper indices denote
replicas.  The analysis of this Hamiltonian is highly non-trivial,
since the displacement enters the argument of the disorder correlator
$\Delta$.  A dimensional analysis shows that $\Delta$ has to be
retained in its full functional form and may not be represented by a
truncated Taylor expansion in $D \leq 4$ \cite{Balents+93}.

Physically, the structure on large scales is determined by a
competition between elastic and pinning energies.  The Flory
argument \cite{Imry+75,Kardar87,Nattermann87} allows one to obtain an
improved value for the roughness exponent $\zeta$ by requiring both
energy contributions to scale with the same exponent on large scales
$|\bz|$.  To be more precise, we rescale $\bz= L \bz'$ and
$\bu(\bz)=L^\zeta \bu'(\bz')$, where $L$ denotes the (variable) length
scale on which we consider the system. Then $\cH_{{\rm el},n} \sim
L^\theta$ with an energy scaling exponent
\begin{eqnarray}
\label{intro.theta}
\theta = 2 \zeta + D-2  .
\end{eqnarray}
The fluctuations of $\bu'$ are determined by an effective temperature
$T'$, which has to be rescaled like the Hamiltonian,
\begin{eqnarray}
T=T' L^{\theta},
\end{eqnarray}
in order to keep the Boltzmann factor $e^{-\cH/T}$ scale invariant.

In {\em pure} systems it is possible to achieve scale invariance not
only of the Boltzmann factor $e^{-\cH_{\rm el}/T}$ but of both
$cH_{\rm el}$ and $T$ by the choice $\zeta=\zeta_{\rm th} \equiv
\frac{2-D}2$ and $\theta_{\rm th}=0$.

At {\em zero} temperature, weak disorder is a relevant perturbation in
$D \leq 4$. Since a short-ranged disorder correlator should
essentially scale like an $N$-dimensional $\delta$-function, the
replicated pinning energy scales as $\cH_{\rm pin}/T \sim L^{D-N \zeta
  - 2 \theta}/{T'}^2$. The exponent of the Boltzmann factor includes
now two terms, which after rescaling behave as $\cH_{\rm el}/T \sim
\cH_{\rm el}'/T' = O(1)$ and $\cH_{\rm pin}/T = O(L^{D-N \zeta - 2
  \theta})$. In the limit of vanishing temperatures, a finite width of
the distribution of $\{\bu'\}$ is only possible if $L^{D-N \zeta - 2
  \theta}= O(1)$, i.e., if
\begin{equation}
\zeta= \zeta_{\rm F} \equiv \frac{4-D}{4+ N},
\label{zeta.F}
\end{equation}
which implies $\theta=2 \zeta + D-2= \frac{2+N}{4+N}(D-D_N)$, where
$D_N \equiv \frac{2N}{2+N}$.  Although $\zeta_{\rm F}$ is an
improvement over $\zeta_{\rm rf}$, this result is (in general) not
exact, since the scaling behaviour of $\cH_{\rm pin}$ was
over-simplified.

On the other hand, at {\em finite} temperatures $e^{-\cH_{{\rm
      pin},n}/T}$ is a relevant perturbation to the Boltzmann factor
$e^{-\cH_{\rm el}/T}$ if $D-N \zeta_{\rm th} - 2 \theta_{\rm th} >0$,
i.e., for
\begin{eqnarray}
  D>D_N \equiv \frac{2N}{2+N},
\label{intro.D_N}
\end{eqnarray}
which is equivalent to $\zeta_{\rm F} > \zeta_{\rm th}$. Note that
$D_N<2$ for $N<\infty$ and $D_\infty=2$. From this observation we
conclude that for $D<D_N$ {\em weak} disorder is irrelevant. Since on
the other hand disorder will certainly become relevant if it is
sufficiently strong, $\Delta> \Delta_c(T)$, one expects in this case a
transition from an unpinned phase for weak disorder to a pinned phase
at strong disorder (which is equivalent to a thermal depinning
transition for increasing temperatures). Contrary to the result for
$\zeta_{\rm F}$, which is an approximate expression for the true
roughness exponent $\zeta$, the result for $D_N$ is exact.

\subsubsubsection{\ref{sec.struct}d Renormalization group analysis}

The algebraic roughness (\ref{intro.W}) of the manifold means that it
is scale invariant on large length scales.  Hence it does not have a
finite correlation length and the manifold can be considered as being
in a critical state with $\zeta$ corresponding to a critical exponent.
In analogy to ordinary critical phenomena, a renormalization group
(RG) analysis is suitable to describe the large-scale features of the
system going beyond perturbation theory and self-consistency
arguments.

In $D\leq 4$ it is not possible to describe pinning by a finite set of
parameters since $\zeta>0$ and a Taylor expansion of $\Delta(\bu)$
would yield terms the relevance of which increases with increasing
order of the expansion.  Therefore one has to use a {\em functional}
renormalization group analysis for the present problem, which was
established by \citeANP{FisherDS86:dr}
\citeyear{FisherDS85:pt,FisherDS86:dr} and \tcite{Balents+93} to first
order in $\epsilon \equiv 4-D$.  On increasing length scales $L
\propto e^l$ the system can be described by a renormalized temperature
and disorder correlator, which flow according to
\cite{FisherDS86:dr,Balents+93}:
\begin{mathletters}
\begin{eqnarray}
\partial_l T &=& - \theta T,
\label{flow.rm.T}
\\
\partial_l \Delta({\bf u}) &=& (\epsilon-4 \zeta)   \Delta({\bf u}) 
+ \zeta u_\alpha \partial_\alpha \Delta({\bf u}) 
\nonumber \\ &&
+ \frac 12\partial_\alpha \partial_\beta \Delta({\bf u})
\partial_\alpha \partial_\beta \Delta({\bf u})
- \partial_\alpha \partial_\beta \Delta({\bf u})
\partial_\alpha \partial_\beta \Delta({\bf 0}).
\label{flow.rm.D}
\end{eqnarray}
\end{mathletters}
Here we introduced a rescaled correlator $\Delta/c_\Delta \to \Delta$
with some non-universal constant $c_\Delta$ that depends on the
short-scale cutoff and the stiffness constant $\st$.

We will see that $\theta=D-2+2\zeta>0$ such that equation
(\ref{flow.rm.T}) implies that the effective temperature vanishes on
large scales.  The system is therefore described asymptotically by a
`zero temperature' fixed point. The fixed-point correlator and
exponent have to be determined numerically from equation
(\ref{flow.rm.D}).

The actual value of the roughness exponent (`random manifold' value)
can be represented introducing a correction factor $\nu(D,N)$ in the
Flory expression (\ref{zeta.F}),
\begin{equation}
\zeta_{\rm rm}=\frac{4-D}{4+\nu(D,N) N}.
\label{zeta.rm}
\end{equation}
In all known cases this correction factor lies in the range $0 \leq
\nu(D,N) \leq 1$.  The findings of the functional RG analysis to order
$\epsilon = 4-D$ \cite{FisherDS86:dr,Balents+93} are equivalent to
$\nu (4,1) \approx 0.800(3)$.  It is interesting to note that the
roughness can be determined exactly for special dimensions:
$\nu(D,\infty)=1$, i.e., the Flory exponent becomes exact for an
infinite number of displacement components \cite{Mezard+90,Mezard+91},
and $\nu(1,1)= 0.5$ corresponding to $\zeta=\frac 23$
\cite{Huse+85:exp}.  From a more involved field-theoretical
self-consistency argument \tcite{Lassig98:prl} obtains $\nu(1,2)=\frac
25$ corresponding to $\zeta=\frac 58$ and $\nu(1,3)=\frac 8{21}$
corresponding to $\zeta=\frac 7{12}$.

Remarkably, there is no renormalization of the elastic constant $\st$
in the Hamiltonian.  The renormalized elastic constant of the manifold
can be determined from its response to tilt field $\muo_{i \alpha}$,
which is coupled to the displacements through
\begin{eqnarray}
{\cal H}_{\mu}=- \int d^D z \ \muo_{i \alpha}  \partial_i
u_\alpha ({\bf z}).
\label{intro.mu}
\end{eqnarray}
In the absence of a pinning potential a constant tilt field leads to a
manifold displacement $\uo_\alpha(\bz) = \muo_{i \alpha} z_i/ \st$
and to a change of the free energy by an amount
\begin{eqnarray}
\overline {\cF[{\bbox \muo}]} - \overline{\cF[0]} 
= - \int d^Dz \ \frac 1{2 \st}  {\muo_{i \alpha}}^2.
\label{shift.F}
\end{eqnarray}
Even in the presence of the pinning potential with a stochastic
translation symmetry, this identity holds exactly. This is due to a
`statistical symmetry' of the pinning energy under a transformation
$\bu(\bz) \to \bu'(\bz) = \bu(\bz) + {\delta \bu}(\bz)$ for an
arbitrary function $ {\delta \bu}(\bz)$
\cite{Goldschmidt+85,Schulz+88,Balents+93}, provided disorder has a
vanishing correlation length in $\bz$ directions, as assumed in
equation (\ref{intro.Delta}).  In particular, the choice $\delta
u_\alpha(\bz)= \muo_{i \alpha} z_i/ \st$ transforms the pinning
potential $V(\bX+\bu(\bz),\bz) \to V(\bX+\bu(\bz) + \delta
\bu(\bz),\bz)= V'(\bX+\bu(\bz),\bz)$, which has the same statistical
properties as the original potential.  This stochastic symmetry is
reflected most obviously by $\cH_{{\rm pin},n}$ in equation
(\ref{H.rep.mani}), which is invariant under the transformation
$\bu^a(\bz) \to \bu^a(\bz) + {\delta \bu}(\bz)$ that is identical in
all replicas.  The non-re\-normali\-za\-tion of $\st$ follows directly
from (\ref{shift.F}), since renormalized elastic constants are defined
by
\begin{eqnarray}
  \frac 1{\st^{\rm eff}}= \frac 1{ND} 
  \frac{\partial}{\partial \muo_{i \alpha}}
  \overline {\langle \partial_i u_\alpha \rangle} 
  = - \frac 1{NDL^D} \frac{\partial^2}{\partial {\muo_{i \alpha}}^2} 
  \overline {\cF[{\bbox \muo}]},
\end{eqnarray}
the change of the free energy and the dependence of $\overline{\cF}$
on $\mu$ is independent of disorder.  Here $L^D = \int d^Dz 1$ is the
system size.

If one starts from a disorder with a finite correlation length also in
$\bz$ direction, there will be a finite renormalization of the elastic
constants due to fluctuations on these small length scales.  Beyond
this scale, the manifold behaves as for vanishing correlation length.
Therefore the asymptotic behaviour on large scales (and in particular
the roughness exponent) will not be affected by such correlations.

The roughness exponent is, however, sensitive to the behaviour of
$\Delta(\bx)$ on large scales, if this correlator is long ranged.
Although this scenario is not of immediate interest for the present
consideration of pinning by point like defects, it is realized, for
example, for pinning of vortex lines by columnar defects or for
magnetic domain walls with random fields. In this case (where $N=1$),
$\Delta(x)-\Delta(0) \propto |x|$ and the roughness exponent has the
`random-field' value \cite{Villain84,Grinstein+84,Bruinsma+84}
\begin{eqnarray}
\label{zeta.rfi}
\zeta_{\rm rfi}=\frac{4-D}3.  
\end{eqnarray}
Surprisingly, this exponent characterizes the manifold in the presence
of a driving force at the depinning threshold, as we will discuss
below.

\subsubsubsection{\ref{sec.struct}e Pinning vs. thermal fluctuations}

In section \ref{sec.struct}d, we have described the structure in the
disorder-do\-mi\-nat\-ed regime, for the `zero-temperature' fixed point.
Now we reconsider the effect of thermal fluctuations in the presence
of disorder.  For this purpose \tcite{FisherDS+91:dp} separated the
displacement correlation function (\ref{intro.W}) into two
contributions, the first one due to disorder
\begin{eqnarray}
W_{\rm pin}({\bf z}-{\bf z}') \equiv 
\overline{\langle {\bf u}({\bf z})-{\bf u}({\bf z}') \rangle^2} ,
\end{eqnarray}
and a second one
\begin{eqnarray}
W_{\rm th}({\bf z}-{\bf z}') 
&\equiv& W({\bf z}-{\bf z}') -W_{\rm pin}({\bf z}-{\bf z}')
\nonumber \\
&=& \overline{\langle [\bu(\bz)-\bu(\bz') - 
\langle \bu(\bz)-\bu(\bz') \rangle ]^2\rangle} 
\nonumber \\
&\sim & \frac T \st |\bz - \bz'|^{2-D}
\label{intro.W.th}
\end{eqnarray}
describing thermal fluctuations. In the absence of disorder, $W_{\rm
  pin}({\bf z}-{\bf z}')=0$.

In particular $W_{\rm th}$ is found to be {\em independent} of
disorder due to the statistical symmetry
\cite{Schulz+88,FisherDS+91:dp} already mentioned above. Thus $W_{\rm
  th}(\bz) \sim |\bz|^{2 \zeta_{\rm th}}$ with the exponent
(\ref{zeta.th}).  Consequently, $W_{\rm pin}(\bz) \sim |\bz|^{2
  \zeta_{\rm rm}}$ in the low-temperature phase, where $\zeta_{\rm rm}
> \zeta_{\rm th}$.

It was argued \cite{FisherDS+91:dp,Hwa+94:prb,Kinzelbach+95} that the
manifold would have (with probability 1) a {\em unique} ground state.
Then $W_{\rm pin}$ essentially characterizes the ground state of the
manifold \cite{Nattermann85:jpc,Huse+85} and $\zeta_{\rm rm}$ would be
the roughness exponent thereof. However, a small fraction of samples
(or, small areas within a sample) will have nearly degenerate excited
states with large displacements relative to the ground state.
Although such excitations are rare, due to the large displacements
they can dominate several disorder-averaged thermodynamic quantities
\cite{Nattermann88,Hwa+94:prb}.  In particular, these rare
fluctuations are responsible for the growth of the thermal width
$W_{\rm th}$ in $D \leq 2$.  Although $W_{\rm th}$ shows exactly the
same behaviour as in the absence of disorder, it is important to
emphasize that the distribution of thermal fluctuations is {\em highly
  non-Gaussian} \cite{FisherDS+91:dp,Hwa+94:prb}. 

For $D>2$ thermal fluctuations have only a {\em finite} width and the
equilibrium state of the manifold globally reflects the ground state.
The broken translation of the Hamiltonian are reflected by the
equilibrium state of the manifold.

In summary, the structure of elastic manifolds in weak disorder
strongly depends on its dimensionalities and shows the following
behaviour (see also figure \ref{fig.D.N}):
\begin{itemize}
\item For $D>4$ the manifold is flat ($\zeta <0$) at all temperatures
  provided disorder is weak. Sufficiently strong disorder will always
  induce roughness.
  
\item For $2 < D \leq 4$ disorder roughens the manifold ($\zeta =
  \zeta_{\rm rm}$) at all temperatures.  The manifold stays close to
  its ground state; the second moment of thermal displacements is
  finite.
  
\item For $2\geq D \geq D_N=2N/(2+N)$ disorder still roughens the
  manifold ($\zeta = \zeta_{\rm rm}$) at all temperatures. Now also
  the second moment of thermal displacements is infinite.
  
\item For $D<D_N$ and sufficiently high temperatures, the manifold is
  entropically driven out of the ground state and shows a structure
  similar to that in the absence of disorder ($\zeta=\zeta_{\rm th}$).
  However, there is also a low-temperature phase where disorder is
  relevant \cite{Cook+89}.  In other terms, the manifold shows a
  temperature-driven depinning transition at a finite temperature.

\end{itemize}

For more details, the interested reader is referred to
\tcite{Halpin+95} and \tcite{Lassig98}.

\begin{figure}
\centering
\epsfig{file=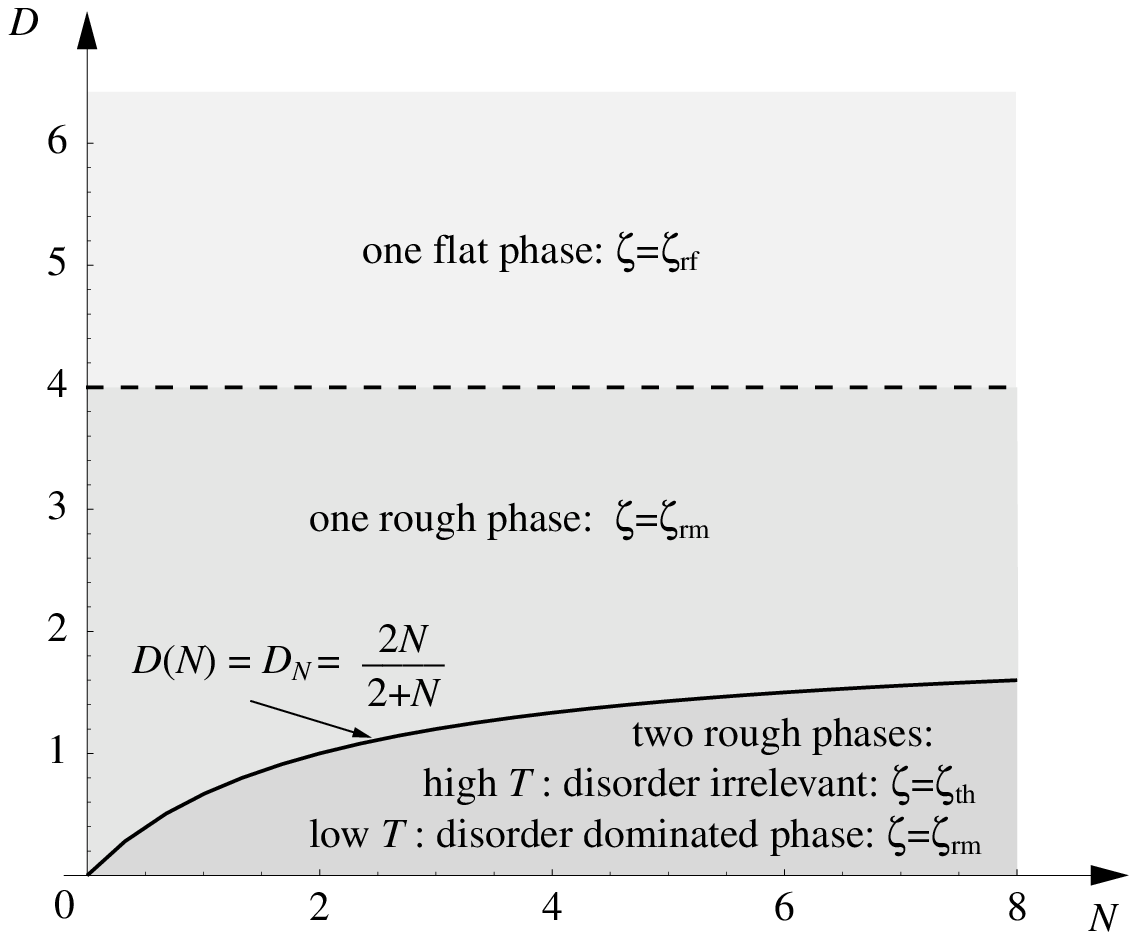,height=8cm}
\fcaption{Illustration of the relevance of disorder for the roughness
  of the manifold in the $(D,N)$-plane (as described in the text).}
{fig.D.N}
\end{figure}

\subsubsection{Thermodynamics}

So far we have described the relevance of disorder for the structure
of the manifold.  In particular, the roughness is qualitatively
increased in the disorder dominated phases.  Could such an increased
roughness be taken as an unambiguous sign for glassiness of the state?
To understand this problem better, we study the possibility of
anomalous scaling properties of thermodynamic quantities such as
energy, entropy, and free energy.

Since the elastic energy and the pinning energy are local quantities,
energy, entropy, and free energy scale extensively with the system
size $L$:
\begin{eqnarray}
\overline{\cal E} \sim L^D, \quad
\overline{\cal S} \sim L^D, \quad
\overline{\cal F} \sim L^D.
\end{eqnarray}
Due to the randomness of pinning there are important sample-to-sample
fluctuations with a scaling (for $T>0$)
\begin{eqnarray}
\overline{(\dE)^2} \sim L^D, \quad
\overline{(\dS)^2} \sim L^D, \quad
\overline{(\dF)^2} \sim L^{2\theta}, \quad
\label{scal.thermo}
\end{eqnarray}
which has been confirmed numerically \cite{FisherDS+91:dp}. Here
$\theta$ is again the energy scaling exponent (\ref{intro.theta}).
The fluctuations of $\dE \equiv {\cal E}-\overline{\cal E}$ and $\dS
\equiv {\cal S}-\overline{\cal S}$ are normal and dominated by
fluctuations on small scales: each sub-volume of the size $L_\xi^D$
contributes independently.  However, the fluctuations of $\dF \equiv
{\cal F}-\overline{\cal F}$ are anomalous and governed by large-scale
contributions.  Clearly, for $T \to 0$ there exists a diverging length
scale below which the fluctuations of $\dE$ scale as those of $\dF$.

In thermodynamic equilibrium the manifold minimizes its free energy.
Therefore energy and entropy are not minimized independently, their
leading order cancels (at $T>0$), and the free energy fluctuations can
be smaller than those of the energy.  One expects $\theta \leq \frac
D2$ in all dimensions, which corresponds to $\zeta \leq \frac {4-D}4$
i.e., $\nu(D,N)\geq 0$ in equation (\ref{zeta.rm}). Indeed, the
exponent $\zeta_{\rm IM}=(4-d)/4$ follows from an Imry-Ma type
argument \cite{Nattermann85:jpc} and is considered to be an upper
bound for $\zeta_{\rm rm}$ in random-bond systems
\cite{FisherDS93:pc}.  At strictly zero temperature ${\cal E}={\cal
  F}$ and both quantities must scale in the same way, $\dE = \dF \sim
L^\theta$.  Therefore temperature plays a crucial role, although it
seems to be irrelevant in the flow equation (\ref{flow.rm.T}).  This
is because temperature is a dangerously irrelevant variable
\cite{FisherDS86:sc}.  In very rare samples there are excited states
which are nearly degenerate with the ground state but which deviate
from it by large displacements.  Such rare but large fluctuations
still give dominant contributions to many thermodynamic
disorder-averaged thermodynamic properties, for a more detailed
discussion see \tcite{FisherDS+91:dp} and \tcite{Hwa+94:prb}.

\subsubsection{Susceptibility}
\label{sec.susc}

In order to find an unambiguous signature of glassy phases,
\tcite{Hwa+94:prl} proposed to examine the susceptibility of the
system with respect to a tilt field.  Although in this section we are
dealing with directed manifolds, we establish the analogous
susceptibility for these simpler systems and show how it is related to
the displacement correlation functions discussed above.

For simplicity we restrict the discussion of susceptibilities to
$N=1$; the general case $N>1$ follows from a straightforward
generalization.  We couple the manifold to an inhomogeneous tilt field
${\bbox \mu({\bf z})}$ via an additional energy contribution as in
(\ref{intro.mu}),
\begin{eqnarray}
{\cal H}_{\mu}=- \int d^D z \ [{\bbox \muo} 
+ {\bbox \mu} ({\bf z})] \cdot \grad_\parallel u({\bf z}).
\label{intro.mu.2}
\end{eqnarray}
As in section \ref{sec.struct}d, ${\bbox \muo}$ is a constant field
applied to measure the tilt response of the system.

The system
\begin{eqnarray}
\cH'&=&\cH_{\rm el}+\cH_{\mu} 
\nonumber \\
&=& \int d^Dz \left\{ \frac \st 2 
\left( \grad_\parallel u  - \frac 1\st 
[{\bbox \muo} + {\bbox   \mu} ]\right)^2 
- \frac 1{2 \st} {[{\bbox \muo} + {\bbox \mu}]^2} \right\},
\label{cal.H.prime}
\end{eqnarray}
without random potential $V$ but with a random tilt ${\bbox \mu} ({\bf
  z})$, has a ground state $\uo ({\bf z})$ which couples only to the
longitudinal part ${\bbox \mu}_L$ of the tilt field and which is
determined by ${\bbox \muo} + {\bbox \mu} ({\bf z})= \st
\grad_\parallel \uo ({\bf z})$.  For a random tilt with zero average
and long-ranged correlations
\begin{eqnarray}
\overline{\mu_i(\bz) \mu_j(\bz')}  \sim
  |\bz-\bz'|^{-\alpha}, \quad \alpha < D,
\label{mu.lrc}
\end{eqnarray}
the manifold has in its ground state an exponent $\zeta= \frac 12
(2-\alpha)$ and will thus be rough if $\alpha<2$.  Nevertheless, since
this toy model is harmonic and trivial (it is bilinear in $u$, and
$\cH'$ is even translation invariant), it cannot be considered as
glass, which motivated \tcite{Hwa+94:prl} to search for unambiguous
signatures of glassiness.

They proposed to examine the sample-to-sample fluctuations of the
linear response susceptibility $\chi_{\alpha \beta}$ of the system for
spatially constant tilt ${\bbox \muo}$, which can be obtained
from the free energy of a particular sample by
\begin{eqnarray}
\chi \equiv  \left. - \frac1{D L^D} 
\frac{\partial^2 \cF[{\bbox \muo}]}
{\partial \muo_\alpha \partial \muo_\alpha}\right|_{{\bbox \muo}=0}.
\label{intro.chi}
\end{eqnarray}
Then it is interesting to consider its sample-to-sample fluctuations
\begin{eqnarray}
\overline{(\Delta \chi)^2} \equiv 
{\overline {\chi^2}} - {{\overline{\chi}}^2}.
\end{eqnarray}
The toy model $\cH'=\cH_{\rm el}+\cH_{\mu}$ that has only random tilt
${\bbox \mu}$ but no random potential, has $\chi = 1/ \st$ for all
realizations of the random tilt because of the statistical symmetry
discussed after equation (\ref{shift.F}). Hence the vanishing of
$\overline{(\Delta \chi)^2}=0$ reflects its trivial nature, i.e., that
random tilt fields only deform the ground state, and thermal
fluctuations {\em around} the ground state are similar to those in the
absence of disorder.

Now we come back to the manifold model in a random potential (we drop
the random tilt but we still assume $N=1$ for simplicity) to
demonstrate that $\overline{(\Delta \chi)^2} >0$, i.e., that
susceptibility fluctuations are a less ambiguous signature of
glassiness than mere roughness.  Due to the statistical symmetry,
$\overline {\cF[{\bbox \muo}]} - \overline{\cF[0]}= - L^D \frac
1{2\st} ({\bbox \muo})^2$ and ${\overline \chi}=1/\st$ {\em exactly}.
We can calculate its susceptibility fluctuations {\em perturbatively}
from the free energy fluctuations at $T=0$,
\begin{eqnarray}
\overline{\Delta \cF [{\bbox \muo}] \Delta \cF [{\bbox \muo}']}
= \int d^Dz \ \Delta \left( \frac 1\st ({\bbox \muo} - {\bbox  \muo}') 
\cdot \bz \right) ,
\end{eqnarray}
involving the potential correlator $\Delta(\bx)$ evaluated for the
tilted manifold in the absence of pinning.  This results in
\begin{eqnarray}
\overline{(\Delta \chi)^2} \propto L^{4-D} \frac{\Delta^{(4)}(0)}{\st^4},
\label{div.chi}
\end{eqnarray}
where we have dropped numerical factors and $\Delta^{(4)}(0)=
\left.\partial_u^4\Delta(u)\right|_{u=0}$.  The dependence of this
result on the system size $L$ shows again that $\overline{(\Delta
  \chi)^2} \neq 0$ in $D\leq 4$. Certainly, we cannot expect this
perturbative result to be quantitatively correct for large $L$, but
qualitatively it reflects the relevance of disorder and the glassiness
of the manifold.  In $D \leq 4$ one expects $\overline{(\Delta
  \chi)^2} $ to be finite. Its correct value has to be calculated
within the RG scheme (see \tcite{Hwa+94:prl}).

We will not further pursue an accurate calculation of the
susceptibility fluctuations here. Instead we establish its connection
to the displacement fluctuations.  To this end we introduce the
response of the local tilt $\partial_\alpha u({\bf z})$ to the field
$\mu_\beta({\bf z}')$ (which is now considered as an external control
parameter rather than a form of disorder)
\begin{eqnarray}
\chi_{\alpha \beta}({\bf z},{\bf z}') &\equiv & \frac {\delta \langle
\partial_\alpha u({\bf z}) \rangle}{\delta \mu_\beta({\bf z}')}
= - \frac{\delta^2}{\delta \mu_\beta({\bf z}') 
\delta \mu_\alpha({\bf z})} {\cal F}[{\bbox \mu}] 
\nonumber \\
&=& \frac 1T 
[\langle \partial_\alpha u({\bf z})\partial_\beta u({\bf z}')\rangle
- \langle \partial_\alpha u({\bf z})\rangle \langle \partial_\beta u({\bf
  z}')\rangle],
\label{chi.loc}
\end{eqnarray}
which is now explicitly expressed as a displacement correlation
function and can be related to correlations of the order parameter
$\psi$ exploiting
\begin{eqnarray}
\partial_\alpha u ({\bf z})= \left. -i \partial_\alpha
\frac{\partial \psi_k({\bf z})}{\partial k} \right|_{k=0} .
\end{eqnarray}
The total susceptibility (\ref{intro.chi}) discussed above can be
obtained as integral of the local susceptibility (\ref{chi.loc}),
\begin{eqnarray}
\chi =  \frac1 {L^D} \int d^Dz \ d^Dz' \  
\chi_{\alpha \alpha}({\bf z},{\bf z}').
\end{eqnarray}
(Note that here is no factor $1/D$ in contrast to (\ref{intro.chi})
since the displacement responds not only to the longitudinal component
of ${\bbox \mu}(\bz)$ but to the entire ${\bbox \muo}$.)  For a
particular sample the local susceptibility explicitly depends on two
space coordinates $\bz$ and $\bz'$.  We introduce a susceptibility $
\chi_{\alpha \beta}({\bf z}-{\bf z}') \equiv L^{-D} \int d^D z_0 \ 
\chi_{\alpha \beta}({\bf z}+{\bf z}_0,{\bf z}'+{\bf z}_0)$ averaged
over the volume ${\bf z}_0 \in L^D$, which now depends only on the
coordinate difference $\bz-\bz'$. Then $\chi$ is conveniently related
to the Fourier transform of $ \chi(\bz)$ through
\begin{eqnarray}
\chi = \left.   \chi_{\alpha \alpha}({\bf k}) \right|_{\bk =0}.
\end{eqnarray}
(For example, the toy model has $ \chi_{\alpha \beta}({\bf k})=
(1/\st) P^L_{\alpha \beta}({\bf k})$ with the longitudinal projector
$P^L_{\alpha \beta}({\bf k}) \equiv k_\alpha k_\beta / k^2$ and hence
$\chi=1/\st$.)

The fact that $\chi$ has sample-to-sample fluctuations in a glassy
phase, $\lim_{L\to \infty} \overline{(\Delta \chi)^2} /
{{\overline{\chi}}^2} \not\to 0$, means that the susceptibility is
{\em not self-averaging} \cite{Aharony+96}. Since $\chi$ is obtained
from a volume average sample-to-sample fluctuations also mean that the
translational average is not equivalent to the average over many
samples.

From the expression for the susceptibility in terms of the
displacement field, equation (\ref{chi.loc}), one recognizes that the
susceptibility essentially measures fluctuations around the ground
state of the system. More precisely, one can relate the susceptibility
to the thermal displacement correlation (\ref{intro.W.th}) by
\begin{eqnarray}
\partial_\alpha \partial_\beta' W_{\rm th}(\bz-\bz')= - 2 T \
\overline{ \chi_{\alpha \beta}(\bz-\bz')} .
\end{eqnarray}
Thus the disorder-averaged susceptibility is related to $W_{\rm th}$.
The fact that $\overline{ \chi}$ is independent of disorder is related
to the disorder independence of $W_{\rm th}$ discussed in section
\ref{sec.struct}e.  A signature of glassiness can therefore appear
only in the sample-to-sample fluctuations of $\chi$.

To conclude the discussion of susceptibility fluctuations as
characteristic for glassiness, we wish to point out that they cannot
be taken as {\em sufficient} criterion for glassiness. This can be
demonstrated by a counterexample with a `pinning' energy
\begin{eqnarray}
  \cH_{\rm pin}=\int d^Dz \ \frac 12 [{\bbox \nu} \cdot
  \grad_\parallel u(\bz)]^2 ,
\end{eqnarray}
where ${\bbox \nu}$ is a vector of fixed length $|{\bbox \nu}|$ but
with random orientation.  This type of disorder actually represents
randomness in the elastic constants and the model is not a glass since
it is Gaussian in the displacement. Nevertheless, it has
sample-to-sample fluctuations
\begin{eqnarray}
\overline{(\Delta \chi)^2}=[(1+ {\bbox \nu}^2/\st)^{1/2} -1 ] 
\ {\overline{\chi}^2}
\end{eqnarray}
that are finite for ${\bbox \nu}^2>0$.


\subsubsection{Barriers}
 
The glassy nature of a system is generally related to an extremely
slow dynamics that is dominated by thermally activated processes in a
complex energy landscape with many meta-stable states.  Glassy systems
have not only a disorder-dominated ground state, but also a huge
number of meta-stable states.  This section is devoted to the
description of the energy landscape.

The dynamical behaviour is determined by `neighbouring' meta-stable
states, which are related to each other by the slip of a restricted
part of the manifold over a certain barrier.  The part of the manifold
is assumed to have a certain length $L_z$ in $z$-directions, the
displacement in this region is of a magnitude $L_x$ and the barrier
height is denoted by $U$, see figure \ref{fig.barrier}.

\begin{figure}
\centering
\epsfig{file=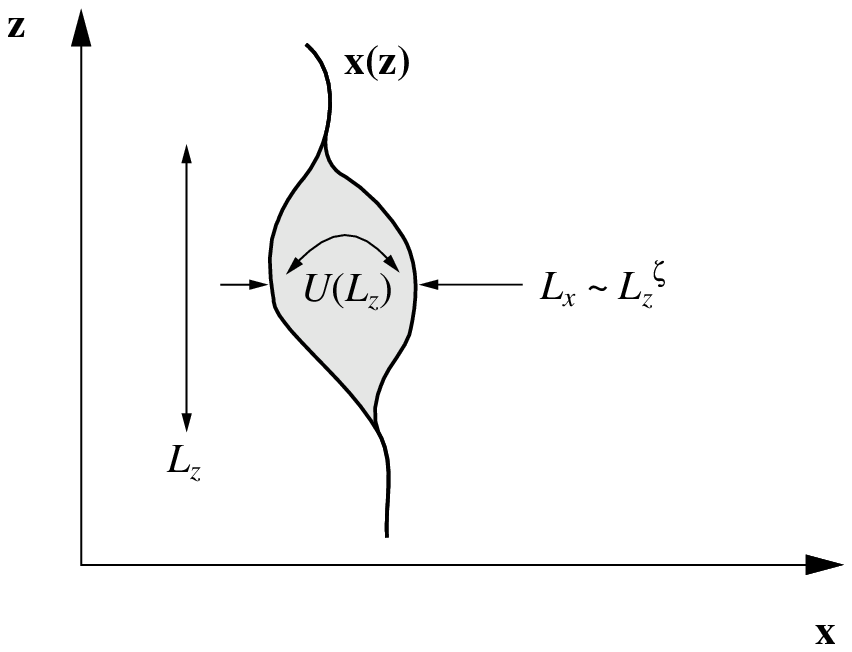,height=6cm}
\fcaption{Illustration of two meta-stable configurations of the manifold
  (solid lines) that are separated by a barrier (shaded area).  The
  characteristic length scales of the barrier are $L_z$ in
  longitudinal direction and $L_x$ in transverse direction. The height
  of the barrier is $U$.}
{fig.barrier}
\end{figure}

A rigorous characterization of the statistics of barriers is very
intricate. Therefore, a scaling picture
\cite{Villain84,Nattermann85:pssb1,Huse+85,Ioffe+87} has been put
forward, where it is assumed that the statistics of barriers is
essentially identical to the statistics of the free energy
fluctuations.  Therefore
the barrier height should scale with the barrier length like
\begin{equation}
U(L_z) \approx U_\xi \left( \frac {L_z}{L_\xi} \right)^\psi , 
\label{scale.barrier}
\end{equation}
where $U_\xi\equiv U(L_\xi) \approx \st \xi^2 L_\xi^{D-2}$ is the
typical height of the smallest possible barrier of the size of the
Larkin length.  The scaling exponent for the barrier height $\psi$ is
assumed to coincide with the exponent $\theta$ from equation
(\ref{scal.thermo}) that describes the free energy fluctuations:
\begin{eqnarray}
  \psi=\theta = 2 \zeta + D-2.
\label{intro.psi.exp}
\end{eqnarray}
This exponent is related through equation (\ref{intro.theta}) to the
roughness exponent $\zeta$ of the manifold, which describes the
scaling relation
\begin{equation}
L_x \approx \xi \left( \frac {L_z}{L_\xi} \right)^\zeta
\label{intro.L_x}
\end{equation}
between the lateral and transverse sizes of the barrier region.

The basic assumption $\psi=\theta$ has been confirmed by analytic
arguments combined with numerical simulations
\cite{Mikheev+95,Drossel+95,Drossel96}.  However, the statistics of
barriers and free energy fluctuations have turned out to possibly be
not {\em strictly} identical: $U(L_z)$ and $\{\overline{[\Delta
  \cF(L_z)]^2}\}^{1/2}$ can differ by factors that are powers of $\ln
L_z$ which do not modify the exponent relation (\ref{intro.psi.exp}).

It is clear that not all barriers of a given size $L_z$ have strictly
the same height but that they are distributed according to a certain
distribution $\cP_{L_z}(U)$.  The above scaling relations are valid
for the {\em `optimal'} (dynamically relevant) barrier height, which
is the average barrier obtained from $\cP_{L_z}(U)$ with an additional
weighting factor for the density of barriers of the size considered
\cite{Vinokur+96}.  In particular in low dimensions the dynamics of
the manifold is not necessarily dominated by the average barrier
height, but by the {\em largest} barriers.  Therefore it is also of
fundamental interest to characterize the distribution $\cP_{L_z}(U)$
for large $U$.

For a string, i.e. $(D,N)=(1,1)$, \tcite{Vinokur+96} found
\begin{eqnarray}
\cP_{L_z}(U) \approx \frac 1{U_\xi} \frac{L_z}{L_\xi} e^{-U/U_\xi} 
\exp[{-(L_z/L_\xi) e^{-U/U_\xi}} ] 
\end{eqnarray}
from a combination of extreme value statistics
\cite{Gumbel58,Galambos78} and a coarse graining approach. Large
barriers are exponentially rare since $\cP_{L_z}(U) \sim \frac
1{U_\xi} \frac{L_z}{L_\xi} e^{-U/U_\xi} $ for $U \to \infty $.
\tcite{Gorokhov+99} found that the free-energy distribution obeys a
similar exponential decay at large {\em negative} values of the free
energies $\cF$.  The free-energy distribution decays much faster at
large positive $\cF$ than at large negative $\cF$, \cite{Gorokhov+99},
which one can expect because the manifold {\em minimizes} the free
energy in equilibrium.

\subsection{Transport properties}
\label{sec.mani.trans}

The glassy nature of systems does not manifest itself only in
equilibrium properties but, maybe even more unambiguously, in its
dynamic behaviour.  The existence of many meta-stable states and the
thermally activated nature of transitions between these states makes
the dynamics of the glass extremely slow.  We describe now the
transport properties of manifolds in a stationary driven state,
leaving aside interesting topics such as relaxational dynamics.

Vortices move dissipatively in the superconducting condensate.  For
small velocities the friction force is proportional to the velocity
with a constant viscous drag coefficient $\eta_0$.  For a single
vortex line in a bulk superconductor \cite{Bardeen+65}
\begin{eqnarray}
\eta_0 \approx \frac{\Phi_0 \Hcii}{\rho_{\rm n} \co^2},
\label{eta.BS}
\end{eqnarray}
where $\rho_{\rm n}$ is the normal state resistivity and $\co$ the
velocity of light.  On large time scales inertial forces can be
neglected in comparison to the friction force.

The equation of motion is then obtained from balancing the friction
force (more precisely: the force density in a $D$-dimensional space;
for brevity forces and force densities are not strictly distinguished)
with the forces arising from the interaction with other vortices, the
pinning force, the driving force $\bF$ and a thermal noise ${\bbox
  \zeta}$,
\begin{eqnarray}
\eta_0 \dot {\bf x} = 
- \frac{\delta {\cal H}[{\bf x}]}{\delta {\bf x}} +
{\bf F} + {\bbox \zeta}.
\label{intro.eqmo}
\end{eqnarray}
The noise is taken as Gaussian distributed with zero average and
variance
\begin{eqnarray}
\langle \zeta_\alpha({\bf z},t) \zeta_\beta({\bf z}',t') \rangle= 
2 \eta_0 T \delta_{\alpha \beta}  \delta({\bf z}-{\bf z}')
\delta(t-t')
\label{corr.zeta}
\end{eqnarray}
such that the equation of motion properly describes thermodynamic
equilibrium according to the fluctuation-dissipation theorem in the
absence of the driving force.

At finite temperatures the manifold will respond to the driving force
with an average velocity ${\bf v}\equiv \overline {\langle \dot {\bf
    x} \rangle}$.  For calculational simplicity it is advantageous to
consider ${\bf v}$ as prescribed and to calculate the driving force
${\bf F}={\bf F}({\bf v})$ required to maintain this velocity.  In the
driven case we define the displacement ${\bf u}(\bz) \equiv {\bf
  x}(\bz)-{\bf X}-{\bf v} t$ such that $\langle {\bf u} \rangle =0$.
This displacement follows the equation of motion
\begin{eqnarray}
  \eta_0 \dot {\bf u} &=& - 
\frac{\delta {\cal H}[\bX+{\bf v}  t +{\bf u}]}
{\delta {\bf u}} +  {\bf F} -\eta_0 {\bf v} + {\bbox \zeta}
\nonumber \\
&=&  \st \grad_\parallel^2 {\bf u} - \grad_\perp
V({\bf X} + {\bf v}t +{\bf u},\bz) 
+  {\bf F} -\eta_0 {\bf v} + {\bbox \zeta},
\label{eqmo.mani}
\end{eqnarray}
which serves as basis for the following analysis [recall that
$\grad_\parallel \equiv (\frac \partial {\partial z_1},\dots,\frac
\partial {\partial z_D})$ and $\grad_\perp \equiv (\frac \partial
{\partial u_1},\dots,\frac \partial {\partial u_N})$].

\subsubsection{Friction}

In the absence of pinning, $V=0$, the manifold moves with constant
velocity $\bv= \bF/\eta_0$.  Pinning tends to slow down the motion of
the system, because the manifold has to overcome barriers and it loses
more time by sliding uphill than it wins by sliding downhill.  Thus,
to establish a certain velocity $\bv$ in the presence of pinning, one
has to apply a force $\bF(\bv)$ which is larger than in the absence of
pinning.  In general, the velocity-force characteristic will be
non-linear but one can define an effective zero-velocity friction
coefficient by
\begin{eqnarray}
\eta_{\alpha \beta} \equiv \left. \frac{d F_\alpha({\bf v})}{d
    v_\beta} \right|_{\bf v=0}.
\end{eqnarray}
We now show that this `renormalized' friction coefficient can diverge
due to the presence of disorder, which is another signature of
glassiness.

The divergence of the effective friction coefficient is examined by
treating disorder perturbatively.  We make use of the linear response
function of the disorder-free system
\begin{eqnarray}
G_{\alpha \beta}({\bf z},t)&\equiv& 
 \frac{\delta \langle u_\alpha(\bz,t)\rangle}{\delta F_\beta(\bN,0)}
= \frac 1 {\eta_0} \left( \frac {\eta_0}{ 4
    \pi \st t} \right)^{D/2} e^{- \eta_0 z^2/2 \st t} 
\ \Theta(t) \delta_{\alpha \beta},
\label{G.free}
\end{eqnarray}
which describes the reaction of the displacement to a locally
perturbating force.

By iterating the equation of motion (\ref{eqmo.mani}) up to $O(V^2)$,
the condition $\overline{\bu}=0$ yields the velocity-force
characteristic and the effective friction coefficient
\begin{eqnarray}
\eta_{\alpha \beta} 
&=&\eta_0 \delta_{\alpha \beta} 
+  \int dt \ t \ G_{\gamma \delta}({\bf z}=0,t) \partial_\alpha
\partial_\beta \partial_\gamma \partial_\delta \Delta({\bf 0})  .
\end{eqnarray}
Inspecting the large-time behaviour of the response function
(\ref{G.free}) for $\bz=\bN$, this coefficient diverges in $D \leq 4$
as long as $ \partial_\alpha \partial_\beta \partial_\gamma
\partial_\delta \Delta({\bf 0}) \neq 0$ for certain indices. The
divergence of $\eta_{\alpha \beta} $ resembles that of the
susceptibility fluctuations (\ref{div.chi}): both couple to the fourth
derivative of $\Delta(\bu)$.  This divergence actually does not mean
that a stationary state with non-vanishing velocity could be
established only by an infinite driving force; it means that at small
velocities $\bv(\bF)$ is no longer linear (Ohmic) but {\em
  sub-linear}.

The qualitative change of the transport characteristic in $D \leq 4$
is consistent with roughening as a qualitative structural change.
This consistency is not just a fortunate coincidence. One can actually
show the relation between dynamic and static features using a
fluctuation-dissipation relation between the correlation and response
function of the free system \cite{Scheidl+98:pm}.  Nevertheless, the
toy model studied in Sec \ref{sec.susc} has a strictly linear
transport characteristic (the same as in the absence of disorder)
despite its roughness.  The transport characteristic is not at all
affected by random bonds coupling to the displacement as in equation
(\ref{intro.mu.2}) because the translation symmetry $\bu \to
\bu+\bx_0$ is not broken.  Thus the qualitative sensitivity of the
transport characteristic to the nature of disorder suggests its use as
a supplementary indicator for glassiness.

If the dynamics of the system is glassy, the transport characteristic
shows several characteristic features, cf. figure \ref{fig.dyn.reg}.
For small driving forces it is important to distinguish between the
cases of zero and finite temperature.  At $T=0$ the driving force has
to exceed a threshold value $F_c$ to set the manifold in motion with a
finite average velocity, since the driving force has to overcome the
average pinning force.  (Strictly speaking, this is true only in the
absence of quantum fluctuations, which can lead to quantum creep at
$T=0$ and $F\leq F_c$, see \tcite{Blatter+94} and references
therein.) This threshold phenomenon is called {\em depinning}.  For
$T>0$ a finite velocity is found even for $|\bF|<F_c$ due to thermal
activation, and the transport characteristic will be sub-linear ({\em
  creep} regime).  At large driving forces $|\bF| \gg F_c$ the effect
of pinning diminishes and the free differential mobility is reached
asymptotically, $d F_\alpha(\bv)/dv_\beta \to \delta_{\alpha \beta}
\eta_0$. This latter regime is called the {\em flow} regime.

\begin{figure}
\centering
\epsfig{file=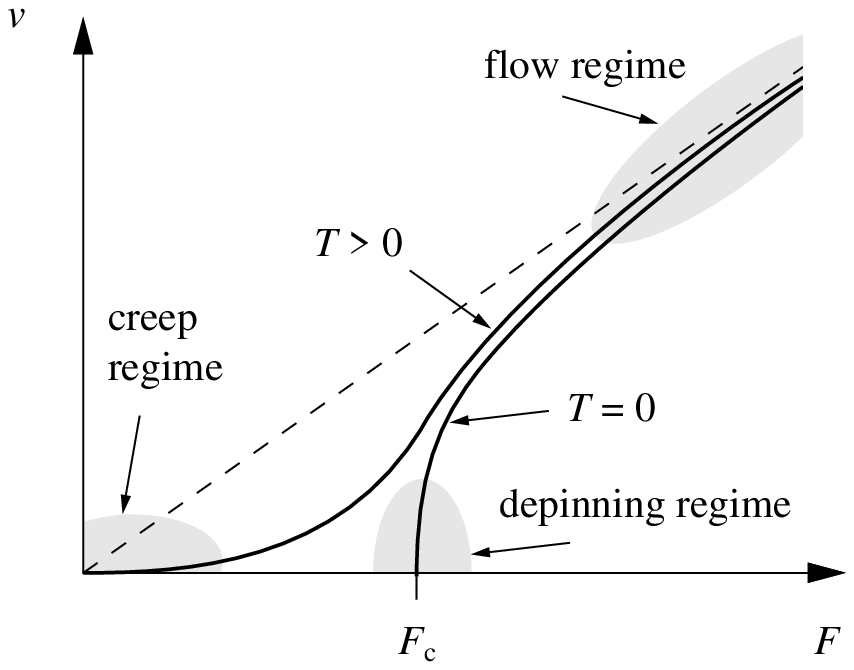,height=6cm}
\fcaption{Illustration of the dynamic regimes (shaded areas) in the transport
  characteristic $v(F)$ of a manifold driven in disorder. In the
  absence of disorder the characteristic is linear (dashed line). In
  the presence of disorder it is non-linear (solid lines). At $T=0$
  the manifold {\em depins} as a critical force $F_c$. At $T>0$
  thermally activated {\em creep} occurs already for $F \ll F_c$. For
  $F \gg F_c$ the manifold {\em flows} essentially as in the absence of
  disorder.}
{fig.dyn.reg}
\end{figure}

The divergence of $\eta_{\alpha \beta}$ implies the break-down of
perturbation theory for a calculation of the transport characteristic.
Although it is difficult to calculate the global form of the
characteristic, it is possible to achieve an analytic description for
small and large velocity.

\subsubsection{Depinning}
\label{sec.depin}

The value of the critical force $F_c$ where motion sets in at $T=0$
can be estimated as follows: On length scales below the Larkin length
$L_\xi$ the manifold performs only displacements smaller than $\xi$,
such that the dependence of the pinning force on the displacement can
be neglected.  In such a volume the local pinning forces add up to
$\bff_\xi= \int_{|z|<L_\xi} d^D z \ [-\grad_\perp V(\bX,\bz)]$. The
forces from different volumes are essentially uncorrelated. Thus
depinning, where the driving force is balanced by pinning force, is
found at \cite{Feigelman83,Bruinsma+84}
\begin{eqnarray}
F_c \simeq L_\xi^{-D} \ {(\overline{\bff_\xi^2})}^{1/2} \simeq
(\Dii)^{1/2} L_\xi^{-D/2} 
\simeq \st \xi L_\xi^{-2}
\sim (\Dii)^{2/\epsilon}, 
\end{eqnarray}
where again $\epsilon \equiv 4-D$ and the pinning force variance
$\Dii$ was given in equation  (\ref{intro.Dii.sec4}).  For fixed but weak
strength of $\Dii$ the critical force $F_c$ vanishes as $D \nearrow 4$
and it is zero for $D>4$.

Thus for weak disorder (de)pinning occurs in $ D \leq 4$ dimensions
only, to which we restrict our analysis.  The velocity is expected to
increase continuously when the driving force exceeds the threshold
value. This observation was made originally by \tcite{Middleton92:prl}
in the context of charge-density waves.  Thus it should be possible to
consider depinning as a {\em continuous transition}, at which the
manifold exhibits critical fluctuations \cite{Nattermann+92}.  The
depinning transition was studied theoretically for directed manifolds
\cite{Nattermann+92,Narayan+93,Ertas+96:prb,Leschhorn+97} and for the
closely related charge-density waves \cite{FisherDS85,Narayan+92:ii}
within a renormalization group approach.  \tcite{Kardar98} and
\tcite{FisherDS98} review depinning and its relevance in a much wider
variety of phenomena.

The phenomenology of the transition is described qualitatively by the
following scenario: as long as the manifold is pinned, it will be in a
rough state because of competition between elastic forces, pinning
forces, and the driving force.  In the absence of the driving force,
the roughness is described by the exponent $\zeta_{\rm rm} =
\zeta_{\rm rm}(0)$ of manifolds in a random potential at equilibrium,
which was described in equation (\ref{zeta.rm}).  A driving force
polarizes the state of the manifold and turns the situation into a
non-equilibrium one, where the exponent is modified to $\zeta_{\rm
  rm}(F)$.  Preliminary numerical simulations \cite{Leaf+99} indicate
that for forces $|\bF| \leq F_c$ below the depinning transition,
$\zeta_{\rm rm}(F)$ continuously increases with the driving force,
until it reaches the value $\zeta_{\rm rm}(F_c) = \zeta_{\rm rfi}$ of
manifolds in a random force field at equilibrium, which was described
in equation (\ref{zeta.rfi}).  This means that the effective force
acting on the manifold has qualitatively changed its character.  Above
depinning, where the manifold moves with a finite velocity, the force
acts on large scales like an effective thermal noise (although we
consider the situation at $T=0$, without thermal noise $\zeta$ in the
equation of motion). Accordingly, the manifold will be flat on large
scales, as long as $D>2$.

For small velocities, the saturation of the displacement correlation
function occurs on a large length scale $\ell_z$, which acts as a
dynamical correlation length that diverges at the transition as
\cite{Middleton92:prl,Nattermann+92,Narayan+93}
\begin{eqnarray}
\ell_z \sim (F-F_c)^{-\nu}  
\end{eqnarray}
with a certain exponent $\nu$.  For this zero-temperature
non-equilibrium transition, the force acts as a control parameter that
plays the role of temperature in a usual finite-temperature
equilibrium transition. Associated with the longitudinal scale
$\ell_z$ there is a transverse scale
\begin{eqnarray}
\ell_x \sim \ell_z^\zeta
\end{eqnarray}
with $\zeta$ being the roughness exponent at depinning.  As the
transition is approached from above, $F \searrow F_c$, there is also a
diverging time scale $t_v$, the correlation time of the pinning force
acting on the manifold,
\begin{eqnarray}
t_v \approx \frac {\ell_x} v.
\label{intro.t_v}
\end{eqnarray}
The dynamical exponent $z$ then can be defined from the scaling
relation between $\ell_z$ and $t_v$,
\begin{eqnarray}
t_v \sim \ell_z^z.
\label{scal.t_v.l_z}
\end{eqnarray}
A further scaling relation describes the continuous onset of motion,
\begin{eqnarray}
v \sim (F-F_c)^\beta  .
\label{scal.v.F}
\end{eqnarray}
From a balance of the elastic and driving force in the equation of
motion (making use of the non-renor\-mali\-zation of the elastic constant)
one derives \cite{Nattermann+92,Narayan+93}
\begin{eqnarray}
\nu = \frac 1 {2-\zeta}.
\label{scal.nu}
\end{eqnarray}
The consistency of equations (\ref{intro.t_v}), (\ref{scal.t_v.l_z})
and (\ref{scal.v.F}) requires the scaling relation
\begin{eqnarray}
\beta=(z-\zeta) \nu.
\label{scal.beta}
\end{eqnarray}

In the general case with more than one displacement component (for
$N>1$), the velocity selects one particular direction out of the
$N$-dimensional space and we may choose the coordinates such that
$\bv=v \delta_{\alpha,1}$.  Due to this selection of a direction by
velocity one has to expect the displacement correlations to be
uniaxially {\em anisotropic}.  We subsequently distinguish
$W_\alpha(\bz,t)=\overline{ \langle [u_\alpha(\bz,t)-
  u_\alpha(\bN,0)]^2 \rangle }$, where $\alpha=1$ for the parallel
component and $\alpha=2,\dots,N$ for the perpendicular component.  The
anisotropy is not a minor quantitative effect, it concerns the scaling
exponents as shown by \tcite{Ertas+96:prb}.  The anisotropic scaling
laws for the displacement components are
\begin{eqnarray}
W_\alpha(\bz,t) &=& |\bz|^{2\zeta_\alpha}
g_\alpha(t/|\bz|^{z_\alpha}),
\end{eqnarray}
with $g_\alpha(0)$ finite and $g_\alpha(y) \sim y^{2
  \zeta_\alpha/z_\alpha}$ for large arguments $y$.  The above scaling
relations (\ref{scal.nu}) and (\ref{scal.beta}) then hold with the
identification
\begin{eqnarray}
\zeta \equiv \zeta_1, \quad  z \equiv z_1 .
\end{eqnarray}
The values of the exponents which can be calculated analytically
within a renormalization group approach, using an expansion in small
$\epsilon = 4-D$ \cite{Nattermann+92,Ertas+96:prb} are:
\begin{mathletters}
\mlabel{dep.exp}
\begin{eqnarray}
\zeta_1 &=& \frac \epsilon  3
\label{zeta.par}
\\
z_1 &=& 2 - \frac {2 \epsilon} 9 + O(\epsilon^2)
\\
\zeta_\alpha &=& \zeta_1 - \frac D2 
= -2 + \frac {5\epsilon} 6 
\\
z_\alpha &=& z_1 + \frac 1 \nu 
= 4 - \frac {5 \epsilon} 9 + O(\epsilon^2)
\end{eqnarray}
\end{mathletters}
and all directions $\alpha \geq 2$ scale identically.  Equation
(\ref{zeta.par}) means that the manifold in a random potential at
depinning has the roughness of a manifold in random force field in
equilibrium, cf. equation (\ref{zeta.rfi}).  \tcite{Narayan+93} proved
that equation (\ref{zeta.par}) is correct to {\em all orders} in
$\epsilon$.

For the depinning of periodic media such as charge-density waves the
same scenario holds as for directed manifolds.  However, they belong
to a different universality class with different values of the
exponents \cite{Middleton92:prl,Narayan+92:i,Narayan+92:ii}.

So far, we have discussed depinning as a phenomenon at $T=0$. The
effect of finite temperature is that potential barriers can be
overcome by thermal activation.  Therefore a finite velocity is
expected for every finite driving force.  Fixing the force at the
$T=0$ threshold value $F=F_c$, velocity increases with temperature
according to a power law
\begin{eqnarray}
v \sim T^{\beta/\tau}  .
\end{eqnarray}
The exponent $\tau$ depends on the type of the pinning potential in a
non-universal way \cite{FisherDS85,Middleton92:prb}.  For the
`ratcheted kick model' \tcite{Middleton92:prb} argued for $\tau=2$
in $D=2,3$.  The actual value of $\tau$ is still controversial  and
subject to investigations \cite{Nowak+98,Roters+99}.

\subsubsection{Creep}

Below the depinning threshold, for $|\bF| \leq F_c$, the dynamics of
the manifold is in the creep regime, where motion is possible only
through thermal activation.  At $|\bF|=F_c$, barriers of {\em all}
sizes are equally relevant and it is precisely for this reason that
depinning appears as a critical phenomenon.  Below depinning, barriers
of a {\em typical} size are prevalent and dominate the thermally
activated dynamics.  In the presence of a driving force the manifold
experiences an effective potential $V_{\rm eff}(\bx,\bz) \equiv
V(\bx,\bz)- \bx \cdot \bF$.  For weak forces the manifold still has
meta-stable states in this effective potential, which are separated by
barriers that also depend on the driving force.

We now focus on the creep regime of very small forces $|\bF| \ll F_c$,
where the barriers are nearly identical to those in equilibrium and
where a scaling picture for the typical, dynamically relevant barrier
has been developed.  The starting point of this picture is the
assumption that barriers of a length $L_z$ and free energy
fluctuations of a system of size $L_z$ have identical scaling as
described in equation (\ref{scale.barrier}).  The typical lengths
$L_z$ and $L_x \sim L_z^\zeta$ are related by equation
(\ref{intro.L_x}). When the manifold overcomes such a barrier, it
gains an energy
\begin{eqnarray}
E_F \approx F L_x L_z^D \approx F \xi L_\xi^{D} 
\left(\frac{L_z}{L_\xi}\right)^{D + \zeta}  
\end{eqnarray}
in the field of the driving force. Therefore, in the effective
potential $V_{\rm eff}$ only those barriers are still effective for
which $U \lesssim E_F$. From $U(L_F) = E_F$ follows
\cite{Ioffe+87,Nattermann87,Feigelman+88,Feigelman+89,Nattermann90} the
size of the largest effective barrier
\begin{mathletters}
\mlabel{creep.formula}
\begin{eqnarray}
L_F &\equiv& L_z(F) \approx  L_\xi 
\left(\frac{F}{F_\xi}\right)^{-1/(D+\zeta -\theta)} 
\sim F^{-1/(D+\zeta -\theta)}
\\
U(F) &\equiv& U(L_F) \approx U_\xi 
\left(\frac {F}{F_\xi}\right)^{-\mu}
\sim F^{-\mu}
\label{def.mu.creep}
\end{eqnarray}
\end{mathletters}
with the creep exponent
\begin{eqnarray}
\mu \equiv \frac \theta{D+\zeta -\theta} = \frac {2 \zeta +
  D-2}{2-\zeta}.
\label{result.mu}
\end{eqnarray}
We have used the abbreviation $F_\xi \equiv U_\xi L_\xi^{-D} \xi^{-1}$
and the identity (\ref{intro.theta}).  The effectiveness of barriers
of a given size $L_z$ is illustrated in figure \ref{fig.rel.barr}.  If
one denotes by $F_{\rm B}(L_z) \equiv U(L_z)/(L_x L_z^D) \sim
L_z^{\zeta-2}$ the threshold force density which is required to
overcome barriers of size $L_z$, the size of the largest effective
barrier is determined alternatively by $F=F_{\rm B}(L_F)$.

\begin{figure}
\centering
\epsfig{file=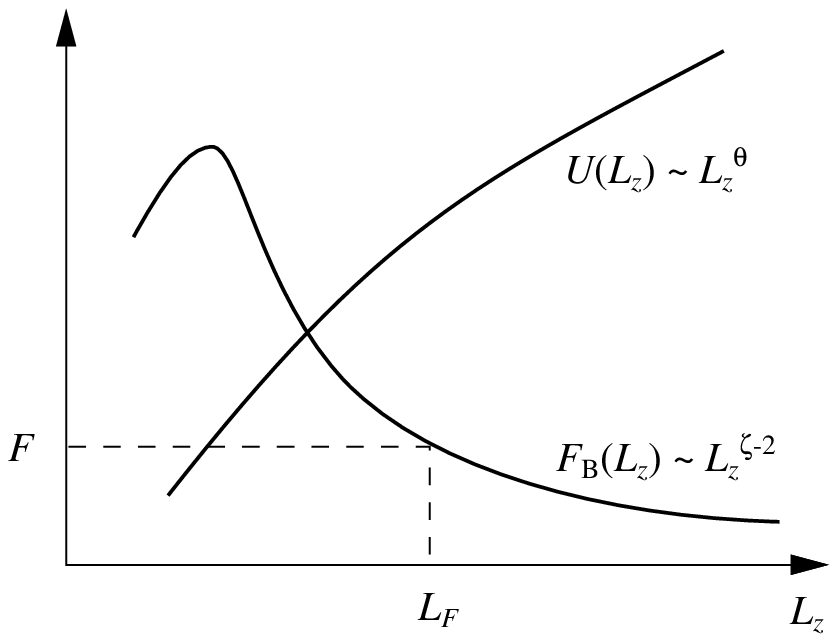,height=6cm}
\fcaption{The height of barriers scales like $U \sim
  L_z^\theta$ with the barrier size $L_z$ (in general, $\theta>0$ at a
  zero-temperature fixed point).  The effective pinning force of these
  barriers scales like $F_B \sim L_z^{\zeta -2}$ (the manifold model
  is valid only as long as $\zeta<1$).  In the presence of a driving
  force $F$ only barriers of a size $L_z < L_F$ are still effective.}
{fig.rel.barr}
\end{figure}

Since the time $t$ to overcome a barrier by thermal activation
increases exponentially with the height $U$,
\begin{eqnarray}
t(U) \sim e^{U/T},  
\end{eqnarray}
it is the {\em largest} effective barrier that dominates the dynamics
and determines the creep velocity
\begin{eqnarray}
v(F) \sim  (F/\eta_0) e^{- U(F)/T}  
\sim  (F/\eta_0) e^{- (U_\xi/T)(F/F_\xi)^{-\mu}  } .
\label{v.creep}
\end{eqnarray}
The pre-exponential factor has been inserted by hand for dimensional
reasons and to match the flow regime.  However, in the creep regime
this factor may be of more complicated form (corresponding to a
logarithmic contribution to $U(F)$) which is not captured by this
scaling argument.

From equation (\ref{v.creep}) we see that the dynamic response will be
exponentially small below depinning, since a disorder-dominated
roughness with exponent $\zeta > \zeta_{\rm th} = (2-D)/2$ implies
$\mu>0$.  The characteristic {\em dynamic} exponent $\mu$ depends only
on the dimension and the {\em static} roughness exponent of the
system.

In order to provide a deeper understanding of these scaling relations
several authors \cite{Radzihovsky98:mm,Balents+98:pc,Chauve+98} have
recently developed a renormalization group analysis for the whole
range of velocities. For $T=0$ their approach correctly reproduces the
properties of the depinning transition and for $T>0$ they obtain a
more precise characteristic $v(F)$, which confirms the exponential
dependence of the characteristic (\ref{v.creep}) to leading order for
small forces but contains also subleading orders.

Experimental observations on magnetic domain walls, charge-density
waves, and VLL are widely consistent with this scaling picture of
barriers.  The most explicit measurement of the roughness and creep
exponent examining magnetic domain wall creep \cite{Lemerle+98} was
made only recently and is in good agreement with the theoretical
results.

\subsubsection{Flow}

The flow regime at large driving force $|\bF| \gg F_c$ does not
usually receive much attention since it is considered trivial and is
expected to be captured within perturbation theory.  However, recently
the flow regime has attracted some attention for driven VLL, where it
turned out to be quite non-trivial and inaccessible to a perturbative
analysis (see section \ref{sec.dyn.phases}) due to an intricate
interplay of pinning forces and driving force.  Therefore we will
briefly sketch this regime also for the manifolds.

Let us consider again zero temperature. There the displacements are
expected to be very small for large driving forces: the manifold can
respond to the pinning force only with certain (scale-dependent)
relaxation times and the pinning force changes rapidly in time for
large $F$. Therefore the effect of pinning effectively gets averaged
out for $F \to \infty$.

Since the disorder induced displacements are small at large $F$, one
may again perform a perturbative analysis as a first step.  According
to the equation of motion (\ref{eqmo.mani}) the effect of disorder is
given to lowest order by a pinning force
\begin{eqnarray}
\bF^{\rm pin}(\bz,t)=- \grad_\perp V(\bX + \bv t,\bz),
\end{eqnarray}
which shakes the manifold as it moves over the pinning potential.  Its
average is zero and its variance is
\begin{eqnarray}
\overline{F^{\rm pin}_\alpha (\bz,t) F^{\rm pin}_\beta (\bN,0)}=
- \partial_\alpha \partial_\beta \Delta(\bv t) \delta(\bz) .
\end{eqnarray}
For qualitative purposes this shaking force might be compared to an
effective thermal noise.  Because the velocity selects a particular
space direction this noise in no longer isotropic.  Effective shaking
temperatures (which were introduced by \tcite{Koshelev+94} for vortex
lattices) for the directions parallel and perpendicular to $\bv
\equiv(v,\bN)$ can be identified from
\begin{mathletters}
\mlabel{T.sh.mani}
\begin{eqnarray}
T^{\rm sh}_\alpha&=& \frac 1{2\eta_0} \int dt d^Dz \ 
\overline{F^{\rm pin}_\alpha (\bz,t) F^{\rm pin}_\alpha (\bN,0)} 
\\
T^{\rm sh}_1&=& 0,
\\
T^{\rm sh}_\alpha &\sim& \frac {\Do}{2 \eta_0 v \xi^{D+1}}, \quad
(\alpha \geq 2).
\end{eqnarray}
\end{mathletters}
These shaking temperatures are to be added to the physical temperature
to give the total effective temperature, which is finite.  Thus, for
$D>2$ the manifold should be flat with exponent $\zeta=\zeta_{\rm
  th}=\frac{2-D}2$ as in the absence of disorder.  This flatness is
expected to persist at any finite velocity, i.e., for all forces above
$F_c$.  However, it is interesting to observe that in the flow regime
the manifold has a larger width in the direction perpendicular to
$\bv$ since $T^{\rm sh}_\alpha > T^{\rm sh}_1$ (with $\alpha \geq 2$).
On the contrary, at depinning the manifold has a larger width in the
direction parallel to ${\bf v}$ since $\zeta_\alpha<\zeta_1$, see
equation (\ref{dep.exp}).  This means that at a certain velocity the
manifold has to {\em reverse the aspect ratio} of its widths.  Apart
from this interesting observation, perturbation theory gives no hints
of qualitative changes of the state with changing velocity such as
dynamic phase transitions.  However, such transitions exist in {\em
  periodic} structures such as VLL (section \ref{sec.dyn.phases}).

\subsection{Summary}
\label{sec.mani.sum}

In this chapter we have reviewed various properties of manifolds in
order as to identify a {\em glassy} state of the system.  Of course,
disorder always quantitatively modifies the state on small length
scales.  Some large-scale properties of the system can be modified
qualitatively by disorder and are hence suitable for the
characterization of glassiness:
\begin{itemize}
\item[(i)] Disorder reduces the {\em positional order} and it
  increases the roughness. Typically, $\zeta > \zeta_{\rm th}$ but
  such an increase is also possible for $\zeta = \zeta_{\rm th}$ if
  $W$ is increased only by logarithmic factors.
\item[(ii)] Disorder leads to large {\em sample-to-sample fluctuations},
  which can be described by susceptibilities.
\item[(iii)] In a renormalization-group description of the large-scale
  physics, there is a {\em disorder-dominated fixed point}. Typically,
  this is a zero-temperature fixed point. In low dimensions ($D\leq
  2$), this may also be a finite-temperature fixed point.
\item[(iv)] Disorder induces thermally activated {\em sub-linear response}
  to a driving force.
\end{itemize}
We consider all these features as generic to identify the glassy
nature of a system.  The insufficiency of individual features can be
shown by examples: An increased roughness alone is not significant as
emphasized by \tcite{Hwa+94:prl} using the bond-disorder model, which
has no broken translation symmetry.  Sub-linear response is a
signature of broken translation symmetry but can also be achieved by
`ordered' potentials.  Similarly, a zero-temperature fixed point can,
in general, represent a low-temperature ordered phase which is not
necessarily a glass.  The sine-Gordon model provides an example for
the last two statements.

  \setcounter{equation}{0}
  \section{Superconducting film in a parallel field}
\label{sec.film.perp}

In treating a many-vortex system we begin with the technically most
simple problem of a superconducting film in a magnetic field {\em
  parallel} to the film plane. The mean-field phase diagram of this
problem was considered by \tcite{Abrikosov64}, who found for $\Hci < H
< \Hcii$ a solution with an equidistant vortex lattice of spacing $a$.
Here $\Hci= \frac 2 \pi \frac{\Phi_0}{s^2} \ln \frac s \xi$ and $\Hcii
=\frac{\sqrt 3} \pi \frac {\Phi_0}{\xi s}$, where $s$ is the thickness
of the film and we assume $\lambda > s/\pi \gtrsim \xi$ (see
\tcite{Sonin92} for films described by the Lawrence-Doniach model).
Such a geometry can be realized experimentally and the vortex physics
was observed recently on mesoscopic scales \cite{Bolle+99}.

In such a planar geometry vortices are not strictly confined within
the film. In principle they may leave and enter the film on a side.
However, such events, as well as a crossing of vortex lines within the
film, always cost energy.  In our theoretical analysis we will neglect
such events.  Then, by construction, vortex lines leave the
superconducting plane only at its boundary, the vortex lattice does
not exhibit topological defects like dislocations and, hence, it
cannot melt unless we reach $\Tc$.  Thermal fluctuations lead
therefore to a {\em quasi-long-range ordered lattice} as in other 2D
crystals \cite{Mermin+66} which persists roughly up to the mean-field
transition temperature $\Tco$.  In the following we consider the
influence of frozen-in disorder on this system. It turns out that the
problem can be mapped onto a magnetic model, the $XY$ model in a
random field, for which many results are known. In addition to the
phase with quasi-long-range order there is a low-temperature {\em
  glassy} phase, which is the result of the balance between thermal
and disorder fluctuations (in this sense, the situation is different
from that for bulk materials, where thermal fluctuations have a much
weaker effect on the formation of the glassy state). Drift of the
vortices due to an external current (perpendicular to the film plane)
leads in the glassy phase to a {\em vanishing linear resistivity}.

\subsection{Mapping on the $XY$ model in a random field}
\label{sec.XY.map}

The mapping of this model onto a two-dimensional $XY$ model in a
random field was first considered by \tcite{FisherMPA89} in his
seminal paper on vortex glasses.  Here we will follow a slightly
different derivation.

To formulate the model, we assume that the superconducting film is in
the $(x,z)$ plane and that the vortex lines are directed along the $z$
axis with an average line spacing $a$. The position of the $n$th
vortex line is denoted by $x_n(z)=X_n + u_n(z)$, where $X_n=n a$ and
$u_n(z) \equiv u(X_n, z)$ are the rest position and the displacement,
respectively. The Hamiltonian can then be written in the form ${\cal
  H}= {\cal H}_{\rm el}+{\cal H}_{\rm pin}$, where
\begin{mathletters}
\begin{eqnarray}
{\cal H}_{\rm el}&=& \int d^2 r \left\{ 
\frac {c_{11}}2 (\partial_x u)^2 + \frac {c_{44}}2 (\partial_z u)^2 \right\},
\label{H.el.2+0}
\\
{\cal H}_{\rm pin}&=& \int d^2 r \ \rho_u({\bf r}) V({\bf r}).
\label{H.pin.2+0}
\end{eqnarray}
\end{mathletters}
Here ${\bf r}\equiv(x,z)$, $c_{11}$ and $c_{44}$ are the compression
and tilt elastic constant, $\rho_u({\bf r})$ the vortex line density and
$V({\bf r})$ the random pinning potential. For the latter we assume
for simplicity
\begin{mathletters}
\begin{eqnarray}
\overline{V({\bf r})}&=&0,
\\
\overline{V({\bf r})V({\bf r}')}&=& \Delta(x-x') \delta(z-z') 
\end{eqnarray}
\end{mathletters}
and, as before, $\Delta(x) \approx \Do/ (\sqrt{2 \pi} \xi) e^{- x^2/2
  \xi^2}$, but $\Do \approx n_{\rm i}^{(2)} f_{\rm pin}^2 \xi^3 s$,
where $n_{\rm i}^{(2)}$ and $f_{\rm pin}$ denote the impurity density
per unit area and the individual force of a single impurity on the
vortex line, respectively \cite{Blatter+94}.

In general, the elastic Hamiltonian will be non-local, reflecting the
dispersion of the elastic constants for wave vectors $|\bk| >
\lambda^{-1}$. Since we, however, are interested in the disordered
case, which is dominated by large scale fluctuations, we ignore this
fact here.  This approximation is justified in particular for weak
disorder, a case in which all relevant length scales are much larger
than $\lambda$.  Thermal and disorder fluctuations will considerably
renormalize $c_{11}$ and $c_{44}$ with respect to their mean-field
values.

If one starts with isolated vortex lines of displacements $u_n$ which
have a stiffness constant $\st$ and a mutual short-ranged interaction
energy $U(a+u_{n+1}-u_n)$, then
\begin{eqnarray}
  c_{11}=a U''(a), \quad c_{44}=\st/a.
\end{eqnarray}
On scales $a \gg \lambda$ mean-field theory yields $U(a)=U_{\rm MF}(a)
\sim e^{-a/\lambda}$.

Thermal fluctuations on scales $|\Delta x| \leq a$ lead to a steric
repulsion \cite{Pokrovsky+79} between the vortex lines, which reads for
$\lambda \to 0$
\begin{eqnarray}
U_{\rm th}(a)=  \frac{T^2}{\st a^2}\frac {\pi^2}6 , 
\label{PT.rep}  
\end{eqnarray}
such that $(c_{11} c_{44})^{1/2}/T=\pi/a^2$ becomes independent of
temperature. As we will see below [cf. (\ref{intro.T_G.1+1})], this
relation is exactly the condition for the glass transition temperature
$\Tg=(c_{11} c_{44})^{1/2} a^2/\pi$. In other words, the relation
(\ref{PT.rep}) maps the system at {\em all} temperatures to the glass
transition temperature $\Tg$.  However, in deriving (\ref{PT.rep})
only a hard-core repulsive interaction between vortex lines was
assumed (indeed, \tcite{Pokrovsky+79} mapped their problem onto
non-interacting fermions). If we include an additional interaction
energy of the order of $U_c$ per contact point of two fluctuating
vortex lines, we get
\begin{eqnarray}
c_{11}^{\rm th}(a) \approx \frac{T^2}{\st a^3} \left({\pi^2}
  +\frac{U_c}T \right).
\end{eqnarray}
This leads to $\Tg \equiv (c_{11} c_{44})^{1/2} a^2/\pi=T(1+ U_c/\pi^2
T)^{1/2}$, which corresponds to $T \gtrless \Tg$ for $U_c \lessgtr 0$.

Disorder leads also to a steric repulsion
\cite{Kardar+85,Nattermann+88:fl} of the form (we drop here
coefficients of order unity because they cannot be determined very
accurately)
\begin{eqnarray}
  U_{\rm pin}(a) \approx \frac{\Do}{T+T_\Delta} \frac 1a,
\end{eqnarray}
where $T_\Delta \approx (\st \xi \Do)^{1/3}$
\cite{Nattermann+88:rep}. This gives 
\begin{eqnarray}
\frac{\Tg}T = \frac {a^2}{\pi T} \sqrt{c_{11} c_{44}} \approx
\frac{T_\Delta} T 
\left(\frac a \xi \frac {T_\Delta}{T+T_\Delta} \right)^{1/2}
\approx \sqrt{\frac{a}{a_c}}.
\end{eqnarray}
The vortex-line distance
\begin{eqnarray}
a_c \approx \xi \frac{T^2 (T+T_\Delta)}{T_\Delta^3}
\end{eqnarray}
denotes a cross-over length scale from a thermal steric repulsion (for
$a \lesssim a_c$) to a disorder dominated steric repulsion (for $a
\gtrsim a_c$).  The formulas for $U_{\rm pin}$ and $a_c$ are not exact
but represent crude interpolation formulas between the limiting cases
$T \ll T_\Delta$ and $T \gg T_\Delta$.

Finally, a contact interaction can also be included in $U_{\rm
  pin}(a)$ such that
\begin{eqnarray}
c_{11}^{\rm pin}(a) \approx \frac{T^2}{\st a^2a_c}\left( 1+ \frac{U_c}T
  \sqrt{\frac{a_c}a} \right).  
\end{eqnarray}
Here we have written $c_{11}^{\rm pin}(a)$ in a form similar to
$c_{11}^{\rm th}(a)$. Comparing these two quantities, it is easy to
see that $c_{11}^{\rm pin}(a) > c_{11}^{\rm th}(a)$ in the region
$a>a_c$, and hence we are below $\Tg$ (provided $U_c>0$).

One should, however, take into account that all these expressions were
derived under the assumption that the bare interaction between vortex
lines is short ranged, i.e., for the case $\lambda \ll a$. In the
opposite case the calculation is more involved and will not be
considered here further.

Next we consider the pinning energy (\ref{H.pin.2+0}).  Using the
Poisson-summation formula, we may rewrite the density $\rho_u({\bf
  r})$ as (\citeNP{Nattermann90,Nattermann+91}; see also appendix
\ref{sec.peri})
\begin{eqnarray}
\rho_u({\bf r})&=&\sum_{n=-\infty}^\infty \delta(x-X_n - u_n(z))
\nonumber \\
&=& \sum_{n=-\infty}^\infty \int_{-\infty}^\infty dX \ 
\delta(X-X_n) \delta(x-X - u_n(z))
\nonumber \\
&=&\frac 1 {a} \sum_{m=-\infty}^\infty \int_{-\infty}^\infty dX \ 
e^{i Q_m X} \delta(x-X-u(X,z))
\nonumber \\
&=& \frac 1 {a} \int_{-\infty}^\infty dX \ (1+2 \sum_{m \geq 1}
\cos(Q_m X))
\delta(x-X-u(X,z))
\nonumber \\
&\approx& \frac 1 {a} \bigg\{ 1-\partial_x u({\bf r}) +
2 \sum_{m \geq 1} \cos(Q_m [x-u({\bf r})])
\bigg\}.
\label{intro.rho}
\end{eqnarray}
$Q_m \equiv 2 \pi m/a$ is a reciprocal lattice vectors of the 2D line
array.  The pinning energy (\ref{H.pin.2+0}) can hence be written as
\begin{mathletters}
\begin{eqnarray}
{\cal H}_{\rm pin}&=&  \int d^2 r \  \bigg\{
-\frac 1 {2 \pi} \partial_x \vphi({\bf r}) V({\bf r})
\nonumber \\ &&
+  \sum_{m \geq 1} 
[h_{1m}({\bf r}) \cos(m \vphi({\bf r})) 
+ h_{2m}({\bf r}) \sin(m \vphi({\bf r})) ]
\bigg\},
\\
h_{1m}({\bf r})&\equiv&\frac 2 {a} V({\bf r}) \cos(Q_m x),
\quad \quad
h_{2m}({\bf r}) \equiv \frac 2 {a} V({\bf r}) \sin(Q_m x),
\end{eqnarray}
\end{mathletters}
where we introduced the phase field $\vphi \equiv 2 \pi u/a$ [which
should not be confused with the phase $\phi$ of the superconducting
order parameter; see equation (\ref{eq.phi.xz}) below]. The disorder
fields $h_{\alpha m}$ ($\alpha, \beta =1,2$) are Gaussian distributed
with
\begin{mathletters}
\begin{eqnarray}
\overline{h_{\alpha m}({\bf r})}&=&0  ,
\\
\overline{h_{\alpha m}({\bf r}) h_{\beta n}({\bf r}')}&\approx& 
2 \frac {\Do} {a^2} f_{\alpha m,\beta n} (Q_m x,Q_n x')
\delta_\xi({\bf r}-{\bf r}').
\end{eqnarray}
\end{mathletters}
On scales $\Delta x \gg a$, $f_{\alpha m,\beta n} (Q_m x,Q_n x')
\approx \delta_{\alpha \beta} \delta_{mn} $ since the rapidly
fluctuating contributions with $m \neq n$ or $\alpha \neq \beta$
average to zero. On these scales, ${\cal H}_{\rm pin}$ can also be
written in the form
\begin{eqnarray}
{\cal H}_{\rm pin} \approx  \int d^2 r \  \bigg\{
-\frac 1 {2\pi} \partial_x \vphi ({\bf r}) V({\bf r})
+  \sum_{m \geq 1} \frac 2a \Do^{1/2}
 \cos(m \vphi({\bf r}) - \aphi_m(\br)) 
\bigg\}.
\label{H.ranpha}
\end{eqnarray}
The random phases $\aphi_m(\br)$ obey
\begin{mathletters}
\begin{eqnarray}
\overline{e^{i \aphi_m(\br)}}&=&0  ,
\\
\overline{e^{i [\aphi_m(\br)- \aphi_n(\br')]}}
&=& \delta_{mn} \delta_\xi({\bf r}-{\bf r}').
\end{eqnarray}
\end{mathletters}
The effect of fluctuations of individual vortex lines on scales
$\Delta x \lesssim a$ was discussed briefly in the first part of this
section.  Since we study the film in a parallel field mainly because
of the possibility of obtaining explicit results for the vortex-glass
state in a case which we understand to a large extent (and not so much
because of its experimental relevance), we will ignore collective
fluctuation effects on small and intermediate scales but concentrate
directly on the large-scale behaviour where typical phase fluctuations
are of the order $2 \pi$. On these scales terms with $m>1$ in
(\ref{H.ranpha}) can be ignored since they are less relevant than that
with $m=1$.

It is convenient to rewrite (\ref{H.el.2+0}) as 
\begin{eqnarray}
{\cal H}_{\rm el}&=& \frac J2 \int d^2 r \ ({\bbox \nabla} \vphi)^2
\label{H.el.2+0.resc}
\end{eqnarray}
by rescaling the $z$ coordinate and introducing the stiffness constant
$J$ according to
\begin{mathletters}
\begin{eqnarray}
\label{rescal}
\tz &\equiv& \left( \frac{c_{11}}{c_{44}} \right)^{1/2} z,
\\
J &\equiv& \left( \frac{a}{2 \pi} \right)^2 \sqrt{{c_{11}}{c_{44}}}.
\label{intro.J}
\end{eqnarray}
\end{mathletters}
This rescaling leaves ${\cal H}_{\rm pin}$ form-invariant apart from a
multiplication with a factor $(c_{44}/c_{11})^{1/4}$. Thus ${\cal
  H}_{\rm el}+{\cal H}_{\rm pin}$ now equals the Hamiltonian of a
random-field (RF) $XY$ model without vortices, which was considered in
the context of magnetic models by a number of authors
\cite{Cardy+82,Goldschmidt+82,Goldschmidt+84,Villain+84}.

After rescaling (\ref{rescal}), ${\cal H}_{\rm pin}$ can be written as
\begin{eqnarray}
{\cal H}_{\rm pin} = \int d^2 r \ \left\{ \tV(\vphi({\bf r}),{\bf r}) -
  {\bbox \mu}({\bf r}) \cdot {\bbox \nabla} \vphi({\bf r}) \right\}
\label{H.pin.rec.2+0}
\end{eqnarray}
with 
\begin{mathletters}
\mlabel{mu.corr}
\begin{eqnarray}
\overline{{\bbox \mu}}&=&\overline{\tV}=0 ,
\\
\overline{ \mu_i({\bf r}) \mu_j ({\bf r}')} &=& J^2 \sigma \delta_{ij}
\delta_\xi ({\bf r}-{\bf r}'),
\label{intro.sigma}
\\
\overline{\tV(\vphi,{\bf r}) \tV(\vphi',{\bf r}')} &=& h^2 \cos(\vphi-\vphi')  
\delta_\xi ({\bf r}-{\bf r}'),
\label{intro.h.rf}
\end{eqnarray}
\end{mathletters}
where we introduced the variances $\sigma$ and $h^2$.  Their bare
(unrenormalized) values are given by
\begin{mathletters}
\mlabel{def.h.sig}
\begin{eqnarray}
h_0^2 &\equiv& 2 \frac{\Do}{a^2} 
\left( \frac {c_{44}}{c_{11}} \right)^{1/2},
\\
\sigma_0 &\equiv & \frac{h_0^2 a^2}{8 \pi J^2} .
\end{eqnarray}
\end{mathletters}
This relation between $\sigma$ and $h^2$ in (\ref{def.h.sig}) holds
only for the unrenormalized quantities. However, under renormalization
these quantities flow independently.  Although (\ref{H.ranpha})
originally contained  only a linear gradient term in the $x$ direction,
we rewrote it here in an isotropic form, anticipating the generation
of $\partial_z \vphi$ terms under the renormalization-group
transformation.

Before we come to the renormalization-group calculation for the model
(\ref{H.pin.rec.2+0}), we will apply a {\em Flory-type analysis} to
it.  To be more general, we will present this analysis in $d$
dimensions. One complication with respect to the problem considered in
section \ref{sec.struct}c is the fact that the correlator of the
pinning Hamiltonian now contains the oscillatory term
(\ref{intro.h.rf}).  For weak disorder $ha \ll J$, a given
configuration of the random phase $\aphi_1(\br) \equiv \aphi(\br)$
implies a certain ground-state configuration $\vphi_0(\br)=
\vphi_0(\br,\{\aphi\})$ which depends on the value of the random phase
everywhere in the system.  Going through the different configurations
of $\{\aphi\}$ will generate a distribution of ground state
configurations $\{\vphi_0\}$ which we assume to be Gaussian
distributed.  Since $ha \ll J$, the correlation of the value of
$\aphi(\br)$ and $\vphi_0(\br)$ at the same position $\br$ will be
weak.  In averaging over the distribution of $\aphi(\br)$ we can
therefore neglect the {\em local} correlations between $\aphi(\br)$
and $\vphi_0(\br)$ in a first approximation.  The typical energy gain
from the pinning term then becomes
\begin{eqnarray}
  E_{\rm pin} \approx 
- \max \left\{ J \sqrt\sigma L^{(d-2)/2} ({\overline{\vphi^2}})^{1/2}
, h L^{d/2} e^{-{\overline{\vphi^2}}/2} \right\} .
\label{E.pin.typ}
\end{eqnarray}
Depending on the dimension $d$, the leading contribution to $E_{\rm
  pin}$ can be the random-tilt or the random-field contribution.  The
dependence of $\overline{\vphi^2}$ on the system size $L$ follows from
equating the absolute value of $E_{\rm pin}$ to the averaged elastic
energy
\begin{eqnarray}
  E_{\rm el} \approx J L^{d-2} \overline{\vphi^2}.
\end{eqnarray}
For $2<d<4$ one finds that the random-field contribution dominates in 
$E_{\rm pin}$, which results in 
\begin{eqnarray}
  \overline{\vphi^2} \approx (4-d) \ln \frac L{L_\xi} 
-2 \ln [\overline{\vphi^2}],
\end{eqnarray}
where we introduced the Larkin length $L_\xi \approx (J/h)^{2/(4-d)}$.
Thus, the Flory argument gives at $T=0$ a logarithmic increase of
$\overline{\vphi^2}$ in all $2<d<4$. We will see in section
\ref{sec.bulk} that a renormalization-group calculation confirms
essentially this result, apart from a (small) modification of the
coefficient of the logarithm.  For $d<2$ on the other hand, $E_{\rm
  pin}$ is dominated by the random-tilt contribution and the balance
of $E_{\rm pin}$ and $E_{\rm el}$ at $T=0$ results in
$\overline{\vphi^2} \approx \sigma L^{2-d}$ in agreement with
renormalization-group calculations \cite{Villain+84}.  We will show
below that in $d=2$ dimensions $\sigma$ is renormalized to large
values such that $E_{\rm pin}$ and $\overline{\vphi^2}$ are dominated
by the random-tilt contribution.

\subsection{Renormalization}

After replication, we get from (\ref{H.pin.rec.2+0}) and (\ref{mu.corr}) 
\begin{eqnarray}
  {\cal H}_n=\sum_{ab}\int d^2 r \ \left\{ 
\frac 12 J (\delta^{ab}-
\frac{J \sigma}T ) {\bbox \nabla} \vphi^a \cdot {\bbox \nabla}
\vphi^b - \frac{h^2}{2T} \cos(\vphi^a- \vphi^b) \right\}.
\label{H.2+0.n}
\end{eqnarray}

The RG flow equations of this model, first found by \tcite{Cardy+82},
are then
\begin{mathletters}
\mlabel{CO.flow}
\begin{eqnarray}
\frac{ dJ}{dl} &=& 0,
\\
\frac {d \sigma}{dl} &=& c_1 \frac {h^4}{T^2 J^2},
\label{flow.sigma.2+0}
\\
\frac{ dh}{dl} &=& \left( 1 - \frac T{\Tg} \right) h - c_2 \frac{h^3}{T^2},
\label{flow.h.2+0}
\end{eqnarray}
\end{mathletters}
with the glass transition temperature 
\begin{eqnarray}
\Tg = 4 \pi J.
\label{intro.T_G.1+1}
\end{eqnarray}
The integration of (\ref{flow.h.2+0}) yields 
\begin{eqnarray}
h^2(L) = h_0^2 \left(\frac La \right)^{2 \taug} 
\left\{ 1 + \frac{h_0^2}{{h^*}^2} \left[\left(\frac La \right)^{2
      \taug} -1 \right] \right\}^{-1}
\end{eqnarray}
with $l=\ln(L/a)$, the fixed-point value ${h^*}^2 = T^2 \taug/c_2$ and
the reduced temperature
\begin{eqnarray}
\taug \equiv 1 - \frac{T}{\Tg}.  
\label{intro.tau}
\end{eqnarray}
It can be shown quite generally that the flow of $\sigma$ does not
feed back into that of $h$ \cite{Hwa+94:prl}.  The constants $c_1$ and
$c_2$ are cut-off dependent, but their ratio
\begin{eqnarray}
\frac{c_1}{c_2^2}= \frac 1 {4 \pi} + O(\taug)
\end{eqnarray}
is universal. The whole calculation is valid only for $\taug \ll 1$.
Considering equivalent models, equivalent RG equations were found by
\tcite{Goldschmidt+82}, \tcite{Rubinstein+83}, \tcite{Goldschmidt+84},
and \tcite{Paczuski+91}.  The parameter flow is shown schematically in
figure \ref{fig.CO.flow}.

\begin{figure}
\centering
\epsfig{file=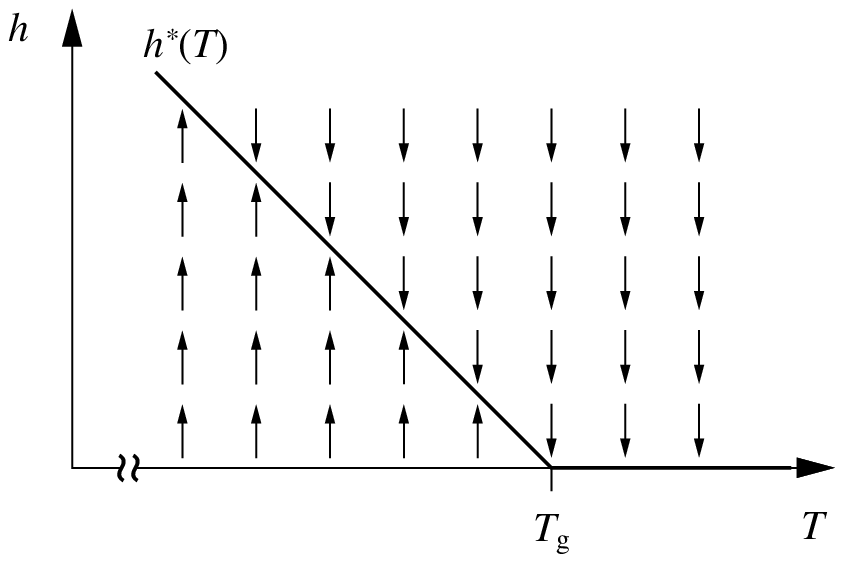,height=6cm}
\fcaption{RG flow in the $(h,T)$-plane. The random-field amplitude $h$
  flows to $h^*=0$ above the glass transition temperature ($T>\Tg$)
  whereas it flows to $h^* \propto (\Tg-T)$ below ($T<\Tg$).}
{fig.CO.flow}
\end{figure}

The quantity $h^2$ approaches a fixed point ${h^*}^2 \approx
{\Tg^2 \taug}/{c_2} $. Integrating (\ref{flow.sigma.2+0}) for $h
\approx h^*$, we find on the length scale $L=a e^l$ 
\begin{eqnarray}
\sigma(L) \approx 4 \pi \taug^2 \ln (L/a) .
\label{s(L).CO}
\end{eqnarray}
If we go back to the renormalized but {\em unrescaled} quantities --
these are the physical quantities we would measure in an experiment on
this scale -- we obtain for the effective parameters on scale $L$
\begin{mathletters}
\begin{eqnarray}
h_{\rm eff} &\approx  & h(L) \frac aL,
\\
\sigma_{\rm eff} &=& \sigma (L).
\end{eqnarray}
\end{mathletters}
Thus, while the effective random-field strength vanishes as $L^{-1}$,
the variance of the coefficient of the linear gradient term grows
logarithmically with length scale $L$.  As a side remark we mention
here that the effective value of the coefficient of the non-linear
term in ${\cal H}_n$ scales as $L^{-2}$ in agreement with the general
result for $\Delta_{\rm eff} \sim h_{\rm eff}^2 \sim L^{d-4}$, cf.
equation (\ref{tD_eff.2}) below.

It is interesting that \tcite{Villain+84} considered this problem at
$T=0$ in a complementary study and found the following set of RG flow
equations (we rewrite here their discrete recursion relations in a
differential form):
\begin{mathletters}
\mlabel{VF.flow}
\begin{eqnarray}
\frac{ dJ}{dl} &=& 0,
\\
\frac {d \sigma}{dl} &=& \tc_1 \frac {h^2}{J^2 + \tc_2 h^2},
\\
\frac{ dh}{dl} &=& \frac{h J^2}{J^2 + \tc_2 h^2},
\label{VF.flow.h}
\end{eqnarray}
\end{mathletters}
with some numerical constants $\tc_1$, $\tc_2$. $h$ flows now to
infinity, $h(L) \approx \sqrt{2/ \tc_2} J \ln L$, such that ${d
  \sigma}/{dl} \approx {\tc_1}/{\tc_2}$.  Thus both calculations, for
$T \lesssim \Tg$ and for $T=0$, give a logarithmic increase of
$\sigma_{\rm eff}(L)$ and a vanishing $h_{\rm eff}(L)$ for $L \to
\infty$. This gives further credibility to the existence of a unique
low-temperature phase.

A central role for the characterization of this phase plays the
displacement correlation function \cite{Goldschmidt+82}
\begin{eqnarray}
W({\bf r}) \equiv 
\overline{\langle [\vphi({\bf r}) - \vphi({\bf 0})]^2 \rangle} = 
\frac T{\pi J} \ln \frac ra + 2 \taug^2 \ln^2  \frac ra.
\label{log.sq}
\end{eqnarray}
Qualitatively, the same result was found by \tcite{Villain+84} at
$T=0$. From (\ref{log.sq}) it is possible to calculate the correlation
function of the order parameter for translational order $\psi_k({\bf
  r})= e^{i k u({\bf r})}$ [cf. equation (\ref{intro.psi_k})]. For
$k=Q_m=2 \pi m/a$ one has $ku = m \vphi$ and hence
\cite{Goldschmidt+82}
\begin{eqnarray}
S(Q_m,\br) &=&
\overline{\langle e^{i Q_m[u({\bf r})-u({\bf 0})]} \rangle} =
\overline{\langle e^{im[\vphi({\bf r})-\vphi({\bf 0})]} \rangle} 
\nonumber \\
&\approx& 
e^{-\frac {m^2}2 \overline{\langle 
[\vphi({\bf r})-\vphi({\bf 0})]^2 \rangle} } \sim 
r^{-m^2 \eta (r)},
\label{intro.eta}
\end{eqnarray}
where
\begin{eqnarray}
\eta(r) = \frac T {2 \pi J} + 2 \taug^2 \ln \frac ra , \quad T<\Tg.
\label{eta.r}
\end{eqnarray}
Thus, in the glassy phase, at $T<\Tg$, the correlation function
$S(Q,\br)$ decays {\em slightly} faster than with a power law.
However, the $\ln^2 r$-behaviour dominates $W(\br)$ only on scales
$|\br| \gg L_{\taug} \approx a e^{2/\taug^2}$ which is large within
the range of validity $\taug \ll 1$ of the RG equations
(\ref{CO.flow}).  The structure factor 
\begin{eqnarray}
\hS(Q+q_x,q_z) \approx \frac 1a \int d^2r \ e^{i(q_x x +q_z z)} \ S(Q,\br)  
\label{intro.S}
\end{eqnarray}
behaves for small $\bq=(q_x, \sqrt{c_{44}/c_{11}} q_z)$ near the first
reciprocal lattice vector $Q=\pm 2 \pi/a$, therefore, as
\begin{eqnarray}
\hS(Q+q_x,q_z) \sim q^{- 2 \taug[1-\taug \ln (aq)]}.
\end{eqnarray}
Contrary to the pure system, where quasi-long-range order is
accompanied by algebraic Bragg peaks, these are smeared out here for
$q \lesssim L_{\taug}^{-1}$.  For $T > \Tg$, the term $\propto \ln r $
in (\ref{eta.r}) is absent and the power law decay of correlations is
regained.

For completeness, and since the pair correlation function $S(Q,\br)$
vanishes faster than a power law, we also consider the positional
glass correlation function $S_{\rm PG}(Q,\br)$. cf. equation
(\ref{eq.D4}).  This was also calculated already by
\tcite{Goldschmidt+82}, who found
\begin{equation}
S_{\rm PG}(Q,\br)\sim |\br|^{-4T/\Tg}
\label{eq.D4a}
\end{equation}
for $|\taug|\ll 1$ on both sides of the transition.  Note that to the
given accuracy of order $\taug$ this glass correlation function decays
as a power law in both phases.  In the high-temperature phase, where
disorder is essentially irrelevant, the lack of a mechanism for
melting in the model -- in particular the absence of dislocations --
prevents $S_{\rm PG}$ from vanishing.  In the low-temperature glassy
phase, on the other hand, thermal fluctuations are still too strong to
allow for true long-range order of the glass correlation function in
this two-dimensional system.  Note, however, that the relative
magnitude of the correlation functions is qualitatively different in
both phases, since $S_{\rm PG}/S^2\to const.$ for $T > \Tg$ but
$S_{\rm PG}/S^2\to \infty$ for $T<\Tg$ and $|\br| \to \infty$.

We finally remark here that for this simple model, which has {\em no
  shear modes}, also the phase-coherent vortex glass correlation
$C_{\rm VG}(\br)$ (\ref{C_VG}) shows an algebraic decay.  Indeed,
since in this layered geometry the phase $\phi$ of the superconducting
order parameter changes by $\pi$ at a vortex line, the latter is
related to the displacement by
\begin{equation}
\phi(x,z)\approx\frac{\pi}{a}\Big(x-u(x,z)\Big) = \frac{\pi x}a -
\frac 12 \vphi(x,z).
\label{eq.phi.xz}
\end{equation}
With (\ref{C_VG}) and (\ref{eq.D4a}) we find therefore
\begin{equation}
C_{\rm VG}(\br)\sim|\br|^{-{T}/{\Tg}}\,,
\label{eq.C_VG.T/Tg}
\end{equation}
i.e., the system shows {\em quasi-long-range vortex-glass order}.

Our result (\ref{log.sq}) for the glassy phase is in contradiction
with a number of results by other authors using Bethe Ansatz
\cite{Tsvelik92,Balents+93:npb}, a variational treatment with one-step
replica-symmetry breaking \cite{Korshunov93:vg,Giamarchi+94}, and
variational methods without replicas \cite{Orland+95}.  In these
studies $W(\br) \propto \ln r$ was also found for the low-temperature
phase, but with a temperature independent coefficient for low $T$.  We
believe that these results are incorrect and in particular that they
demonstrate the flaws of the Gaussian variational method, which gives
only Flory-like results for the correlation function. Indeed,
\tcite{Bauer+96} showed for an $N$-component version of our model
(\ref{H.2+0.n}), that the coefficient of the $\ln^2 r$-term in
$W(\br)$ vanishes as $1/N^3$ for large $N$. This explains the absence
of the $\ln^2 r$-term in the variational calculations which give exact
results only in the limit $N \to \infty$.

Numerical studies give a controversial picture: while in the
investigations of \tcite{Batrouni+94} and \tcite{Cule+95} only a $\ln
r$-behaviour of $W(\br)$ was found, later studies
\cite{Lancaster+95,Marinari+95,Zeng+96,Rieger+97} were able to detect
a $\ln^2 r$-behaviour. However, the coefficient of the $\ln^2 r$ term
found in the finite-temperature simulations is much smaller than the
RG prediction (\ref{log.sq}), even if one takes into account that the
true coefficient $2 \taug^2$ is smaller by a factor $\frac 14$ than
assumed originally. The difficulty in observing the latter behaviour
may partially be explained by a large crossover length.  More
recently, \tcite{Zeng+99:2} were able to confirm in their simulation
both the $\ln^2 r$ dependence and the $\taug^2$ prefactor of the
correlation $W(\br)$.

The faster decay of correlations of the order parameter for
translational order in the low-temperature phase clearly indicates the
influence of disorder.  However, the resulting phase is not
necessarily a glassy phase.  Indeed, as was remarked by
\tcite{Hwa+94:prl} and mentioned in connection with our discussion of
manifolds, a harmonic toy model (\ref{H.el.2+0.resc}),
(\ref{H.pin.rec.2+0}) with $\tV(\vphi, \br )\equiv 0$ but [compare
(\ref{mu.lrc})]
\begin{eqnarray}
\overline{\mu_i(\br) \mu_j(\br')} = \delta_{ij}
J^2 \sigma f(\br-\br'), \quad f(\br) \sim
  |\br|^{-\alpha}, \quad \alpha < d
\end{eqnarray}
already results in
\begin{eqnarray}
W(\br)=\overline{\langle[\vphi(\br)-\vphi(\bN)]^2\rangle}  
\sim \sigma |\br|^{2- \alpha}.
\end{eqnarray}
Although the roughness exponent $\zeta= \frac 12 (2-\alpha)$ can be
larger than zero in $d=2$ dimensions, and hence $S(Q,\br)$ would decay
like a stretched exponential function, the system is certainly {\em
  not} glassy since it has indeed a continuum of ground states which
differ by the value of a constant $\vphi_1$ but have the same energy.
Thus, there is no pinning of the vortex-line lattice in this toy
model.

We conclude from this exercise that a stronger decay of the
correlation function $S(Q,\br)$ than that resulting from thermal
fluctuations is a necessary but not sufficient requirement for the
existence of a glassy phase. It is clear that the existence of many
meta-stable states and, hence, the anharmonicity of the model is
decisive.

\subsection{Susceptibility}

As a better signature of glassiness \tcite{Hwa+94:prl} proposed to
look at the response to a change $\delta {\bf H}= (\delta H_x ,\delta
H_z ) $ in the applied field. This change leads to a new term $- (
\Phi_0/8 \pi^2) \int d^2r \ \delta {\bf H} \cdot {\bbox \nabla} \vphi$
in the Hamiltonian. One can now consider the susceptibility
\begin{eqnarray}
  \chi_{ij} \equiv \frac \partial {\partial h_j} \langle \partial_i
  \vphi \rangle .
\end{eqnarray}
In isotropic systems $\chi_{ij}= \chi \delta_{ij}$ and  the
magnetic permeability is found as
\begin{eqnarray}
  \mu_{\rm mag} = \chi \Phi_0^2/(16 \pi^3) .
\label{intro.mu_mag}
\end{eqnarray}
It was then shown that the sample-to-sample fluctuations of $\chi$ or
$\mu_{\rm mag}$ in the low-temperature phase fulfill the relation
\begin{eqnarray}
  \overline{(\Delta \chi)^2} \equiv \overline{(\chi - \overline
    \chi)^2} = C \taug {\overline \chi}^2, \quad (T < \Tg),
\label{susc.fluct.1+1}
\end{eqnarray}
where $\overline \chi =1/J$ is independent of the disorder as a result
of an underlying statistical tilt symmetry \cite{Schulz+88}.  $C$ is a
universal boundary- and geometry-dependent coefficient and $\taug
\approx c_2 (h^*/\Tg)^2$ [cf. equation (\ref{intro.tau})] is a measure
of the effective non-linearity of the model on large length scales. On
the contrary, $\overline{(\Delta \chi)^2} \to 0$ for $T> \Tg$ and for
large systems.  The non-vanishing sample-to-sample fluctuations of the
susceptibility for $T < \Tg$, which may be tested already at a simple
sample by rotating the direction of the magnetic field, are therefore
a better way to define glassiness.  Although early numerical
simulations by \tcite{Batrouni+94} were not able to confirm prediction
(\ref{susc.fluct.1+1}), more recent simulations by \tcite{Zeng+99:2}
are in agreement with (\ref{susc.fluct.1+1}).

For completeness we mention in addition a study by \tcite{Sudbo93}, in
which he considers the response $D \sim \chi_{xx}$ of the vortex-line
array to a tilt of the magnetic field. By mapping the system of
hard-core repelling vortex lines onto non-interacting 1D fermions in a
{\em time dependent} random potential, he identifies a possible glassy
phase of vortex lines with the localization of fermions. For electrons
in a time {\em independent} random potential one expects a vanishing
average charge stiffness $\overline D$ in the localized phase, in
contrast to his Bethe-Ansatz calculation, which gives $\overline D
\neq 0$. However, as mentioned before, $\overline D \neq 0$ is the
consequence of an underlying statistical tilt symmetry (which exists
only for a fermion potential which is random both in space and in time
direction) and not a signature of the absence of a glassy phase
\cite{Hwa+93:prl}.

\subsection{Dynamics}

The physically most convincing way to demonstrate glassy properties of
a system is to consider the dynamics. Before we begin this topic, we
will make a small digression to discuss the appearance of electrical
resistivity from the motion of vortex lines. Under the influence of an
external current density $\bj$ a Lorentz force with density
\begin{eqnarray}
  \bF = \frac 1{\co} \bj \wedge \bB
\end{eqnarray}
acts on the vortex-line array. $\bB$ denotes its magnetic induction
and $c$ the velocity of light.  The force density leads under
dissipative conditions to a steady-state motion of vortex lines of
velocity $\bv=\bv(\bF)$, which generates an electric field $\bE=\bB
\wedge \bv /\co$. If $\bv$ and $\bF$ are parallel to each other,
$\bE$ and $\bj$ are as well. The resistivity of the vortex-line array
follows, therefore, from
\begin{eqnarray}
\rho(j)=\frac {dE}{dj} = \frac {B^2}{\co^2} \ \frac {dv}{dF} .
\end{eqnarray}
A simple situation exists if the relation $\bv(\bF)=\bF/\eta$ is
linear with a friction coefficient $\eta$, for which
$\rho=B^2/(\co^2\eta)$ holds.  An example for a linear relation is the
{\em Bardeen-Stephen flux flow} where $\eta= B \Hcii/(\co^2 \rhon)$
with the normal resistivity $\rhon$ \cite{Bardeen+65}.

We consider now the influence of disorder in the 2D case with the
magnetic field parallel to the film plane.  The equation of motion of
the vortex-line array under the influence of a driving force density
$F$ in $x$ direction (the external current is assumed to be
perpendicular to the film) reads
\begin{mathletters}
\begin{eqnarray}
  \eta_0 \dot \vphi &=& - \frac{\delta {\cal H}}{\delta \vphi} + f +
  \zeta,
\\
\langle \zeta ({\bf r},t)  \zeta ({\bf r}',t') \rangle &=& 2 \eta_0 T
\delta({\bf r}-{\bf r}') \delta(t-t') ,
\end{eqnarray}
\end{mathletters}
with $f \sim F$ and $\eta_0 \sim B \Hcii/\rhon$. 
 
From a dynamical RG one finds \cite{Goldschmidt+85,Tsai+92,Tsai+94:I}
a renormalization of the friction coefficient according to
\begin{eqnarray}
  \label{flow.eta}
  \frac{d \eta}{dl} = 4 c_2 \sqrt{\cg} \frac {h^2}{\Tg^2} \eta,
\end{eqnarray}
where $ \cg \equiv e^{2\gE} /4 $ and $\gE$ is Euler's
constant, i.e., $2\sqrt{\cg} \approx 1.78$.   $c_2$ was introduced in
equation (\ref{CO.flow}c).   The integration of (\ref{flow.eta}) gives
for the effective friction constant on the scale $L=a e^l$
\begin{eqnarray}
\eta(L) \approx \eta_0 \left[1+ \frac{c_2}\taug \left(\frac
    {h_0}{\Tg}\right)^2 \left((L/a)^{2 \taug} -1 \right) \right]^{2 \sqrt
  {\cg}} \sim L^{z-2}
\label{eta.L}
\end{eqnarray}
for $T< \Tg$, where we used (\ref{CO.flow}c).  Because of the absence
of a trivial rescaling in (\ref{flow.eta}), $\eta(L)$ coincides with
the effective friction coefficient $\eta_{\rm eff}(L)$.

The relation $\eta(L) \sim L^{z-2}$ defines the dynamical critical
exponent $z$ with
\begin{eqnarray}
z= \left\{
    \begin{array}{ll}
      2+ 4 \sqrt {\cg} \taug & \textrm{ for } T<\Tg
\\
      2 & \textrm{ for } T \geq \Tg.
    \end{array}
\right.
\label{intro.z}
\end{eqnarray}
from (\ref{eta.L}).

For $T>\Tg$, $\taug<0$ and the asymptotic ($L \to \infty$) value
$\eta_\infty$ of $\eta$ is
\begin{eqnarray}
\eta_\infty(\taug) \approx \eta_0 \left(1 + c_2 \frac{h_0^2}{\Tg^2}
  |\taug|^{-1} \right)^{2 \sqrt {\cg}} \sim |\taug|^{-2 \sqrt {\cg}}.
\end{eqnarray}
From this one finds for the linear resistivity close to $\Tg$
\begin{eqnarray}
\rho_\infty \approx \rhon \frac B \Hcii 
\left(1 + c_2 \frac{h_0^2}{\Tg^2}
  |\taug|^{-1} \right)^{-2 \sqrt {\cg}}
\sim |\taug|^{2 \sqrt {\cg}}.
\end{eqnarray}
In other words, we obtain for the depinned solid phase a linear
(Ohmic) resistivity which vanishes at $\Tg$ as a power law in
$|\taug|$ with the exponent $2\sqrt{\cg}$.

For $T < \Tg$, $\taug>0$ and $\eta(L)$ increases with increasing
length scale until $L \approx L_{\rm max} \approx a (J/F)^{1/2}$ is
reached where the RG flow is stopped \cite{Tsai+92}. The response is
therefore non-linear with ($F \ll J$)
\begin{eqnarray}
\eta_{\rm eff} (\taug>0,F) \approx \eta_0  
 \left[1+ \frac{c_2}\taug \left(\frac
    {h_0}{\Tg}\right)^2 \left((J/F)^{\taug} -1 \right) \right]^{2 \sqrt
  {\cg}},
\end{eqnarray}
which gives a {\em non-linear} resistivity
\begin{eqnarray}
\rho_{\rm eff}(j) \approx \rho_n \frac{B} \Hcii 
\left[1+ \frac{c_3}\taug \left(\left(\frac{j_c}j\right)^{\taug} -1 \right) 
\right]^{-2 \sqrt  {\cg}} \sim j^{({z-2})/2},
\label{non-lin.1.1.2}
\end{eqnarray}
where $j_c$ has the meaning of a zero-temperature critical current
density, $c_3= O(1)$ is a constant and $j \ll j_c$. Apparently, the
linear resistivity $\rho_{\rm eff}(j \to 0)$ vanishes, i.e., the
system in the glassy phase is a {\em true superconductor}.

For $\taug \searrow 0$ we obtain in particular at the vortex-glass
transition
\begin{eqnarray}
\rho_{\rm eff}(j) \sim   \rho_{\rm n} \frac{B} \Hcii  
\left[\ln \frac {j_c}j \right]^{-2 \sqrt {\cg}},
\end{eqnarray}
i.e., the system is a true superconductor also at $\Tg$.

The transition to the high-temperature phase is therefore continuous
at $\Tg$ where $\rho(0)=0$.  The results of the RG calculation were
derived under the assumption $|\taug| \ll 1$ and no definite
conclusions can be drawn for $T \ll \Tg$. However, it is very likely
that the dynamical critical exponent $z$ diverges as $T$ goes to zero.
The increase of $z$ with decreasing $T$ was confirmed numerically by
\tcite{Lancaster+95}.

It is interesting to remark that one can obtain the current-voltage
relation $\dot \vphi \sim E \sim j^{z/2}$ also from an argument which
is a modification of a scaling argument first used by
\tcite{FisherDS+91:tf}. \tcite{FisherDS+91:tf} assume that the
vortex-glass transition is continuous and characterized by the
divergence of the characteristic length and time scales, $\xig$ and
$\txig \sim \xig^z$, respectively. Their original argument is as
follows: We express $E$ as $- \dot \bA/\co$. Since in the
GL-Hamiltonian $\bA$ appears in the combination $(\grad - (e/\co)
\bA)$, $\bA$ should then scale at the glass transition as an inverse
length $\xig^{-1}$ and hence $\bE$ as $\dot \bA \sim \xig^{-1}
\txig^{-1} \sim \xig^{-1-z}$.  If the critical properties are
described by a finite-temperature fixed point, the free energy density
$\bj \cdot \bA$ scales as $\xig^{-d}$ and hence $j \sim \xig^{-d+1}$.
The current-voltage relation has, therefore, in the critical region
the form
\begin{eqnarray}
E \xig^{1+z} = \Phi_d (j \xig^{d-1}).
\end{eqnarray}
Since $\xig$ has to drop out of this relation for $T \to \Tg$, we
obtain $\Phi_d(u) \sim u^{(1+z)/(d-1)}$ and hence
\begin{eqnarray}
E \sim j^{(1+z)/(d-1)}
\end{eqnarray}
at $\Tg$, which is the result of \tcite{FisherDS+91:tf}. 

In the present case of a two-dimensional system with $\bB$ parallel to
the plane this argument has to be modified.  Since only the $z$
component $B_z = \partial_x A_y - \partial_y A_x$ of the magnetic
field is non-zero and because of the absence of any $y$ dependence in
this geometry, the relevant component of $\bA$ is $A_y$. In
particular, there is no $(\partial_y- \frac e{\co} A_y)$ term in
${\cal H}_{\rm GL}$ and hence we conclude $A_y \sim \xig^0$, $E \sim
\xig^{-z}$, $j\sim \xig^{-2}$ which results in
\begin{eqnarray}
E \xig^{z} = \Phi_2 (j \xig^2).
\end{eqnarray}
At the transition, this gives $E \sim j^{z/2}$, in agreement with the
RG result $E\sim j$ (apart from a logarithmic correction) {\em at} the
vortex glass transition temperature, where $z=2$.

In the high-temperature phase ($\taug < 0$) where $\Phi_2 \sim j
\xig^2$ and $z=2$, we get $E \sim j$, such that the correlation length
drops out of the result. The same is true for $T< \Tg$, where the RG
result gives $E \sim j^{z/2}$. From this one has to conclude that {\em
  both phases are critical} such that the transition is not
accompanied by a diverging correlation length.

For completeness, we remark here that \tcite{Toner91} and
\tcite{Nattermann+91} tried to calculate the current-voltage relation
from an estimate of the energy barrier $E_B(j)$ between different
meta-stable states. Both authors use the result from the statics for
$\sigma(L)$, equation (\ref{s(L).CO}). While \tcite{Toner91} finds
$E_B(j) \sim \taug [\ln(j_c/j)]^{1/2}$, \tcite{Nattermann+91} get
$E_B(j) \sim \taug^2 \ln(j_c/j)$.  Together with the general creep
formula (\ref{v.creep}), the latter result gives a power law for the
current-voltage relation, but with a wrong result for the exponent
$z-2$, which is proportional to $\taug$ and not to $\taug^2$.  This
discrepancy is related to the fact that for $T \lesssim \Tg$ the
physics is described by a finite-temperature fixed point at which
energy barriers are not well defined.

We conclude from this section that below the temperature $\Tg$ the
vortex-line array in an impure superconducting film in a parallel
field exhibits a glassy phase with vanishing linear resistivity.  This
phase is the most accurately studied example of a vortex glass
(although of mainly academic interest). The spatial correlations of
the vortex-line array exhibit a decay slightly faster than algebraic.
The vortex-glass transition is not accompanied by a diverging
correlation length (as assumed in the scaling argument of
\tcite{FisherDS+91:tf} and as seen in many experiments for bulk
materials) since both phases are critical.  Because of the absence of
shear modes, below $\Tg$ the system is both a positional and a
phase-coherent vortex glass (in the sense discussed in section
\ref{subsec.infl.of.disorder}).

  \setcounter{equation}{0}
  \section{Superconducting film in a perpendicular field}
\label{sec.SF.perp.filed}

So far we examined in chapter \ref{sec.ran.man} a single vortex as an
example of a directed manifold and in chapter \ref{sec.film.perp} a
periodic array of vortex lines in a film.  Both cases are simple in
the sense that the vortex conformations can be described by a
single-valued displacement field, or in other terms, that these
systems are free of {\em topological defects}. The possible presence
of such defects implies that the {\em elastic approximation}, which
was the basis of our analysis in the previous chapters, may break
down.  We now turn to the simplest system that allows for topological
defects: an array of point-vortices in a superconducting film induced
by a perpendicular magnetic field.

We consider here only the case of a {\em thin} film, where the film
thickness $s$ is much smaller than the bulk correlation length $\xi$.
Then the condensate wave function can be considered as constant in the
direction perpendicular to the film, i.e., it is a two-dimensional
(2D) degree of freedom.  Nevertheless, the screening mechanism remains
three-dimensional, since the electromagnetic field still propagates
into the third dimension.

\subsection{Pure system}

To set the stage for an analysis of pinning effects in this
two-dimensional system, we briefly summarize important properties of
the pure system (see in particular \tcite{Doniach+79},
\tcite{Huberman+79}, \tcite{FisherDS80}).

The structure and interaction of vortices in thin films were first
studied by Pearl \citeyear{Pearl64,Pearl65}.  Several distinct
features are to be mentioned. (i) Because of screening, the creation
of a vortex costs a {\em finite} amount of energy.  This is in
contrast to vortices in 2D superfluids, the energy of which diverges
logarithmically with the system size due to the absence of screening.
(ii) Every vortex has a magnetic moment $M$ that diverges proportional
to the system size $L$.  Thus the lower critical field vanishes in
infinite films, $\Hci \propto 1/L$. (iii) The vortex pair interaction
potential is
\begin{mathletters}
\mlabel{V.Pearl}
\begin{eqnarray}
U({\bf x}) &=& 2 s \eo U^{(0)}(x/\Ls)
\\
U^{(0)}(x) &\sim& 
\left\{
\begin{array}{ll}
- \ln x, &\quad x \ll 1
\\
\frac 1x, &\quad x \gg 1
\end{array}
\right.
\end{eqnarray}
\end{mathletters}
where
\begin{eqnarray}
s \eo \equiv s \left(\frac{\Phi_0}{4 \pi \lambda} \right)^2
\label{intro.eps0}
\end{eqnarray}
is the energy scale and $\lambda$ is the bulk penetration length.  A
further important length scale is the screening length
\begin{eqnarray}
  \Ls \equiv 2 \lambda^2/s,
\label{intro.L_s}
\end{eqnarray}
which may be macroscopically large (in the range of millimeters).
Screening sets in only on scales beyond $\Ls$, i.e., this scale is
intimately related to the {\em inhomogeneity} of $B$.

The importance of the screening length becomes evident already at {\em
  zero} magnetic field ($H=0$).  Only in the limiting case $\Ls \to
\infty$ the vortex self energy diverges logarithmically with the
system size and individual vortices cannot be created by thermal
activation.  Then vortex-antivortex pairs interact logarithmically on
large scales and dissociate according to the entropy-driven
Kosterlitz-Thouless (KT) mechanism \cite{Kosterlitz+73} at a
temperature
\begin{eqnarray}
  T_{\rm KT} \approx \frac { s \eo}2 = \frac {\Phi_0^2}{16 \pi^2 \Ls}.
\label{intro.T_KT}
\end{eqnarray}
In this expression we neglect renormalization effects which lead only
to a quantitative reduction of the transition temperature.  However,
in the limit $\Ls \to \infty$ the transition temperature vanishes like
$T_{\rm KT} \propto 1/ \Ls$ and the system becomes critical only at
zero temperature.  At large but finite $\Ls$ a crossover in the
vicinity of the small but finite $T_{\rm KT}$ survives.

The electric resistivity of the film can be understood in terms of
vortex-antivortex pairs under the influence of a transport current
which drives vortices and antivortices in opposite directions
\cite{Halperin+79}.  For finite $\Ls$ and $T>0$ the film has Ohmic
resistivity at sufficiently small current densities, since
vortex-antivortex pairs have to overcome only a {\em finite} 
energy barrier to dissociate.  Thermal activation therefore leads to a
finite dissociation rate and a resistivity proportional to the driving
current.  At larger current densities, where the vortex interaction is
probed on scales $r \ll \Ls$ (i.e., on this scale the attractive
vortex interaction is balanced by the driving force), the logarithmic
vortex interaction implies a power law for the current-voltage
relation, which was predicted \cite{Halperin+79} and observed
experimentally \cite{Epstein+81,Kadin+83} already long ago.  More
recent experimental \cite{Repaci+96}, numerical \cite{Simkin+97}, and
analytical \cite{Pierson+99p} work was devoted to examining
finite-size effects on the KT transition, which includes the case of a
superconducting film regarding $\Ls$ as `intrinsic' finite-size scale.

Although the case of zero field would be of interest on its own, we
subsequently consider only {\em finite} fields, $H>0$, which induce a
finite average vortex density.  Then one can think of the vortex
system as a superposition of a neutral subsystem composed of vortex
dipoles and a subsystem of excess vortices oriented in the direction
parallel to $\bH$.

We now focus on the subsystem of excess vortices and assume that the
effect of the dipole subsystem amounts to a screening of the
interaction between the excess vortices.  This interaction is expected
to be purely repulsive even in the presence of screening by dipoles
such that the ground state of the excess vortices should be a
triangular lattice.  Thermal fluctuations lead to the creation and
annihilation of extra excess vortices and thereby create vacancies and
interstitials in the lattice.  Such fluctuations cost finite minimum
energy in contrast to vortex displacements.  Therefore we restrict the
subsequent analysis to displacement fluctuations and examine their
influence on the positional and orientational order of the lattice.

In the literature the vortex lattice is, in general, not examined on
the basis of the physically given vortex interaction (\ref{V.Pearl}).
Instead, simplified model interactions are used.  Since certain
physical features depend on the choice of the model interaction, we
attempt to classify them by distinguishing the following cases
according to the behaviour of the potential on large length scales:
\begin{description}
\item[`Short-range' case:] $U$ is short-ranged, such that $\int
  d^2x \ U(\bx)<\infty$.  Only in this case the interaction energy of
  the vortex lattice is extensive and all elastic constants are
  finite.
\item[`Long-range' case:] The interaction $U$ decays on large length
  scales but it is long-ranged, such that $\int d^2x \ U(\bx)=\infty$
  and the interaction energy is not extensive.  This case includes the
  Pearl interaction.  The system is incompressible, i.e., it has a
  {\em singular} compression modulus.
\item[`Logarithmic' case:] The interaction is logarithmic on large
  length scales, $U(\bx) \propto \ln x$.  This case, which also
  represents an incompressible system, includes the so-called
  two-dimensional Coulomb gas and the superconducting film in the
  limiting case when $\Ls$ is larger than the system size.
\end{description}
We now briefly summarize the elastic properties of the system and the
presence of phase transitions in these cases.

{\bf Short-range case}.  Among these three cases, this one has been
examined most since it can be represented by a well-defined elasticity
theory with finite elastic constants.  In general, the elastic
Hamiltonian can be written in Fourier space as
\begin{eqnarray}
{\cal H}_{\rm el}[{\bf u}] = 
\frac 12 \int_{BZ} \frac{d^2 q}{(2 \pi)^2}
[c_{11}({\bf q}) q^2  | \bu_L({\bf q})|^2 
+ c_{66}({\bf q}) q^2| \bu_T({\bf q})|^2 ].
\label{H.el.2.2.2}
\end{eqnarray}
The displacement field is defined in position space only as a function
of the discrete positions $\bX$ of the undistorted lattice. In Fourier
space it reads
\begin{eqnarray}
{\bf u}({\bf q})&=& a^2 \sum_\bX e^{-i {\bf q} \cdot {\bf X} }  \ 
{\bf u}({\bf X})
\end{eqnarray}
with the area $a^2=\Phi_0/B$ per vortex.  The displacement can be
decomposed into a longitudinal and a transverse component
\begin{eqnarray}
{\bf u}_L({\bf q})= {\bf P}^L ({\bf q}) \cdot {\bf u}({\bf q}),
\quad
{\bf u}_T({\bf q})= {\bf P}^T ({\bf q}) \cdot {\bf u}({\bf q})
\label{ul.ut}
\end{eqnarray}
by means of the projectors\label{intro.P} $P^L_{\alpha \beta}({\bf
  q})=q_\alpha q_\beta /q^2$ and $P^T_{\alpha \beta}({\bf
  q})=\delta_{\alpha \beta} - P^L_{\alpha \beta}({\bf q})$.  The
energy of longitudinal and transverse displacements is given by the
compression modulus $c_{11}$ and the shear modulus $c_{66}$,
respectively.  In general both moduli depend on the wave vector $\bq$
because of the non-locality of the interaction.

The elastic moduli are known explicitly for the special interaction
$U^{(0)}(x) = {\rm K}_0(x)$ with the Bessel function ${\rm K}_0$,
which decays exponentially beyond the screening length.  This is the
interaction of straight vortex lines in a three-dimensional
superconductor apart from a replacement of the screening length of the
bulk material, $\lambda$, by the screening length of the film, $\Ls$.
Thus one finds from the bulk moduli \cite{Blatter+94} within the
continuum isotropic approximation
\begin{mathletters}
\mlabel{2d.moduli}
\begin{eqnarray}
c_{66}({\bf q}) &=& 2s\eo \frac{B}{8 \Phi_0},
\\
c_{11}({\bf q}) &=& 2s \eo  
\frac{2 \pi \Ls^2}{1+\Ls^2 q^2} \frac{B^2}{\Phi_0^2}.
\end{eqnarray}
\end{mathletters}
In this chapter we use $B$ to parameterize the vortex density
$B/\phi_0$ per unit area rather than as average magnetic induction.

In the case with finite elastic constants, the pure system can have
two distinct phase transitions (for a review, see e.g.
\tcite{Strandburg87}): a {\em melting} transition between a solid and
a hexatic liquid phase, and a transition between the hexatic and an
isotropic liquid
\cite{Kosterlitz+73,Halperin+78,Nelson78,Young79,Nelson+79}.  In this
scenario, called the KTHNY scenario after the aforementioned authors,
the solid has quasi-long-range positional order and long-range
orientational order; the hexatic liquid has short-range positional
order but still quasi-long-range orientational order, and finally in
the isotropic liquid orientational order is also short-ranged.  These
transitions have been discussed in the particular context of
superconducting films by \tcite{Doniach+79}, \tcite{Huberman+79} and
\tcite{FisherDS80}.

The melting transition is driven by the entropic unbinding of
dislocation pairs.  In a triangular lattice, which is the type of
lattice formed by vortices, dislocations have a pair interaction
\begin{eqnarray}
U_{\rm disloc}(x) \simeq \frac{2 }{\pi \sqrt 3}
\frac{c_{66}(c_{11}-c_{66})}{ c_{11}} \ln (x/a) 
\end{eqnarray}
\cite{Young79,FisherDS80}, where $c_{ii}\equiv c_{ii}(\bq=\bN)$. The
given expression represents only the leading term for large distances;
at smaller distances additional terms are present.  Thus, according to
the Kosterlitz-Thouless scenario, dislocation pairs unbind at a
temperature
\begin{eqnarray}
\label{T_M.2D}
  \Tm \approx \frac 1{2  \pi \sqrt 3} \frac{\Phi_0}B 
\frac{c_{66}(c_{11}-c_{66})}{c_{11}},
\end{eqnarray}
where we have ignored the renormalization of the elastic constants.
The quantitatively correct value of $\Tm$ is hard to determine
analytically, since it depends on the dislocation core energy, i.e.
the discrete vortex structure of the core, which is difficult to
include in analytic treatments.

It is interesting to note that the melting transition can be captured
by a modified Lindemann criterion.  The traditional Lindemann
criterion $\langle \bu^2(\bX) \rangle \leq \cli^2 \atr^2$ for the
stability of the lattice is not applicable, since in two dimensions
$\langle \bu^2(\bX) \rangle=\infty$ at any finite temperature.
However, considering the {\em relative} displacement of neighbouring
vortices (separated by a basis vector ${\bf b}$ with $|{\bf b}|=\atr$)
instead of the absolute displacement, one obtains
\begin{eqnarray}
\cli^2 \atr^2 &\leq& \langle [\bu(\bX)-\bu(\bX+\bb)]^2 \rangle
\nonumber \\
&\approx& \frac 12 \int_\bq (\bq \cdot \bb)^2 W_{\alpha \alpha}(\bb)
\nonumber \\
&\approx& \frac{\atr^2 T}{4 a^2 c_{66}} ,
\label{Linde.2D}
\end{eqnarray}
where we have assumed $c_{11} \gg c_{66}$ \cite{Scheidl+98:li}. A
comparison of (\ref{T_M.2D}) and (\ref{Linde.2D}) allows the
determination of the Lindemann parameter
\begin{eqnarray}
  \cli \approx \left(\frac 1 {8 \pi \sqrt 3}\right)^{1/2} \approx 0.15
  \ .
\end{eqnarray}
This value may be a good reference value to estimate the stability of
higher-dimensional systems with quasi-long-range positional order
(such as the bulk vortex lattice with pinning) to the proliferation of
topological defects.

The second transition from a hexatic liquid to an isotropic liquid is
driven by the unbinding of disclination pairs.  Such pairs also have a
logarithmic interaction,
\begin{eqnarray}
U_{\rm discl}(x)=\frac{\pi K_A}{18} \ln (x/a)
\end{eqnarray}
with the Frank constant $K_A$ \cite{Nelson+79}.  Thus disclination
pairs unbind according to the KT mechanism at a temperature
\begin{eqnarray}
  \Th \approx \frac{\pi K_A}{72} .
\label{T_H}
\end{eqnarray}
Again, the quantitative value of $\Th$ is very difficult to determine
analytically.  Partially, the difficulty is due to the strong
temperature dependence of $K_A=K_A(T)$, which diverges above the
melting transition.  However, this divergence ensures that $\Th \geq
\Tm$, i.e., that the lattice is more stable against disclinations than
against dislocations.  Thus, in order to study the stability of the
elastic (solid) phase, it is sufficient to focus on the proliferation
of dislocations.

{\bf Long-range case}.  The actual interaction (\ref{V.Pearl}) between
vortices in a film is long-ranged.  If we ignore the regime on length
scales below $\Ls$ and take $U(x)=2s \eo \Ls /x$, the interaction is
the electrostatic one for a Wigner crystal.  In this case the elastic
moduli are \cite{Bonsall+77}
\begin{mathletters}
\begin{eqnarray}
c_{66}({\bf q}) &\approx& 0.25 \cdot 2 s \eo \Ls 
\left( \frac{B}{\Phi_0} \right)^{3/2},
\\
c_{11}({\bf q})&=&   2 s \eo \Ls   \frac{2 \pi}q
\left( \frac{B}{\Phi_0} \right)^2.
\end{eqnarray}
\end{mathletters}
As in the short-ranged case, the shear modulus is essentially local
and, in particular, $c_{66}({\bf q}=0)$ is finite.  However,
$c_{11}({\bf q}) \propto q^{-1}$, which means that the vortex lattice
is incompressible.  Thus conventional local elasticity theory does not
apply.  Nevertheless, by formally taking $c_{11} \to \infty$, most of
the results obtained for the short-ranged case can be transferred to
this long-ranged case.  Since the dislocation interaction remains
logarithmic on large scales one still expects a melting transition at
a {\em finite} temperature according to equation (\ref{T_M.2D}).

{\bf Logarithmic case}. Finally, the logarithmic case
is of interest, since it describes the superconducting film in the
limit $\Ls \to \infty$.  In this limit, the elastic moduli can be
obtained directly from equations (\ref{2d.moduli}),
\begin{mathletters}
\begin{eqnarray}
c_{66}({\bf q}) &=& 2s \eo \frac{B}{8 \Phi_0},
\\
c_{11}({\bf q}) &=& 2s \eo 
\frac{2 \pi}{q^2} \frac{B^2}{\Phi_0^2}.
\end{eqnarray}
\end{mathletters}
Again, $c_{66}$ is finite on large scales, but the longer range of the
interaction has even further increased the incompressibility.
Nevertheless, a melting transition should be present according to the
formal argument used already for the long-ranged case.

The analytical description of the two distinct transitions
(solid-to-liquid and hexatic-to-isotropic) assumes that the
topological defects are very dilute at the transition, i.e., that they
have a sufficiently large core energy.  However, for small core
energies the KT transition -- which is thermodynamically of infinite
order -- may become a first-order transition (FOT)
\cite{Minnhagen87,Thijssen+88}.  Thus it is not clear whether the 2D
melting must have a unique nature or whether there can to be two {\em
  separate} transitions for the loss of positional and orientational
order.  There has been controversial evidence for both the KTHNY
scenario and a FOT scenario.  Whereas earlier simulations always
seemed to support the FOT scenario, it is now believed that this might
be an artifact for finite-size effects and more recent simulations
\cite{Bagchi+96,Perezgarrido+98,Alonso+99,Jaster99} provide evidence
for the KTHNY scenario, which seems to be particularly well
established for particles with short-ranged interactions (see, e.g.,
\tcite{Chaikin+95}).

For the long-ranged case and even more for the logarithmic case, the
thermodynamic nature of the transition is not unambiguously
understood.  Conventional thermodynamics already runs into trouble
since the energy of the system is no longer extensive.  To close this
section on the pure system, we quote some references to provide the
reader with some references into the literature concerning the
long-ranged systems rather than to give a comprehensive overview.

\tcite{Calliol+82} examine the long-ranged case with $U(x) \propto
1/x$ and find crystallization at $T\approx \frac 1{140} 2 s \eo$
(without being able to resolve the question of whether it is a single
transition or a sequence of two transitions).  This value of $\Tm$ is
smaller than the one given in equation (\ref{T_M.2D}) by a factor of
approximately 5, which in principle can be ascribed to a
renormalization of elastic constants that is not taken into account in
equation (\ref{T_M.2D}).

Recent simulations \cite{Moore+99} of the logarithmic case suggest
that the solid might be destroyed at any finite $T$.  Many simulations
have been performed on the model in the lowest Landau-level (LLL)
approximation, which also results in a logarithmic vortex interaction
due to the absence of screening.  On the one hand there is evidence
suggestive of a first-order melting \cite{Hu+93,Kato+93,Sasik+94}
using the LLL approximation, but on the other hand there are also
claims of the absence of the solid phase at finite temperatures
(\citeANP{Oneill+92} \citeyearNP{Oneill+92,Oneill+93},
\citeANP{Yeo+96:prl} \citeyearNP{Yeo+96:prl,Yeo+96:prb,Yeo+97}).
\tcite{Moore+99} argue that in the logarithmic case dislocations
screen the interaction of disclinations such that disclination pairs
unbind and destroy crystalline order at any finite temperature.
However, in the long-range case numerical simulations are likely to be
affected by finite-size effects and the choice of boundary conditions
even for very large system sizes.

On the experimental side there is evidence for the KTHNY scenario in
superconducting films \cite{Wordenweber+86,Berhguis+90,Theunissen+96}.
In other systems experimental evidence supports the FOT scenario (see
\tcite{Chaikin+95} and \tcite{Perezgarrido+98} for a summary), which
partially may be ascribed to the presence of additional
symmetry-breaking fields.

\subsection{Disordered vortex lattice without dislocations}

We now address the question again of how quenched impurities affect
the solid vortex lattice.  Hereby we focus on the case with
short-ranged interactions. Then the vortex interactions can be
represented by the elastic Hamiltonian (\ref{H.el.2.2.2}), provided
displacements are small (i.e., the phase is solid).  Since we aim
mainly at the properties of the system on length scales larger than
$\Ls$, we may ignore the dispersion of the elastic constants, the
value of which will be renormalized by fluctuations on small scales,
as argued in section \ref{sec.film.perp}.  Although the incompressible
vortex lattice in a film does not strictly belong to the short-ranged
case, it should be well captured by the formal limit $c_{11} \to
\infty$.

The effect of disorder on the solid vortex lattice of the film in a
perpendicular field turns out to be very similar to that of the film
in a parallel field, as treated in the previous chapter.  The main
complication consists in the fact that the displacement field has two
(instead of one) components and, consequently, the different elastic
response of shear and compression modes.  Our presentation of this
generalization closely follows \tcite{Carpentier+97} and
\tcite{Carpentier99}.

According to the lines sketched in Appendix \ref{sec.peri} the
effective contribution of pinning to the Hamiltonian can be written in
the form
\begin{eqnarray}
{\cal H}_{\rm pin} &=& \sum_\bX V(\bX+\bu(\bX))
\nonumber \\ 
&=&\int d^2x \left\{
\tV(\bu(\bx),\bx) - 
\frac 12 \mu_{\alpha \beta}(\bx)
[\partial_\alpha u_\beta(\bx) + \partial_\beta u_\alpha(\bx)]
\right\}
\label{H.pin.2.2.2}
\end{eqnarray}
with an effective potential $\tV(\bu(\bx),\bx)= \rho_\bu(\bx) V(\bx)$
of zero average and 
\begin{mathletters}
\label{H.pin.2.2.2.corr}
\begin{eqnarray}
\overline{\tV(\bu,\bx) \tV(\bu',\bx')}&\approx&\tD(\bu-\bu')
\delta(\bx-\bx'),
\\
\tD(\bu) &=& \rho_0 \sum_\bX \Delta(\bX+\bu),
\end{eqnarray}
\end{mathletters}
where the effective correlator $\tD$ is periodic due to a sum over the
undistorted lattice positions $\bX$ and $\Delta(\bx)=\overline{V(\bx)
  V(\bN)}$ is the correlator of the original pinning potential.

From the correlator of $\tV$ we have separated a random field ${\bbox
  \mu}$, the correlation of which can be written as
\begin{eqnarray}
  \overline{\mu_{\alpha \beta}({\bf x}) \mu_{\gamma \delta} ({\bf
      x}')} &=& [(\sigma_{11} c_{11}^2-2\sigma_{66} c_{66}^2)
  \delta_{\alpha \beta} \delta_{\gamma \delta} 
\nonumber \\ &&
+ \sigma_{66} c_{66}^2(\delta_{\alpha \gamma} \delta_{\beta \delta} +
  \delta_{\alpha \delta} \delta_{\beta \gamma} )] \Qtr^{-2}
  \delta({\bf x}-{\bf x}').
\end{eqnarray}
For the unrenormalized model $\sigma_{66}=0$ and $\sigma_{11}=\Qtr^2
\rho_0^2 \Do/T^2$ arises from the coupling of the random potential $V$
to the divergence of the displacement field (cf. equation
(\ref{app.manip.rho})).  The length $\Qtr$ of the smallest
non-vanishing reciprocal lattice vector (RLV) is related to the vortex
spacing via $\Qtr^2 = 16 \pi^2/3\atr^2 \label{intro.Qtr}$.  Equation
(\ref{H.pin.2.2.2}) is the generalized analog of equation
(\ref{H.pin.rec.2+0}) for the film in a parallel field.

\tcite{Carpentier+97} and \tcite{Carpentier99} studied this model
retaining only the contribution from the smallest non-vanishing RLV to
the correlator, i.e., approximating $\tD(\bu) \approx \rho_0^2
\sum_{\bQ: |\bQ|=\Qtr} $ $\hD(\bQ) \cos[\bQ \cdot \bu]$.  Then the
Fourier coefficients $\hD(\bQ)$ act like a random-field amplitude, cf.
equation (\ref{intro.h.rf}), for which we introduce an effective
variance with bare value $g=\rho_0 \hD(\Qtr)/T^2$.

Under renormalization, not only the field amplitude $g$ evolves, but
the random potential induces also bond disorder, i.e., a flow of
$\sigma_{11}$ and $\sigma_{66}$. \tcite{Carpentier+97} and
\tcite{Carpentier99} found the flow equations
\begin{mathletters}
\begin{eqnarray}
\partial_l c_{11} &=& 0 ,
\\
\partial_l c_{66} &=& 0 ,
\\
\partial_l \sigma_{11} &=& b_{11}(\alpha) g^2 ,
\\
\partial_l \sigma_{66} &=& b_{66}(\alpha) g^2 ,
\\
\partial_l g &=& 2 \left(1 - \frac{T}{\Tg} \right) g
- \bg(\alpha) g^2 ,
\label{flow.g}
\end{eqnarray}
\end{mathletters}
with the coefficients
\begin{mathletters}
\begin{eqnarray}
b_{11}(\alpha) &\equiv& \frac{3 \pi}4 \frac{T^2 \Qtr^4}{c_{11}^2}  
[ 2I_0(\alpha) +I_1(\alpha)] ,
\\
b_{66}(\alpha) &\equiv& \frac{3 \pi}4 \frac{T^2 \Qtr^4}{c_{66}^2} 
[ 2I_0(\alpha) - I_1(\alpha)] ,
\\
\bg(\alpha) &\equiv& {2 \pi}  [2I_0(\alpha/2) - I_0(\alpha)] ,
\end{eqnarray}
\end{mathletters}
that depend through the Bessel functions $I_{0,1}$ on the parameter
\begin{eqnarray}
\alpha &\equiv& \frac {T \Qtr^2}{4 \pi} \left(\frac 1{c_{66}}- \frac
  1{c_{11}} \right),
\end{eqnarray}
which measures the difference between the lattice response to shear
and compression.  From the linear term in equation (\ref{flow.g}) one
recognizes that disorder is relevant only at temperatures below the
glass transition temperature $\Tg$. Its value is given by
\begin{eqnarray}
\Tg &\equiv& \frac  {8 \pi}{\Qtr^2}  
\frac {c_{11} c_{66}}{c_{11}+c_{66}}
=\frac {\sqrt 3}{\pi} \frac {\Phi_0} B 
\frac {c_{11} c_{66}}{c_{11}+c_{66}} 
\label{T_G.2D}
\end{eqnarray}
\cite{Giamarchi+95}.

In principle, there is a finite renormalization of the elastic
constants on small length scales because of the finiteness of the
disorder correlation length $\xi$.  On large scales, where the pinning
energy correlator effectively factorizes as in equation
(\ref{H.pin.2.2.2.corr}), the renormalization of the elastic constants
is absent due to a statistical symmetry analogous to the tile symmetry
discussed in section \ref{sec.struct}d.  The short-scale
renormalization is neglected here, since it does not influence the
large-scale properties of the vortex lattice.  However, to be accurate
the transition temperature is given by equation (\ref{T_G.2D}) using
the {\em renormalized} value of the elastic constants.

The fluctuations of the displacement field can be represented by the
correlation function in the form
\begin{eqnarray}
W_{\alpha \beta}({\bf x}) &\equiv& \overline{\langle 
[u_\alpha({\bf x})-u_\alpha({\bf 0})]
[u_\beta({\bf x})-u_\beta({\bf 0})]\rangle}
\nonumber \\
&=& W_L({\bf x}) P^L_{\alpha \beta} ({\bf x}) + 
W_T({\bf x})  P^T_{\alpha \beta}({\bf x})
\end{eqnarray}
with projectors $P^L_{\alpha \beta}(\bx) \equiv x_\alpha
x_\beta/\bx^2$ and $ P^T_{\alpha \beta} \equiv \delta_{\alpha \beta} -
P^L_{\alpha \beta}$.

Above the glass transition, $T>\Tg$, the renormalization of the
elastic constants and of the strengths $\sigma_{11}$ and $\sigma_{66}$
of the random `tilt' fields are finite such that $W(\bx) \sim \ln
(x/a)$ and the correlation function $S(\bQ,\bx-\bx')$ decays
algebraically,
\begin{mathletters}
\begin{eqnarray}
S(\bQ,\bx-\bx')&=& \overline{\langle e^{-i {\bf Q} \cdot 
[{\bf u}({\bf x})-{\bf u}({\bf x}')]}\rangle}
 \sim  \left( \frac{|{\bf x}-{\bf x}'|}a \right)^{-\eta_Q},
\\
\eta_Q &=& \frac{Q^2}{2 \pi \Qtr^2} \left( 
\frac {T \Qtr^2} {c_{11}} +  \frac {T \Qtr^2} {c_{66}} 
+ {\sigma_{11}} + {\sigma_{66}} \right).
\end{eqnarray}
\end{mathletters}
Thus there is quasi-long-range order as in the absence of disorder, and
the effect of disorder (as seen in the static structure factor)
amounts to an increased effective temperature.

Below the glass transition, $\taug \equiv 1-T/\Tg >0$, $g$ flows to a
certain finite fixed-point value and generates an {\em unlimited}
increase $\sigma_{11} \propto \sigma_{66} \propto \ln(L/a)$ with
increasing length scale $L \propto e^l$.  Therefore the roughness is
larger than logarithmic and, to leading order in $\taug$, is given by
\cite{Carpentier+97}:
\begin{mathletters}
\begin{eqnarray}
W_T(\bx) \sim W_L(\bx) &\sim& \frac {w(\alpha)}{\Qtr^2}\taug^2 \ln^2 (x/a),
\\
W_T(\bx) - W_L(\bx) &\sim& \frac {\tw(\alpha)}{\Qtr^2} \taug^2 \ln (x/a),
\\
w(\alpha)&\equiv&6 \frac{2 I_0(\alpha)(1+\alpha^2/4)-\alpha I_1(\alpha)}
{[2I_0(\alpha/2)-I_0(\alpha)]^2},
\\
\tw(\alpha)&\equiv&6 \frac{I_1(\alpha)(1+\alpha^2/4)-2 \alpha I_0(\alpha)}
{[2I_0(\alpha/2)-I_0(\alpha)]^2}.
\end{eqnarray}
\end{mathletters}
Precisely at the transition the parameter
\begin{eqnarray}
\alpha (\Tg) &\equiv &2 \frac{c_{11}-c_{66}}{c_{11}+c_{66}} 
\end{eqnarray}
depends on the ratio of the elastic constants only.  The main
structural feature, the increase $W(\bx) \propto \ln^2(x/a)$, is thus
similar for magnetic fields parallel and perpendicular to the film,
cf. equation (\ref{log.sq}).

A further similarity is found in the dynamical properties, where a
dynamical exponent \cite{Carpentier+97}
\begin{eqnarray}
z=2 + 3 e^\gE
\frac{(2+\alpha)[(2-\alpha)/(2+\alpha)]^{(2-\alpha)/4}}
{2I_0(\alpha/2)-I_0(\alpha)} \ \taug
\end{eqnarray}  
(with the Euler constant $\gE$) is found on the basis of the
overdamped equation of motion
\begin{mathletters}
\begin{eqnarray}
\eta_0 \partial_t u_\alpha &=& - \frac{\delta {\cal H}}{\delta u_\alpha}
+ \zeta_\alpha,
\\
\langle \zeta_\alpha({\bf x},t) \zeta_\beta({\bf x}',t')\rangle &=&
2 \eta_0 T \delta_{\alpha \beta} \delta({\bf x} -{\bf x}') \delta(t-t').
\end{eqnarray}
\end{mathletters}
Consequently, the vortex lattice responds with a velocity 
\begin{eqnarray}
v \propto   F^{z/2}
\label{non-lin.0.2.2}
\end{eqnarray}
to a driving force $F$.  Since $F$ is proportional to the current
density and $v$ is proportional to the electric field, the linear
electric resistivity vanishes in the glassy phase because of $z>2$ in
agreement with the film in a parallel field, cf. equation
(\ref{non-lin.1.1.2}).

\subsection{Disordered vortex lattice with dislocations}

The analysis of the previous subsection was based on the elastic
approximation, {\em excluding topological defects} such as
dislocations.  In order to check whether the glassy features found
above actually describe the vortex system, it is crucial to examine the
relevance of dislocations.

By comparing the melting temperature and the glass-transition
temperatures, equations (\ref{T_M.2D}) and (\ref{T_G.2D}), we
immediately run into a dilemma because $\Tm/\Tg \approx
[1-(c_{66}/c_{11})^2]/6 < 1$.  At $T>\Tm$ the pure system is liquid
and the effects of disorder cannot be studied starting from the
elastic approximation (disorder will certainly not stabilize the
ordered phase).  On the other hand, if $T<\Tm$, then also $T<\Tg$ and
disorder generates $W(\bx) \propto \ln^2(x/a)$.  Unfortunately, this
increased roughness due to the divergent renormalization of the bond
disorder makes dislocation pairs unbind, because on large scales the
gain of pinning energy is larger than the attraction energy of
dislocation pairs. (The situation is analogous to the $XY$ model,
where bond disorder beyond a finite critical strength induces
vortices, see e.g. \tcite{Nattermann+95}.)

Consequently, on length scales beyond a scale $\Ldis$, where disorder
induces the dissociation of dislocation pairs, dislocations are
relevant at {\em all} temperatures and the elastic description breaks
down and there can be neither a true melting transition nor a true
glass transition.  Only on length scales below $\Ldis$ the system can
be described within the elastic approximation.  This scale is of the
order of the vortex spacing slightly above $\Tm$ but it can become
very large below $\Tm$ if disorder is very weak.  In this case melting
survives as a {\em crossover} phenomenon, i.e., as a rapid decrease of
$\Ldis$ if the temperature is increased above a renormalized $\Tm^* <
\Tm$.

The length scale $\Ldis$ can be estimated on the basis of the above RG
flow equations \cite{Carpentier+98}.  For qualitative purposes, one
can examine $\Ldis$ from the simpler random-field $XY$ model. Using
further approximations \tcite{Giamarchi+95} and \tcite{Ledoussal+98p}
found that for temperatures slightly below $\Tm$ topological defects
appear beyond
\begin{eqnarray}
\Ldis  \sim R_a e^{c_1 [(\Tm/T-1) \ln(R_a/a)]^{1/2}}.
\label{val.Ldis}
\end{eqnarray}
Here $R_a$ is the length scale on which the displacement becomes of
the order of the vortex spacing, $W(R_a)=a^2$, and $c_1$ is a
numerical constant.  The last equation overestimates $\Ldis$ at low
temperatures, where $\Ldis$ saturates at a finite value for
sufficiently weak disorder. This value is obtained from equation
(\ref{val.Ldis}) after replacing $T$ by an effective value $T^* < \Tm$
that depends only on the disorder strength
\cite{Carpentier+98,Ledoussal+98p}.

Given the finiteness of $\Ldis$ even at very low temperatures, one has
to draw the conclusion that in a superconducting film with
perpendicular field an elastic vortex glass, where dislocations would
be irrelevant on asymptotically large scales, {\em does not exist}.
Evidence for this fact stems also from numerical calculations
\cite{Shi+91,Gingras+96,Middleton98,Zeng+99}.  The resulting physical
situation can be thought of as a vortex lattice broken into
crystallites of size $\Ldis$.  As an immediate consequence of this
fragmentation, the system has only {\em short-ranged} positional order
(but the correlation length diverges with vanishing disorder strength
for $T<\Tm$).  A finite correlation length also implies finite pinning
energies for these crystallites.  Therefore thermal activation leads
to a {\em linear} current-voltage relation for sufficiently weak
currents.

  \setcounter{equation}{0}
  \section{Bulk Superconductor}
\label{sec.bulk}

In this chapter we consider the weakly disordered {\em bulk}
superconductor in an external field.  We will first treat the system
in the elastic approximation by neglecting dislocations. Later we show
that this is indeed justified in a certain region of the phase
diagram.

\subsection{Elastic approximation}

For not too large fields, $H \ll \Hcii$, we may use the London
approximation to describe the vortex-line lattice (VLL).  Then the
elastic Hamiltonian can be written in Fourier space as
\begin{eqnarray}
{\cal H}_{\rm el} = 
\frac 12 \int_{BZ} \frac{d^d q}{(2 \pi)^d}
\{[c_{11} q_\perp^2 + c_{44} q_\parallel^2] |\bu_L({\bf q})|^2 
+ [c_{66} q_\perp^2 + c_{44} q_\parallel^2] |\bu_T({\bf q})|^2 \},
\label{H.el.3d.q}
\end{eqnarray}
involving only wave vectors $\bq=(\bq_\parallel, \bq_\perp)$ within
the first Brillouin zone (BZ).  The displacement field
$\bu=\bu_L+\bu_T$ has been decomposed into its longitudinal and
transverse component with respect to $\bq_\perp$ as in equation
(\ref{ul.ut}).  We extended the model from three to general dimension
$d$ since we will perform below an expansion around $d=4$ dimensions
below.  In this generalization $\bq_{\perp}$ stands for the
two-dimensional component of the wave vector perpendicular to the
generalized magnetic field and $\bq_{\parallel}$ stands for the
$(d-2)$-dimensional parallel component.

In principle, the elastic constants are scale (momentum) dependent.
However, since we are interested in the case of {\em weak} disorder,
which becomes relevant only on large scales ($L \gg \lambda$), we may
neglect the dispersion of the elastic constants.  In the case of
strong disorder, in which relevant length scales (like the Larkin
length) are smaller than $\lambda$, the consideration of the
dispersion of the elastic constants can be achieved in principle (see
\tcite{Kierfeld98}).

Next we include the interaction between the randomly distributed
impurities and the vortex lattice. As in the previous chapters for
simplicity we assume that the impurities result in a random potential
$V(\br)$, which has zero average and is Gaussian correlated with
two-point correlations
\begin{eqnarray}
\overline{V(\br) V(\bN)}= \Delta(\bx) \delta^{(d-2)}(\bz).
\label{corr.V}
\end{eqnarray}
The weight of this correlator is $\Do= \int d^2x \Delta(\bx) = f_{\rm
  pin}^2 n_{\rm i}^{(d)} \xi^6$, where $f_{\rm pin}$ denotes the
pinning force of an individual impurity, $n_{\rm i}^{(d)}$ is the
impurity density, and $\xi$ the maximum of the coherence and disorder
correlation length \cite{Blatter+94}. The pinning energy is then
\begin{eqnarray}
  \cH_{\rm pin} &=& \int d^2x d^{d-2} z \ \rho_\bu(\br) V(\br)
\nonumber \\
&=& \sum_\bX \int d^{d-2} z \ V(\bX + \bu(\bX,\bz),\bz) 
= \int d^dr \ \tV(\bu(\br),\br)
\label{H.pin.3D}
\end{eqnarray}
where $\rho_\bu(\br)=\sum_\bX \delta^{(2)} (\bx-\bX-\bu(\bX,\bz))$
denotes the vortex-line density (per unit area) and $\bX$ is a vector
of the Abrikosov triangular lattice. $\tV(\bu(\br),\br) \equiv
\rho_\bu(\br) V(\br)$ is the pinning energy density (per unit volume).

It is interesting to note that the energy of the distorted VLL has
certain symmetries.  The pinning energy density $\tV(\bu,\br) \equiv
\rho_\bu(\br) V(\br)$ is apparently invariant with respect to the
transformation
\begin{eqnarray}
\bu(\bX,\bz) \to \bu(\bX',\bz) + \bX'-\bX,
\label{relabel}
\end{eqnarray}
i.e., by relabeling the vortex lines.  Such  relabeling also leaves
the elastic Hamiltonian invariant and this symmetry is a symmetry of
the total Hamiltonian $\cH = \cH_{\rm el} + \cH_{\rm pin}$.

$\cH_{\rm el}$ possesses a second symmetry related to a rotation of
the displacement field and a simultaneous exchange of shear and
compression modulus.  To demonstrate this symmetry, it is useful to
introduce
\begin{eqnarray}
\cs \equiv \sqrt{c_{11} c_{66}}, \quad
\gamma \equiv \frac{c_{66}}{c_{11}}
\label{intro.gamma}
\end{eqnarray}
and to rescale $\bz \to \bz' = \bz\sqrt{\cs / c_{44}}$, which leads to
\begin{eqnarray}
\cH_{\rm el} &=& \frac 12 \cs \left(\frac{c_{44}}{\cs}\right)^{d-2}
\int d^2x d^{d-2} z' \Big\{ 
\gamma^{-1/2} (\grad_\perp \cdot\bu)^2 
\nonumber \\ &&
+ \gamma^{1/2} (\grad_\perp \wedge \bu)^2 
+  (\grad_\parallel' \bu)^2 \Big\}.
\label{H.el.gamma}
\end{eqnarray}
As far as the thermodynamic properties of the pure system are
concerned, these are invariant under the transformation $\gamma \to
\gamma^{-1}$, since the Hamiltonian is invariant under the
simultaneous transformation ($\bu \equiv (u_1,u_2)$)
\begin{eqnarray}
  \gamma \to \gamma^{-1}, \quad u_1\to u_1'=-u_2, \quad u_2\to
  u_2'=u_1
\label{trafo.gamma}
\end{eqnarray}
(which also amounts to transforming the longitudinal displacement
component into the transverse one and vice versa) and the partition
function is invariant under the rotation of coordinates. The
correlation functions will change sign if they include an odd power of
$u_2$.  We note, however, that the symmetry (\ref{trafo.gamma}) breaks
down as soon as dislocations are present, since the interaction energy
of a dislocation pair is not symmetric with respect to the
transformation $\gamma \to \gamma^{-1}$ and a full treatment including
both $c_{66}$ and $c_{11}$ is required for studying the general case.
In the presence of pinning this symmetry is absent even as a
statistical symmetry in a strict sense.  This can be seen from the
disorder-averaged replicated pinning energy. However, we will find
below that this symmetry is restored on large scales $R \gg a$.

\subsection{The Bragg glass}
\label{sec.BG}

Historically, three different approximations have been discussed in
treating the pinning Hamiltonian, which turn out to be valid on (i)
small, (ii) intermediate and (iii) large length scales.

(i) In his pioneering work on pinning in type-II superconductors,
\tcite{Larkin70:t} expanded (\ref{H.pin.3D}) for small displacements
$\bu$
\begin{eqnarray}
  \cH_{\rm pin} \approx \int d^{d-2}z \sum_\bX \{ V(\bX,\bz) 
+ \bu(\bX,\bz) \cdot \grad_\perp V(\bX,\bz) + O(u^2)\} .
\label{H.pin.3D.Lark}
\end{eqnarray}
$\grad_\perp V(\bX,\bz)$ is a random pinning force per length which
acts on the vortex line of rest position $\bX$.  Although it was
proved by \tcite{Efetov+77:t} that, using perturbation theory, the
lowest order (linear) term in $\bu$ in the expansion of $ \cH_{\rm
  pin} $ (under certain conditions) already gives the exact result for
the thermodynamic quantities, it became clear later that perturbation
theory breaks down in random systems with many energy minima
\cite{Villain+83,FisherDS85:pt}. Thus we can use the expansion
(\ref{H.pin.3D.Lark}) only as long as $|\bu|$ is smaller than the
typical distance between energy minima of $V(\br)$ which is of the
order of $\xi$. This restricts the validity of the expansion
(\ref{H.pin.3D.Lark}) to length scales small compared to the Larkin
length.  In $d=3$ one finds from first-order perturbation theory
\cite{Larkin70:t}
\begin{eqnarray}
W(\br) &=& \overline{ [\bu(\bx,z)-\bu(\bN,0)]^2}
\nonumber \\
&\approx& 
\frac {\rho_0 \Dii}{\pi c_{44}^{1/2}}
\left\{ \frac 1 {c_{11}^{3/2}}(\bx^2+z_l^2)^{1/2} + \frac 1{c_{66}^{3/2}}
(\bx^2+z_t^2)^{1/2} \right\} ,
\label{W.Lark}
\end{eqnarray}
where we introduced 
\begin{eqnarray}
z_t^2=\frac{c_{66}}{c_{44}} z^2, \quad z_l^2=\frac{c_{11}}{c_{44}}z^2.
\label{def.z_t}
\end{eqnarray}
The coefficient $ \rho_0 \Dii \equiv -\frac 12 \left. \grad_\perp^2
  \Delta(\bx)\right|_{\bx=\bN} \approx f_{\rm pin}^2 n_{\rm i}^{(3)}
\xi^2/a^2$ has the meaning of the density of the fluctuations of the
pinning forces in the volume fraction occupied by the vortex lines.  As
already mentioned, this result is valid only as long as $W(\bx,z)
\lesssim \xi^2$, which defines the Larkin lengths $L_\xi$ and $R_\xi$
parallel and perpendicular to the magnetic field respectively:
\begin{mathletters}
\mlabel{def.L.R_xi}
\begin{eqnarray}
L_\xi &\approx& \frac {\pi \xi^2}{\rho_0 \Dii} 
  \frac{c_{11} c_{44} c_{66}}{c_{11}+c_{66}},
\\
R_\xi &\approx&  \frac {\pi \xi^2}{\rho_0 \Dii} 
\frac{c_{44}^{1/2} (c_{11} c_{66})^{3/2}}{c_{11}^{3/2}+c_{66}^{3/2}}.
\end{eqnarray}
\end{mathletters}
In general dimensions perturbation theory gives a roughness exponent
[cf. equation (\ref{zeta.rf})]
\begin{eqnarray}
\zeta=\zeta_{\rm rf} \equiv \frac{4-d}2.
\end{eqnarray}
From equation (\ref{W.Lark}) one finds for the correlation function of
translational order
\begin{eqnarray}
S(\bQ,\br)= \overline{e^{i \bQ \cdot[\bu(\br)-\bu(\bN)]}},
\label{Psi.Lark}  
\end{eqnarray}
where $\bQ$ is a reciprocal lattice vector, an {\em exponential} decay
with the correlation length $(\atr^2/4 \pi^2 \xi^2) L_\xi$ and $(\atr/
4 \pi \xi^2) R_\xi$, respectively. Note, however, that the regime of
the true exponential decay is not reached because of the restrictions
$x < R_\xi$ and $ z < L_\xi$.

(ii) On length scales larger than $L_\xi$ and $R_\xi$, respectively,
the increase of $W(\br)$ will continue, but with a smaller roughness
exponent since perturbation theory is known to overestimate the
influence of disorder.

As long as the displacement of the vortex lines is much smaller than
the lattice spacing $a$ of the VLL (but larger than $\xi$), each
vortex line sees its own random potential which cannot be reached by
other vortex lines. Hence, since $|\bu-\bu'| \ll a$,
\begin{eqnarray}
\overline{V(\bX+\bu,\bz) V(\bX'+\bu',\bz')} &=& \Delta(\bX-\bX'+\bu-\bu')
\delta(\bz-\bz') 
\nonumber \\
&\simeq& \Do \delta_{\bX,\bX'} 
\delta_\xi^{(2)}(\bu-\bu') \delta(\bz-\bz').
\end{eqnarray}
The pinning energy density $\tV(\bu,\br) \approx \rho_0 V(\bx+
\bu,\bz)$ then obeys the approximate relation
($\rho_0=B/\Phi_0=a^{-2}$)
\begin{eqnarray}
\overline{\tV(\bu,\br) \tV(\bu',\br')} \simeq 
\rho_0 \Delta(\bu-\bu')
\delta_a^{(2)}(\bx-\bx') \delta(\bz-\bz'),
\end{eqnarray}
which agrees with that of the {\em random-manifold model} of chapter
\ref{sec.ran.man} \cite{Feigelman+89,Bouchaud+91,Bouchaud+92:prb}.  In
the random-manifold regime the roughness exponent $\zeta_{\rm rm }$
cannot be calculated exactly.  A crude estimate is given by the Flory
result $\zeta_{\rm F}=(4-d)/6 \approx 0.167$ according to equation
(\ref{zeta.F}). [Note that explicit values for exponents, which no
longer depend on $d$, are given for $d=3$.] \tcite{Emig+99} calculated
$\zeta_{\rm rm}$ from a functional renormalization-group treatment and
found $\zeta_{\rm rm }\approx 0.175$ which varies only weakly with
$\gamma=c_{66}/c_{11}$ (see below). The decay of $S(\bQ,\br) \approx
\exp [- Q^2 W(\br)/2]$ in this regime is therefore of {\em stretched
  exponential} form.

(iii) Finally, for $L \gg L_a \approx L_\xi (a/\xi)^{1/\zeta_{\rm rm}}
\label{intro.L_a}$ (or $R \gg R_a \approx R_\xi (a/\xi)^{1/\zeta_{\rm
    rm }} \label{intro.R_a}$) the vortex line displacement becomes of
order $a$.  Clearly, in the approximation for the perturbative regime
and the manifold regime used so far, $\tV(\bu,\br)$ does not fulfill
the invariance property (\ref{relabel}). However, it is precisely {\em
  this} property which determines the physics in the regime of very
large length scales $L \gg L_a$ ($R \gg R_a$). In this regime the
effect of disorder is very weak since displacements of vortex lines
larger than $a$ are not very favourable because there is already a
vortex line within a distance $a$ of any impurity.  To keep this
feature in our pinning Hamiltonian, we rewrite the vortex-line density
as
\begin{eqnarray}
\rho_\bu(\br) &\!=\!& \sum_\bX \delta(\bx-\bX-\bu(\bX,\bz))
= \int d^2 x' \sum_\bX \delta(\bX-\bx') \delta(\bx-\bx'-\bu(\bx',\bz))
\nonumber \\
&\!=\!& \int d^2x' \ \delta(\bx-\bx'-\bu(\bx',\bz))
\rho_0 \sum_\bQ e^{i \bQ \cdot \bx'} 
\nonumber \\
&\!\approx\!& \rho_0 \frac 1 {|1 + \partial_\alpha u_\alpha|}
\sum_\bQ e^{i \bQ \cdot[\bx-\bu(\br)]} .
\end{eqnarray}
The pair correlator of the pinning energy density
$\tV(\bu,\br)=\rho_\bu(\br) V(\br)$, which is the only quantity which
enters the following calculation, is then
\begin{eqnarray}
\overline{\tV(\bu,\br) \tV(\bu',\br')} &\simeq& 
\rho_0^2 \Delta(\bx-\bx') \delta(\bz-\bz')
\frac 1 {|1 + \partial_\alpha u_\alpha|}
\frac 1 {|1 + \partial_\beta' u_\beta'|}
\nonumber \\
&& \times 
\sum_{\bQ,\bQ'} e^{i \bQ \cdot (\bx-\bx') + i (\bQ+\bQ')\cdot \bx' -i
  \bQ \cdot \bu - i \bQ' \cdot \bu'} .
\label{tV.tV}
\end{eqnarray}
To exploit (\ref{tV.tV}) further, a few remarks are in order: 

(i) The terms in the sum over $\bQ'$ with $\bQ+\bQ' \neq \bN$ on the
right hand side of (\ref{tV.tV}) are rapidly oscillating on scales
$|\Delta \bx'| \gg a$ and therefore average to zero. (For weak
disorder, where $L_\xi \gg a$, this is even true in the perturbative
regime.) 

(ii) The denominators lead to terms of the form $\sigma
(\partial_\alpha u_\alpha^a)( \partial_\beta u_\beta^b)$ in the
replica Hamiltonian (density). In $d>2$ dimensions $\frac {d
  \sigma}{dl}=(2-d) \sigma + \dots$ such that these terms renormalize
to zero for weak disorder. We therefore omit them in the following.

(iii) We approximate
\begin{eqnarray}
\overline{\tV(\bu,\br) \tV(\bu',\br')} \simeq 
\tD(\bu-\bu') \delta(\br-\br').
\label{corr.tV}
\end{eqnarray}
Hence we get 
\begin{eqnarray}
\tD(\bu)= \rho_0^2 \sum_\bQ \hD(\bQ) e^{i \bQ \cdot \bu} =
\rho_0 \sum_\bX \Delta(\bu + \bX),
\label{intro.tD}
\end{eqnarray}
where $\hD(\bQ)$ is the Fourier transform of $\Delta(\bx)$,
\begin{eqnarray}
\hD(\bQ)=\int d^2x \  \Delta(\bx) \ e^{-i \bQ \cdot \bx}.  
\label{intro.hD}
\end{eqnarray}
Apparently, the correlator $\tD(\bu)$ is invariant under the
transformation $\bu \to \bu + \bX$, where $\bX$ is an arbitrary vector
of the Abrikosov lattice.  If $\Delta(\bx) \simeq (\Do/2 \pi \xi^2)
\exp(-\bx^2/2 \xi^2)$ then $\hD(\bQ)=\Do \exp(-\bQ^2 \xi^2/2 )$ and
\begin{eqnarray}
\tDii= -\frac 12 \left. \grad_\perp^2 \tD(\bu)\right|_{\bu=\bN} =
\frac 12 \rho_0^2 \sum_\bQ \hD(\bQ) \bQ^2 \approx \rho_0 \Do/ 2 \pi
\xi^4\label{intro.tDii}.
\end{eqnarray}

To obtain the disorder averaged configuration of the VLL on a
particular length scale, one has to take into account the
renormalization of $\tD(\bu)$ by fluctuations on shorter length
scales.  This can be done systematically by a functional
renormalization group (FRG) \cite{FisherDS86:dr} for the replica
Hamiltonian $\cH_n$ resulting from equations (\ref{H.el.3D}),
(\ref{H.pin.3D}), and (\ref{corr.tV}).  After the disorder averaging
we obtain
\begin{eqnarray}
\cH_n & \simeq & \frac 12 \sum_a \int d^2x d^{d-2} z 
\left\{ c_{11} (\grad_\perp \cdot   \bu^a)^2 
+ c_{66}  (\grad_\perp \wedge \bu^a)^2 
+ c_{44} (\grad_\parallel \bu^a)^2  \right\}
\nonumber \\&&
- \frac 1{2T} \sum_{ab} \int d^d r \ \tD(\bu^a-\bu^b).
\label{Ham}
\end{eqnarray}
Because of the statistical invariance of the replica Hamiltonian under
a shift of $\bu$ by an arbitrary vector field inducing a compression,
shear and/or tilt of the VLL, there is no renormalization of the
elastic moduli \cite{Hwa+94:prl}.  Therefore the temperature obeys the
exact flow equation $dT/dl=-(d-2)T$ leading to a $T=0$ fixed point for
$d>2$.  Notice, however, that in the original model, (\ref{H.el.3d.q})
with (\ref{H.pin.3D}), the statistical invariance is not fulfilled
exactly on length scales smaller than $a$, leading to a small
renormalization $c_{ii} \rightarrow \tilde{c}_{ii}$ of the elastic
constants which will be considered from now on as effective parameters
(similar effects have been considered in some detail in section
\ref{sec.XY.map}).  This renormalization is negligible for the
calculation of the leading asymptotic behaviour on largest scales.  In
the above approximation, $\cH_n$ {\em including the pinning energy} is
invariant under the transformation (\ref{trafo.gamma}). Therefore this
transformation will be present as symmetry in all thermodynamic
quantities (such as the the displacement correlation) calculated from
(\ref{Ham}).

The FRG was applied to (\ref{Ham}) in the case of a {\em scalar}
displacement field $u$ first by \citeANP{Giamarchi+94}
\citeyear{Giamarchi+94,Giamarchi+95}. Here we follow closely the
derivation of \tcite{Emig+99} who treat the general case of a {\em
  vector} field $\bu$.  In the present FRG only coordinates are
rescaled, $\br \to \exp(dl)\br$, to keep the cutoff $\Lambda$ fixed
with $\Lambda dl$ the infinitesimal width of the momentum shell.
Because of the dispersion of the elastic constants on scales smaller
than the penetration depth $\lambda$, we have to choose here
$\Lambda\approx 2\pi/\lambda$.  Fluctuations on smaller scales can be
ignored if the Larkin length $L_{\xi}$ is much larger than $\lambda$,
i.e., for weak disorder. For larger disorder one has to take into
account the dispersion of $\tilde{c}_{11}$ and $\tilde{c}_{44}$, which
will result in a more complicated cross-over but which will {\em not}
affect the asymptotic behaviour (see also section IV.D in
\tcite{Giamarchi+95}).  The flow equation for $\tD(\bu)$ can then be
derived along the lines discussed in detail in Refs.
\cite{FisherDS86:sc,Balents+93}.  In contrast to previous cases,
\tcite{Emig+99} took into account the existence of a longitudinal and
a transverse part in the elastic propagator. With the replacement
\begin{mathletters}
\mlabel{resc.tD}
\begin{eqnarray}
\frac {C}{2 a^2} \tD(\bu) &\to&  \tD(\bu)
\\
C &\equiv& \int_{|\bq|=\Lambda} \left\{ \frac 1{(\tc_{11} q_\perp^2 
+ \tc_{44} q_\parallel^2)^2} + 
 \frac 1{(\tc_{66} q_\perp^2 + \tc_{44} q_\parallel^2)^2} \right\}
\nonumber \\
&=& \frac 1{8 \pi^2}
\frac {1 + \gamma}{\tc_{44} \tc_{66}} \Lambda^{-\epsilon}
\end{eqnarray}
\end{mathletters}
one obtains to lowest order in $\epsilon=4-d$ a flow equation for the
renormalized and rescaled correlator on scale $L=\lambda e^l$ (as in
most of the previous chapters we suppress here the RG-flow variable
$l$, i.e., we express $\tD(\bu,L) \equiv \tD_l(\bu)$ as $\tD(\bu)$)
\begin{eqnarray}
  \label{RGflow}
\frac{d \tD(\bu)}{dl} &=& \epsilon \tD(\bu) +\frac{\atr^2}{2}
\Bigg\{\left(\partial_1^2 \tD \right)^2
+\left(\partial_2^2 \tD \right)^2+2\left(
  \partial_1\partial_2 \tD \right)^2
\nonumber\\
&&+2\tDii \left(\partial_1^2 + \partial_2^2\right)  \tD
-\frac{\delta}{4}\left[\left(\partial_1^2 \tD-\partial_2^2 \tD\right)^2+
  4\left(\partial_1 \partial_2 \tD\right)^2\right]\!\!\Bigg\} \ \ \ \ \ \
\label{flow.tD}
\end{eqnarray}
with the dimensionless parameter 
\begin{eqnarray}
\tDii \equiv \left.-\partial_1\partial_1 \tD(\bu) \right|_{\bu=\bN}
=  \left.-\partial_2\partial_2 \tD(\bu)\right|_{\bu=\bN}
\label{tDii.nochmal}
\end{eqnarray}
and $\partial_1 = \partial/\partial u_1 $ etc.. Both the last equality
as well as $\left. \partial_1\partial_2 \tD(\bu)\right|_{\bu=\bN} =0$
follow from the requirement of hexagonal symmetry for $\tD(\bu)$. The
anisotropy parameter is given by
\begin{equation}
\delta(\gamma)=1-\frac{2\ln(\gamma)}{\gamma-\gamma^{-1}}
= 1 - \frac 2C \int_\bq 
\frac 1{(\tc_{11} q_\perp^2 + \tc_{44} q_\parallel^2)
(\tc_{66} q_\perp^2 + \tc_{44}   q_\parallel^2)},
\label{def.delta(gamma)}
\end{equation}
i.e., $0\le \delta \le 1$ for any ratio
$\gamma=\tilde{c}_{66}/\tilde{c}_{11}$.

In the special case $\tilde c_{11}=\tilde c_{66}$ (i.e., $\delta=0$)
the flow equation (\ref{RGflow}) reduces to that of
\tcite{Balents+93}. If $\tilde c_{11}\rightarrow \infty$, as often
assumed for VLLs, $\delta=1$.

To obtain the renormalized function $\tD(\bu)$ on large length scales
$L$ including the fixed point $\tD_\infty(\bu)$ for $L\to \infty$, one
has to integrate equation (\ref{RGflow}).  With the bare correlator of
equation (\ref{intro.tD}) showing the full symmetry of the triangular
lattice -- translation, sixfold rotational axis, six mirror lines --
and the flow of equation (\ref{RGflow}) preserving these symmetries as
it ought to, the set of possible solutions is restricted to functions
with the full lattice symmetry on every length scale.  One can solve
(\ref{flow.tD}) in a straightforward manner by rewriting the
functional flow equation as a set of non-linear ordinary flow
equations for the Fourier coefficients $\hD(\bQ)$, cf. equation
(\ref{intro.hD}).  Rather than solving for the fixed point
$\hD_\infty(\bQ)$ directly, \tcite{Emig+99} numerically integrated the
flow equations exploiting the remaining point group symmetries. It
turns out that there is convergence to a fixed point (the pinning
force correlator of which is illustrated in figure \ref{fig_fp}) from
a large basin of attraction.  Of special interest is the flow of
$\tDii_l$, since the renormalized propagator is proportional to
\begin{eqnarray}
\frac{\tD_{\rm eff}^{(2)}(q^{-1})}{q^4} 
\sim q^{-d+2  \zeta},
\label{tD_eff}
\end{eqnarray}
where 
\begin{eqnarray}
\tD_{\rm eff}^{(2)}(q^{-1})=
\tD^{(2)}_{l=\ln(\Lambda/q)} (\Lambda/q)^{d-4},
\label{tD_eff.2}
\end{eqnarray}
which determines the roughness exponent $\zeta$ of the VLL. As shown
in figure \ref{fig_co}, $\tDii_{l=\ln(\Lambda/q)}$ exhibits three
scaling regions. 

(i) On scales $L=q^{-1}<L_\xi$, $\tDii_{\ln(\Lambda L)} \sim (\Lambda
L)^{4-d}$ which reproduces the result of \tcite{Larkin70:t} for the
perturbative regime.

(ii) In the region $L_\xi<L<L_a$, $\tDii_{\ln(\Lambda L)} \sim
(\Lambda L)^{2\zeta_{\rm rm}(\gamma)}$, where $\zeta_{\rm rm}(\gamma)$
is the roughness exponent of the random-manifold regime.  Numerically,
$\zeta_{\rm rm}(\gamma)$ ranges from $0.1737$ for $\gamma=0$ to
$0.1763$ for $\gamma=1$ and is continuously increasing in this
interval (see figure \ref{fig_exp}).

(iii) Finally, for $L_a<L$, $\tDii$ approaches a fixed point
$\tDii_\infty$ which determines the asymptotic Bragg glass regime. The
numerical value of $\tDii_\infty$, which is of order $\epsilon=4-d \ll
1$, still depends on $\gamma=c_{66}/c_{11}$.

\begin{figure}
\centering
\epsfig{file=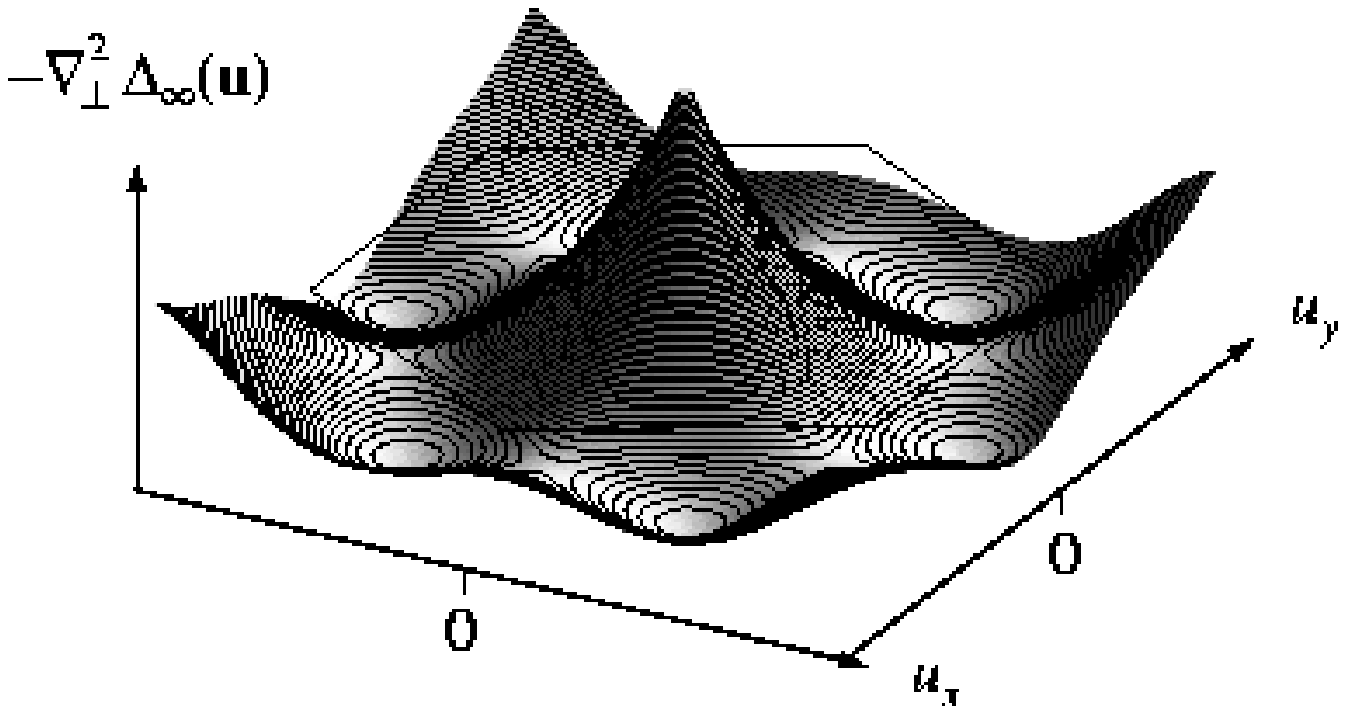,height=6cm}
\fcaption{Illustration of the pinning-force correlator $-\grad_\perp^2
  \Delta_\infty(\bu)$ at the RG fixed point after
  \protect\tcite{Emig+99}. The cusp-like non-analyticity at $\bu=\bN$
  is a characteristic of the Bragg-glass phase.  The hexagon
  represents the Wigner-Seitz cell of the vortex lattice.}
{fig_fp}
\end{figure}

\begin{figure}
\centering
\epsfig{file=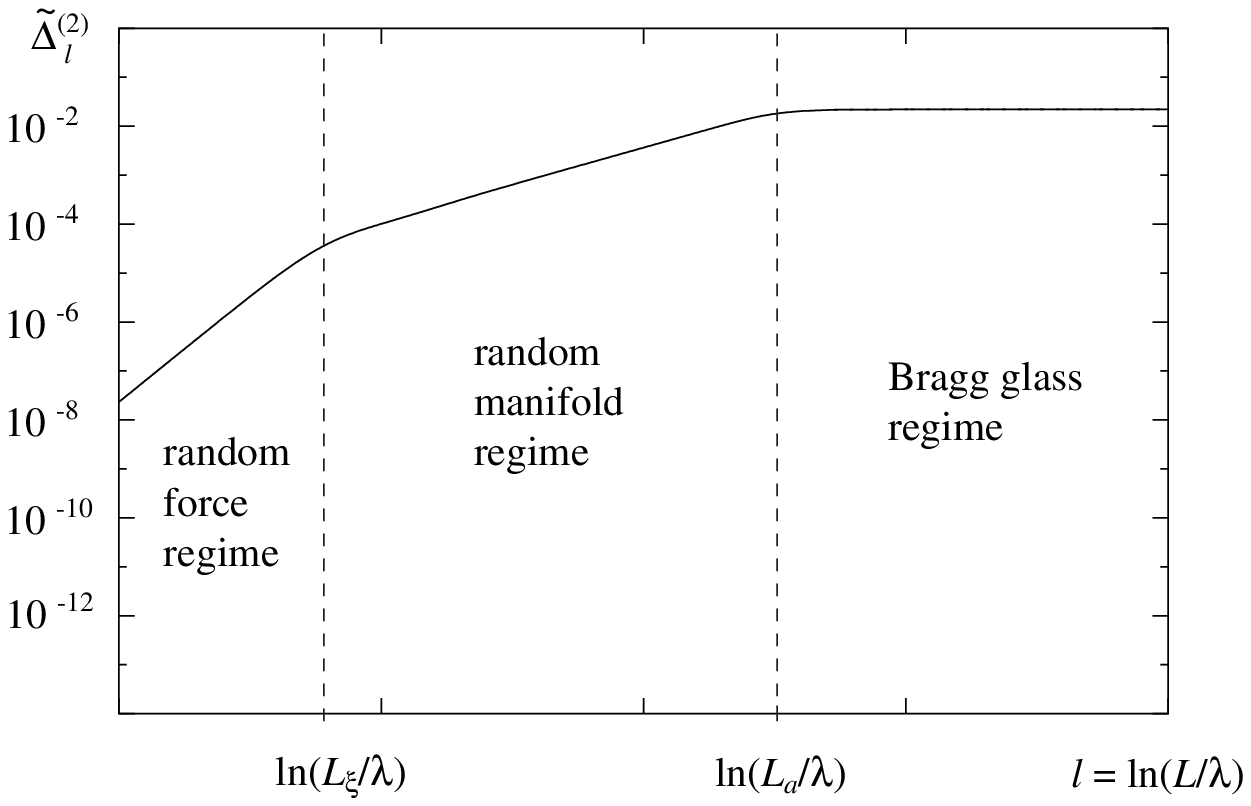,height=6cm}
\fcaption{RG flow of $\tDii_l$ through three regimes 
  \protect\cite{Emig+99}: The random-force regime with a roughness
  exponent $\zeta_{\rm rf}=(4-d)/2$ for length scales $L \lesssim
  L_\xi$, the random-manifold regime with a non-universal exponent
  $\zeta_{\rm rm} \approx (4-d)/6$ for length scales $L_\xi \lesssim L
  <\lesssim L_a$, and the asymptotic Bragg-glass regime with
  logarithmic roughness is reached for $L \gtrsim L_a$. For weak
  disorder and $\xi \ll a$, the width of the crossover regions is
  small compared to the width of the regimes.}
{fig_co}
\end{figure}

\begin{figure}
\centering
\epsfig{file=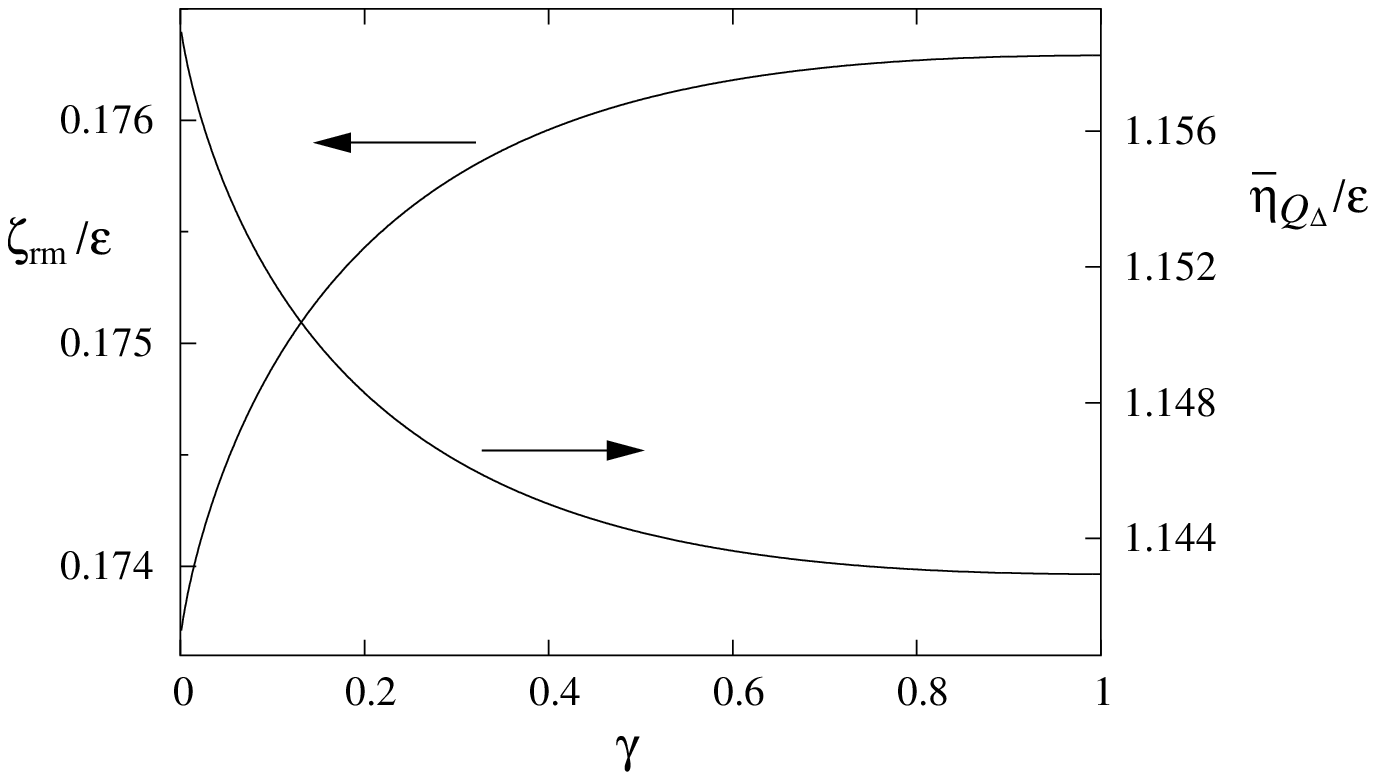,height=6cm}
\fcaption{Non-universal variation of the roughness exponents 
  $\zeta_{\rm rm}$ of the random-manifold regime and the correlation
  function exponent $\bar\eta_{\bf \Qtr}$ with anisotropy parameter
  $\gamma=c_{66}/c_{11}$ \protect\cite{Emig+99}.}
{fig_exp}
\end{figure}

With the numerical value for $\tDii_\infty$ at hands, the explicit
form of the displacement correlations
\begin{eqnarray}
W_{\alpha
  \beta}(\br)=\overline{\langle[u_\alpha(\br)-u_\alpha(\bN)]
  [u_\beta(\br)-u_\beta(\bN)]\rangle} 
\label{def.W_ab}
\end{eqnarray}
in the Bragg-glass phase is given by \cite{Emig+99}
\begin{mathletters}
\mlabel{displace_ab}
\begin{eqnarray}
\label{displace_11}
W_{11}(\br)&=&\frac{\tDii_\infty(\gamma)\atr^2}{1+\gamma}\Bigg\{
\ln\!\left(\frac{x^2+z_{t}^2}{L_a^2}\right)
+\gamma \ln\!\left(\frac{x^2+z_{l}^2}{L_a^2}\right)
\nonumber \\ &&
+\frac{x_2^2-x_1^2}{x^2}\left[1-\gamma-\frac{z_t^2}{x^2}
\ln\!\left(\!1+\frac{x^2}{z_t^2}\!\right)+\gamma\frac{z_l^2}{x^2}
\ln\!\left(\!1+\frac{x^2}{z_l^2}\!\right)\right]\!\!\Bigg\} ,
\quad \quad \\
\label{displace_12}
W_{12}(\br)&=&\frac{2\tDii_\infty(\gamma)\atr^2}{1+\gamma}
\frac{x_1 x_2}{x^2}
\Bigg\{\gamma-1-\gamma\frac{z_l^2}{x^2}\ln\left(1+\frac{x^2}{z_l^2} \right)
\nonumber \\ &&
+\frac{z_t^2}{x^2}\ln\left(1+\frac{x^2}
{z_t^2} \right)\Bigg\},
\end{eqnarray}
\end{mathletters}
and $W_{22}(\br)$ follows from $W_{11}(\br)$ by permuting $\tilde
c_{11}$ and $\tilde c_{66}$.  To lowest order in $\epsilon=4-d$, these
correlations lead to the translational order correlation function
\begin{eqnarray}
S(\bQ,\br) &\equiv&  \overline{\langle
\exp(-i\bQ \cdot [\bu(\br)-\bu(\bN)]) \rangle}
\nonumber \\
&\sim&  g_\bQ L_{a}^{\bar\eta_\bQ}(x^2+z_{t}^2)^
{-\frac{\bar\eta_\bQ}{2(1+\gamma)}}
(x^2+z_{l}^2)^{-\frac{\bar\eta_\bQ}{2(1+1/\gamma)}}
\label{CK-corr}
\end{eqnarray}
with the non-universal $\gamma$-dependent exponent 
\begin{eqnarray}
\bar\eta_\bQ=\tDii_\infty(\gamma)(\atr Q)^2
\end{eqnarray}
and the geometrical prefactor
\begin{eqnarray}
g_\bQ  &=&  \exp\left\{ 
\frac{\tDii_\infty(\gamma)(\atr Q)^2}{1+\gamma}\left[
      (\hat \bx \cdot \hat \bQ)^2-\frac{1}{2}\right]\! \, \right. 
\nonumber \\ 
 &&\times \, \left.
    \left[\!\left(\!1- \frac{z_t^2}{x^2}
        \ln\!\left(\!1+\frac{x^2}{z_t^2}\right)\!\!\right)
      -\gamma\left(\!1-\frac{z_l^2}{x^2}
        \ln\!\left(\!1+\frac{x^2}{z_l^2}\right)\!\!\right)\!\right]
         \right\}, 
\label{eq.gQ}
\end{eqnarray}
which completely describes the angular dependencies of the
translational order. Note that the factor $g_\bQ$ goes to $1$ in the
limit $z\to\infty$. Therefore, in this limit the dependence of
$S(\bQ,\br)$ on the reciprocal lattice vector $\bQ$ remains only in
the exponent $\bar\eta_\bQ$. Moreover, it is interesting to note that
the exponents of the algebraic decay in equation (\ref{CK-corr})
depend on the elastic moduli as soon as $z$ is finite even without
taking into account the non-universality of the exponent
$\bar\eta_\bQ$ itself.  If one ignores the non-trivial $\gamma$
dependence of $\bar\eta_\bQ$, in the case $z=0$ the above formulas
reduce to those found in \citeANP{Giamarchi+94}
\citeyear{Giamarchi+94,Giamarchi+95}.  A logarithmic roughness of
$W(\br)$ was already predicted earlier from scaling arguments
\cite{Nattermann90} and found also from a variational treatment with
replica-symmetry breaking by \tcite{Korshunov93:vg} and
\citeANP{Giamarchi+94} \citeyear{Giamarchi+94,Giamarchi+95}.  However,
this method is not able to capture the non-universality of
$\bar\eta_\bQ$.

The $\gamma$-dependence of the exponent $\bar\eta_{\bf \Qtr}$ for a
smallest reciprocal lattice vector $\Qtr$ is depicted in figure
\ref{fig_exp} and varies numerically from $\bar\eta_{\bf \Qtr}=1.143$
for $\gamma=0$ to $\bar\eta_{\bf \Qtr}=1.159$ for $\gamma=1$.  In
isotropic superconductors at low temperatures, where vortex lines
interact via central forces, one has $0\le\gamma\le 1/3$. $\gamma\sim
1/3$ for $\lambda\le a$, i.e., for fields close to $\Hci$, and $\gamma
\rightarrow 0$ for $H \rightarrow \Hcii$.  For most of the field
region $\gamma \approx \Phi_0/16\pi\lambda^2B$.  Thus, an increase of
the external field from $\Hci$ to $\Hcii$ should result in an increase
of $\bar\eta_{\bf Q}$ and a decrease of $\zeta_{\rm rm}$.  At higher
temperatures, where the vortex line interaction is renormalized
considerably by thermal fluctuations, as well as in anisotropic
superconductors, the above inequality for $\gamma$ may no longer be
fulfilled.  Clearly, in the latter case our starting Hamiltonian
(\ref{Ham}) would also have to be modified.

The non-universality of $\bar\eta_\bQ$ could, in principle, be tested
by neutron scattering experiments at changing external fields.  On the
experimental side, Bragg peaks have indeed been observed in BSCCO for
$H \lesssim 500$G by \tcite{Cubitt+93}.  More recently, \tcite{Kim+99}
have used the decoration technique to determine the structural
properties of the vortex lattice in BSCCO and confirmed the existence
of the perturbative and the random-manifold regime with $\zeta_{\rm
  rf} \approx 0.22$.  On larger length scales their data exhibited
non-equilibrium features such that the true asymptotic regime was not
reached.  So far, the resolution is, however, too weak to determine
the $\gamma$ dependence of $\bar\eta$.  Contact to the neutron
scattering experiment can be made by the structure factor
\begin{eqnarray}
\hS(\bk_\perp, k_z) &=& \sum_\bX \int dz \ 
\overline{\langle e^{i \bk_\perp \cdot
  [\bX+\bu(\bX,z)-\bu(\bN,0)] + i k_z z} \rangle}  
\nonumber \\
&\approx& \sum_\bX \int dz \ e^{i (\bk_\perp \cdot \bX + k_z z)} 
e^{- \frac 12 k_\alpha  k_\beta W_{\alpha \beta}(\bX,z)},
\end{eqnarray}
where we have used the Gaussian approximation (which is correct to
order $\epsilon$) for the distribution of $\bu$ and the definition
(\ref{def.W_ab}) for $W_{\alpha \beta}$. With $\bk_\perp = \bQ +
\bq_\perp$ and $|\bq_\perp| \ll |\bQ|$ we get
\begin{eqnarray}
\hS(\bQ+\bq_\perp,k_z) \approx \rho_0 \int d^3r \ e^{i(\bq_\perp
  \cdot \bx + k_z z)} S(\bQ,\br). 
\end{eqnarray}
$\hS(\bk)$ describes the divergence of the scattered intensity with
vanishing $\bq$ in the vicinity of a reciprocal lattice vector $\bQ$.
The above integral is dominated for small $\bq$ by the large scale
$S(\bQ,\br)$, provided $\bar\eta_{\bf Q}(\gamma)<3 $. It is thus the
sub-algebraic growth of the displacements (\ref{displace_ab}) that
gives rise to the Bragg peaks, hence the name `Bragg glass'.  In the
special cases $\gamma=0$ and $\gamma=1$ one obtains
\begin{eqnarray}
\hS(\bQ+\bq_\perp,k_z) \sim 
\left(q_\perp^2 + \frac {c_{44}}{c_{66}} k_z^2\right)^{
[-3+\bar\eta_{\bf Q}(\gamma)]/2}.
\end{eqnarray}

To summarize the situation in impure bulk superconductors: it turns
out that these still show a quasi-long-range ordered `Bragg-glass'
phase which is described by a {\em non-universal} power-like decay of
the order parameter correlations. In particular, the decay-exponent
$\bar\eta_{\bf Q}$ depends on the ratio $\gamma={\tilde
  c}_{66}/{\tilde c}_{11}$ of the elastic constants, similar to $2D$
pure crystals at their melting temperature. For weak disorder we find
a crossover of the structural correlation functions $S(\bQ,\br)$ from
a Larkin regime, where perturbation theory applies and $S(\bQ,\br)$
decays exponentially, to the random-manifold regime with a stretched
exponential decay of $S(\bQ,\br)$ and eventually to the asymptotic
Bragg-glass regime.

In addition to the disorder-averaged positional correlation function
$S(\bQ,\br)$ it is interesting to also consider the glass correlation
functions $S_{\rm PG}(\bQ,\br)$ and $C_{\rm VG}(\bQ,\br)$.  In the
framework of the functional RG in $d=4-\epsilon$ dimensions, discussed
above, it is easy to show that to {\em first} order $\epsilon$ one
finds \cite{Bogner+00}
\begin{eqnarray}
S_{\rm PG}(\bQ,\br) \approx S_{\rm PG}^{(0)}(\bQ,\br) \left[1 +
\epsilon \sum_{m \geq 1} c_m \left( \frac T {r^2} \right)^{2m} 
+ O(\epsilon^2)\right],
\label{S.PG.3D}
\end{eqnarray}
where $S_{\rm PG}^{(0)}(\bQ,\br) = [S^{(0)}(\bQ,\br) ]^2$.
$S^{(0)}(\bQ,\br) = \exp[- \frac 12 Q_\alpha Q_\beta W_{{\rm th},
  \alpha \beta}(\br)]$ denotes the pair-correlation function for TLRO
of the {\em pure} system, which is finite for $r \to \infty$ if $d>2$
and is reduced with respect to unity merely by the standard
Debye-Waller factor. For large distances $|\br|$ the leading
correction in (\ref{S.PG.3D}) comes from the term with $m=1$. Since
$c_1>0$, the disorder increases slightly the glass order with respect
to the pure system as has to be expected.  With $S_{\rm PG}(\bQ,\br) /
[S(\bQ,\br) ]^2 \to \infty$ for $r \to \infty$, we conclude that the
Bragg-glass regime is indeed a positional glass.  Using the relation
(\ref{delta.varphi(bq)}) it is also easy to show to order $\epsilon$
that $C_{\rm VG}(\br)$ decays exponentially in $d<4$ dimensions (and
algebraically for $d=4$).  Expanding all corrections from the disorder
with respect to $\epsilon=4-d$ we get
\begin{eqnarray}
  C_{\rm VG} (\br)\approx e^{-\frac 12 \langle ( \delta \phi(\br)-\delta
    \phi(\bN))^2 \rangle_{\rm th}} \left[1 +
\epsilon \sum_{m \geq 0} \tilde c_m T^2 \left( \frac T r \right)^{2m} 
+ O(\epsilon^2)\right],
\end{eqnarray}
where
\begin{eqnarray}
  \langle ( \delta \phi(\br)-\delta
    \phi(\bN))^2 \rangle_{\rm th} = 2 \left( \frac{2
      \pi}{a^2}\right)^2 T \int_\bq \frac{q_\perp^2 [1-\cos(\bq \cdot
    \bx)]}{q^4 (c_{66} q_\perp^2 + c_{44} q_\parallel^2)} \sim \frac 1
  \epsilon |\bx|^\epsilon
\end{eqnarray}
denotes the phase fluctuations of the order parameter in the pure
system (see \tcite{Moore92}).  Although the leading correction
($m=0$) from the disorder increases $C_{\rm VG}(\br)$, it does not
compensate its exponential decay originating from strong thermal
fluctuations in the pure system.  Whether this result remains
qualitatively correct if higher order terms in $\epsilon$ are taken
into account remains an open question.  For the moment we conclude
that to order $\epsilon$ there is no phase-coherent vortex-glass order
in the Bragg-glass phase.

Our findings are not necessarily in contradiction to the result of
\tcite{Dorsey+92} who, starting from a random-$\Tc$ GL model and using
mean-field theory, found a transition to a phase with phase-coherent
glass order.  However, this calculation -- once fluctuations are taken
into account -- proves the existence of a vortex glass transition with
a diverging susceptibility $\chi_{\rm VG} = \int d^dr \ C_{\rm
  VG}(\br)$ only in $d=6-\epsilon$ dimensions ($\epsilon \ll 1$).  In
this region ($d>4$) our calculation also gives both $S_{\rm
  PG}(\bQ,\br)$ and $C_{\rm VG}(\br)$ non-zero for $r \to \infty$.

\subsection{Stability of Bragg glass}

In this chapter we have been treating so far the vortex-line lattice
in the {\em elastic approximation}, i.e., we have disregarded
topological defects such as dislocations or disclinations.

In the following we will consider the stability of the Bragg-glass
phase with respect to dislocations.  This problem was considered by
several authors
\cite{Giamarchi+95,Kierfeld+97,Gingras+96,Carpentier+96,%
  Ertas+96:pc,Giamarchi+97:prb,FisherDS97,Kierfeld98,Koshelev+98}.
Since the disorder seen by the dislocation is already partially
screened due to the elastic deformations (a fact which was overlooked
in the early discussion of the elastic approximation by
\tcite{FisherDS+91:tf}), we start this section with a brief discussion
of the {\em effective} disorder strength $\tD_{\rm eff}(R)$ acting on
scale $R$ in the various regimes.  We will thereby ignore all
numerical coefficients of order unity and restrict ourselves to the
case $d=3$.

For simplicity, we restrict ourselves here to the case $c_{11} \gg
c_{66}$ such that we can assume $\grad_\perp \cdot \bu=0$.  It is then
convenient to rewrite the Hamiltonian using the coordinate
$z_t=({c_{66}/c_{44}})^{1/2} z$ introduced in equation
(\ref{def.z_t}). The elastic part of the Hamiltonian is now isotropic
with $c_{44}$ and $c_{66}$ replaced by their geometric mean $({c_{44}
  c_{66}})^{1/2}$.  The same transformation in the pinning part
changes $\tD(\bu)$ into $({c_{44}/ c_{66}})^{1/2} \tD(\bu)$. It is
also convenient to introduce the (anisotropic) distance $R_t$,
\begin{eqnarray}
R_t^2=\bx^2+z_t^2.
\label{intro.r_t}
\end{eqnarray}
$\tDii_{\rm eff}(R)$ was introduced already in equation
(\ref{tD_eff.2}).  We use here (\ref{tD_eff}) to find explicit
expressions for the mean square displacement $\overline{\bu^2}(R_t)$.
To obtain an estimate for $\overline{\bu^2}(R_t) \approx W(R_t)/2$ on
the scale $R_t$ one equates the elastic energy $E_{\rm el} \approx
({c_{44} c_{66}})^{1/2} \ \overline{\bu^2} R_t$, with the fluctuation
of the pinning energy
\begin{eqnarray}
E_{\rm pin} \approx \left(\frac{c_{44}}{c_{66}} \right)^{1/4}
R_t^{3/2} \left(\tDii_{\rm eff}(R_t) \overline{\bu^2}\right)^{1/2}.  
\label{E.pin.fluct}
\end{eqnarray}
$\tDii_{\rm eff}(R_t)$ denotes the curvature of the effective
potential correlator at $\bu=0$ on the length scale $R_t$. The
estimate for the pinning energy corresponds to the use of perturbation
theory applied on the {\em renormalized} random potential. In this way
we obtain
\begin{eqnarray}
  \overline{\bu^2} \approx c_{44}^{-1/2} c_{66}^{-3/2} \tDii_{\rm
    eff}(R_t) R_t = 
\xi^2 \frac{\tDii_{\rm eff}(R_t)}{\tDii} \frac{R_t}{R_\xi},
\label{u2.vs.rt}
\end{eqnarray}
where $\tDii \equiv \tDii(R_\xi)$ denotes the bare value of
$\tDii(R_t)$.  

Now we consider the different regimes already discussed in section
\ref{sec.BG}.

(i) In the {\em perturbative} regime $R_t < R_\xi \approx \xi^2
c_{44}^{1/2} c_{66}^{3/2} / \tDii$, $\tDii_{\rm eff} \approx \tDii $
in equation (\ref{u2.vs.rt}).

(ii) In the {\em manifold} region, $R_\xi < R_t < R_a$, where
$\overline{\bu^2} \approx \xi^2 (R_t/R_\xi)^{2 \zeta_{\rm rm}}$, we
find from equation (\ref{u2.vs.rt})
\begin{eqnarray}
\tDii_{\rm eff}(R_t) \approx \tDii \cdot 
\left( \frac{R_t}{R_\xi}\right)^{2\zeta_{\rm rm }-1}  .
\label{mani.reg.tdiieff}
\end{eqnarray}

(iii) Finally, in the {\em asymptotic} Bragg-glass region, $R_a <
R_t$, we get from $\overline{\bu^2} \approx a^2 \ln (R_t/R_a)$
\begin{eqnarray}
  \tDii_{\rm eff}(R_t) \approx \tDii \frac{R_\xi}{R_t}
  \frac{a^2}{\xi^2}
\approx   \frac{a^2 c_{44}^{1/2} c_{66}^{3/2}}{R_t}.
\label{tDii.eff.R_t}
\end{eqnarray}
Thus, $\tDii_{\rm eff}(R_t)$ is independent of the bare disorder
strength.  Here we have to ignore the logarithm since it originates
from the summation over many different length scales.  

We use now these expressions to obtain an estimate for the energy of a
{\em dislocation loop}.  For simplicity, we assume an isotropic (i.e.,
almost circular) loop of linear size $R_t$. In the pure system, its
elastic energy is
\begin{eqnarray}
E_{\rm el}^{\rm loop} \approx b^2 (c_{44} c_{66})^{1/2} R_t \ln (R_t/a_1),  
\label{E.el.loop.pure}
\end{eqnarray}
where $|\bb|=\atr$ and $\bb$ denotes the Burgers vector of the
dislocation.  The logarithm results from the fact that in a distance
$R_t$ from the dislocation centre the displacement $\bu_{\rm disloc}$
produced by the dislocation line obeys $|\grad \wedge \bu_{\rm
  disloc}| \sim \frac {|\bb|} {2 \pi R_t}$.  Integration over the
plane perpendicular to the dislocation line then yields $\sim \atr^2
(c_{44} c_{66})^{1/2} \ln (R_t/a_1)$ for the energy per unit length,
where $a_1$ acts as a small distance cut-off.

In a random system, the situation is more complicated. Since the
displacement $\bu_{\rm disloc}$ created by a dislocation line obeys
the saddle-point equation $\delta {\cal H}/\delta{\bu}=0$ as well as
$\oint d\bu = \bb$, the elastic energy of the dislocation now depends
also on the disorder.  Since the situation apparently is rather
involved, we will adopt here a simplified picture using the following
observation: The displacement in the neighbourhood of a dislocation
has two sources: Its very existence leads in the distance $R_t$ from
the dislocation to a (essentially tangential) displacement of order
$a$. The disorder, on the other hand, creates displacements (both in
the absence or in the presence of a dislocation) of order $a$ or
larger only on scales $R_t >R_a$. In calculating the energy of a loop
of linear size $R_t> R_a$ in the disordered case we estimate the
elastic energy of the dislocation loop therefore by
\begin{eqnarray}
E_{\rm el}^{\rm loop} \approx a^2 (c_{44} c_{66})^{1/2} R_t \ln (R_a/a_1),  
\label{E.el.loop}
\end{eqnarray}
since distortions originating from the random potential are small
compared to $a$ on scales $R \lesssim R_a$ but are dominating on
larger scales.  So far we have ignored the dispersion of the elastic
constant $c_{44}$, which becomes relevant on scales small compared to
$\lambda$. On these scales the lattice is softer by a factor
$(R_t/\lambda)^2$.  In order to take this effect into account, we
replace the small scale cut-off $a_1$ in (\ref{E.el.loop}) by $(a_1^2
+ \lambda^2)^{1/2}$. 

The distortions on scales larger than $R_a$ are dominated by the
disorder and lead to an energy gain of the order (again omitting the
factor $\ln(R_t)$ in $\overline{\bu^2}$, since it comes from the
summation over displacements on different length scales)
\begin{eqnarray}
  E_{\rm pin}^{\rm loop} \approx - \left( \frac{c_{44}}{c_{66}}
  \right)^{1/4} R_t^{3/2} 
\left(\tDii_{\rm eff} \ \overline{\bu^2} \right)^{1/2} 
\approx -a^2 (c_{44}   c_{66})^{1/2} R_t
\end{eqnarray}
where we used equations (\ref{E.pin.fluct}) and (\ref{tDii.eff.R_t}).
The total loop energy of a dislocation loop can therefore be written
as
\begin{eqnarray}
 E^{\rm loop} \approx a^2 (c_{44} c_{66})^{1/2} R_t \left( \ln \left(
     \frac{R_a^2}{a_1^2 + \lambda^2}\right) - c_1 \right),
\label{E.loop}
\end{eqnarray}
where $c_1$ is a constant of order unity. (It is easy to see by
analogous arguments, but using equations (\ref{u2.vs.rt}) or
(\ref{mani.reg.tdiieff}) for $\Delta_{\rm eff}(R_t)$, that dislocation
loops of size $R_t<R_a$ always have a positive energy.)

From (\ref{E.loop}) we conclude that the system is stable against the
formation of a dislocation loop as long as
\begin{eqnarray}
R_a \gtrsim c_2 (a_1^2+\lambda^2)^{1/2}
\label{stab.loop}  
\end{eqnarray}
with $c_2 \approx e^{c_1}$. As shown by \tcite{Kierfeld98} for strong
disorder (i.e., $R_\xi \leq a$), this relation can be rewritten in the
form of a Lindemann criterion.  Indeed, in the manifold regime (note
that the Larkin regime vanishes for strong disorder), where
\begin{eqnarray}
  \frac{\overline{\bu^2}(R_a)}{\overline{\bu^2}(a)} \approx \left(
  \frac{R_a^2}{a^2+\lambda^2} \right)^{\zeta_{\rm rm}}
\end{eqnarray}
and $\overline{\bu^2}(R_a) \approx a^2$ one obtains from
(\ref{stab.loop})
\begin{eqnarray}
{\overline{\langle[\bu(a)-\bu(0)]^2\rangle}}^{1/2} \lesssim \cli \ a,
\label{bragg.stab.li}
\end{eqnarray}
which is the Lindemann criterion appropriate for disordered systems in
which the mean-square displacement of a single vortex line diverges.
$\cli= c_2^{-\zeta_{\rm rm}}$ denotes the Lindemann number. The main
conclusion from this consideration is that the elastic,
dislocation-free Bragg glass is {\em stable} as long as the criterion
(\ref{stab.loop}) or, for strong disorder, equation
(\ref{bragg.stab.li})] is fulfilled.

A few concluding remarks are in order:
\begin{itemize}
\item[(i)] The criterion (\ref{bragg.stab.li}) [or, equivalently,
  equation (\ref{stab.loop})] was also derived by \tcite{Kierfeld+97}
  and \tcite{Carpentier+96} via a variational treatment for a layered
  system with the magnetic field parallel to the layers. In this model
  only dislocations with Burgers vectors parallel to the layers are
  allowed.  The most important result is the determination of the
  Lindemann number $\cli \approx 0.14$. We refer the reader to these
  papers for further details.
\item[(ii)] \tcite{Ertas+96:pc} took equation (\ref{bragg.stab.li}) as
  a starting point for a discussion of the onset of irreversibility
  and entanglement of vortex lines in high-$T_c$ materials.
\item[(iii)] \tcite{Kierfeld98}, on the basis of equation
  (\ref{bragg.stab.li}), and including the dispersion of the elastic
  constants in detail, calculated the stability boundaries of the
  Bragg-glass phase for YBCO and BSCCO. The resulting phase diagram is
  depicted schematically in figure \ref{fig_phadi_kier}. Since the
  shear modulus decreases for large and small fields, respectively,
  there are two corresponding stability boundaries.
\item[(iv)] \tcite{FisherDS97} undertook a much more detailed
  study of the stability of a defect loop in the random-field $XY$
  model which corresponds to the layered model mentioned in (i).
  Although the details of his energy estimate for the dislocation loop
  are slightly different from those presented here (using a statistical
  tilt symmetry, he estimates $E_{\rm el}^{\rm loop}$ by equation
  (\ref{E.el.loop.pure}) but includes also rare fluctuations in the
  energy gain from the disorder, which also include logarithmic
  corrections), his final conclusions concerning the stability of the
  defect-free phase are essentially identical to those presented here.
\item[(v)] \tcite{Ryu+96:i} numerically investigated the
  field-driven transition from a dislocation-free to a
  dislocation-dominated phase in an impure layered superconductor and
  found good quantitative agreement with the estimates made above as
  well as with the neutron-scattering data of \tcite{Cubitt+93}, who
  found the disappearance of Bragg peaks in BSCCO for $H\gtrsim$500G.
\item[(vi)] Most recently, \tcite{Kierfeld+99} used a random-stress
  model of the form (\ref{cal.H.prime}) as an effective model to
  describe dislocation lines in an impure superconductor. The
  correlations of the random-stress field ${\bbox \mu}(\br)$ on
  different length scales are chosen in such a way that the correct
  roughness exponents in the manifold and the Bragg-glass regime,
  respectively, are reproduced.  Considering in particular dislocation
  {\em lines} with main orientation parallel to the magnetic field,
  which enter and leave the sample at its surface, \tcite{Kierfeld+99}
  found a global phase diagram, which includes -- besides the elastic
  Bragg-glass phase with vanishing dislocation density -- an amorphous
  vortex-glass and a vortex-liquid phase separated by a first-order
  transition line which terminates in a critical point (see figure
  \ref{fig_phadi_kier}).
\end{itemize}
  
\begin{figure}
\centering
\epsfig{file=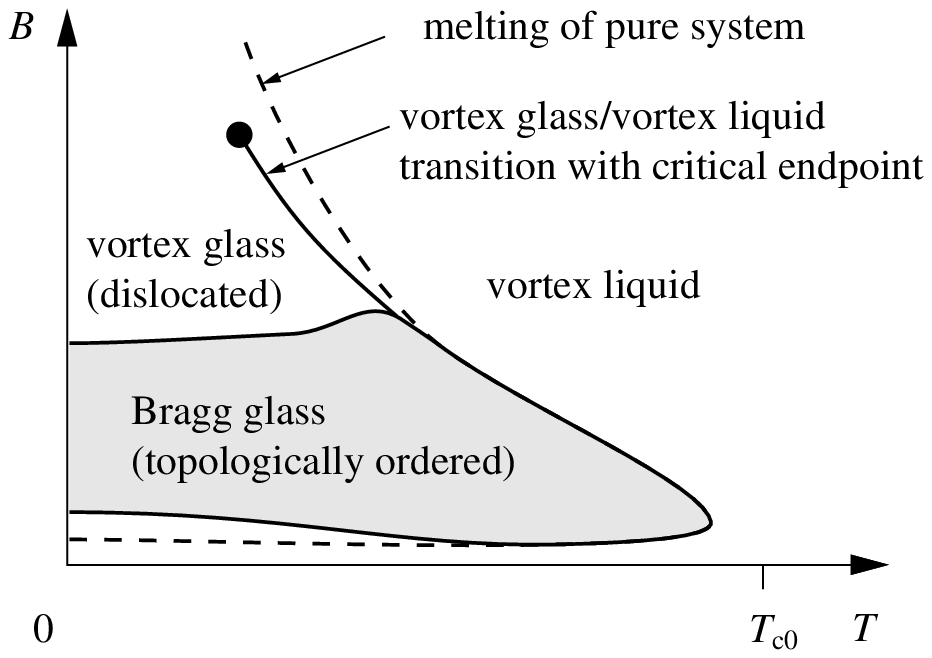,height=6cm}
\fcaption{Schematic phase diagram of the vortex system according to 
  \protect\tcite{Kierfeld98} and \protect\tcite{Kierfeld+99}. In the
  absence of disorder the vortex lattice melts into the vortex liquid
  at low and high magnetic fields (dashed line). Due to the presence
  of disorder the vortex lattice becomes the Bragg glass in a reduced
  stability region (shaded area), where the system is stable to the
  proliferation of dislocations.  At low temperatures and low or high
  fields, disorder induces the proliferation of dislocations for
  energetic reasons and the phase becomes the vortex glass.  The phase
  transition from the vortex lattice to the vortex liquid has a
  critical end point.}
{fig_phadi_kier}
\end{figure}

  \setcounter{equation}{0}
  \section{Vortices driven far from equilibrium}
\label{sec.dyn.phases}

As the main result of the previous chapter we found that weak pinning
reduces the positional order of the vortex lattice in a bulk
superconductor from long-range order to quasi-long-range order.  The
discussion was restricted to thermodynamic equilibrium.  However,
since the electric resistivity is one of the most relevant physical
properties of superconductors, it is important to study the {\em
  driven non-equilibrium} situation.  Since the efficiency of pinning
is related to the structure of the vortex system, it is of particular
interest to characterize this structure.  In addition, since the
melting transition of the vortex lattice shows up as a pronounced
shoulder in the transport characteristic (see, e.g., \tcite{Safar+92},
\tcite{Kwok+92}), it is desirable to locate this transition in the
non-equilibrium situation.

The VLL shows the three regimes of {\em creep}, {\em depinning}, and
{\em flow} in its transport characteristic, which were already
described for manifolds in chapter \ref{sec.mani.trans} (cf. figure
\ref{fig.dyn.reg}).  We recall that in a superconductor the vortex
velocity is proportional to the electric field and the driving force
is proportional to the electric current density.  So far, we have
mainly focused on the {\em creep} regime when we addressed transport
properties in the previous chapters.  In this regime, which is {\em
  close to equilibrium}, the dynamical response is determined within
the scaling approach by the structure of the VLL {\em in equilibrium}.
In particular, the logarithmic roughness of the elastic vortex glass
(described by a roughness exponent $\zeta=0$) results in a drift
velocity
\begin{eqnarray}
 v(F) \sim  (F/\eta_0) \ e^{- U(F)/T} ,
\end{eqnarray}
where the effective barrier height $U(F)$ scales with the driving
force $F$ according to
\begin{eqnarray}
U(F) \sim F^{-\mu}
\end{eqnarray}
with the creep exponent \cite{Nattermann90}
\begin{eqnarray}
  \mu=\frac{d-2}2
\end{eqnarray}
[cf. equations (\ref{result.mu}) to (\ref{v.creep}) with the
appropriate substitution for the dimension of the manifold].  In $d<2$
one thus expects $\mu <0$, which means that the system has a {\em
  linear} transport characteristic at small driving forces. Only in $d
\geq 2$ we can expect to find {\em true} superconductor with a
vanishing {\em linear} resistivity.  The ($d=2$)-dimensional case is
marginal since it has the creep exponent $\mu=0$.  There the effective
barrier height depends logarithmically on the driving force and
results in a power-law transport characteristic as found in equations
(\ref{non-lin.1.1.2}) and (\ref{non-lin.0.2.2}) for the
superconducting film in parallel and perpendicular field.

In this chapter we will focus on the flow regime, which is {\em far
  from equilibrium} but accessible to a theoretical analysis because
it is, roughly speaking, close to the {\em pure} case since disorder
is dynamically `averaged out'.  We will elucidate this point later on.
Although the flow regime might appear to be of no special interest at
first sight, only recently it turned out to be quite non-trivial,
including phenomena such as non-equilibrium phase transitions.

To perform the theoretical analysis of a stationary state at average
velocity $\bv$ it is convenient to define the displacement
$\bu(\bX,\bz,t) \equiv \bx(\bX,\bz,t) -\bX - \bv t$ by subtracting the
average temporal displacement, such that $\bu$ still can be considered
as a small quantity.  $\bX$ is then the ideal vortex position measured
in a frame moving with velocity $\bv$.  In analogy to manifolds, the
equation of motion for the vortex lattice,
\begin{eqnarray}
\label{eqmo.VL}
\eta_0 \dot \bu = \bF^{\rm int} 
+\bF^{\rm pin}  + \bF - \eta_0 \bv + \bzeta,
\end{eqnarray}
is over-damped.  In comparison to equation (\ref{eqmo.mani}) for a
single vortex line, equation (\ref{eqmo.VL}) includes the vortex
density $\rho_0$ as additional factor, such that the latter equation
is an equation for the force {\em density} in the $d$-dimensional
space.  Accordingly, the left-hand side is the Bardeen-Stephen
friction force $\eta_0 \dot \bu $ \cite{Bardeen+65}, with the friction
coefficient $\eta_0 \approx B \Hcii/\rho_{\rm n} \co^2$ referring to
the force per unit volume, whereas the previous expression
(\ref{eta.BS}) refers to the force per unit length of a vortex line.
$\bF^{\rm int}$ is the force acting on a vortex due to its interaction
with other vortices.  In a harmonic approximation for the {\em
  elastic} phase one has $\bF^{\rm int} \approx \bF^{\rm el}$.  The
instantaneous elastic force $\bF^{\rm el}$ depends linearly on $\bu$
and reads
\begin{mathletters}
\mlabel{intro.dispers}
\begin{eqnarray}
F^{\rm el}_\alpha (\bq,t) &=& \Gamma_{\alpha \beta}(\bq) u_\beta(\bq,t),
\\
\Gamma_{\alpha \beta}(\bq) &=&  \sum_p [c_p q_\perp^2 
+ c_{44} q_\parallel^2 ]P^p_{\alpha \beta}(\bq_\perp) ,
\end{eqnarray}
\end{mathletters}
where $\Gamma$ represents the elastic dispersion, $p=L,T$ stands for
the longitudinal/transverse polarization, $c_p \equiv c_{11},c_{66}$
respectively, and $q_\perp^2=q_x^2+q_y^2$.  As far as we consider a
general dimension $d$, $\bx$ and $\bq_\perp$ are two-dimensional and
$\bz$ and $\bq_\parallel$ are $(d-2)$-dimensional.  $\bF^{\rm pin}$ is
the pinning force {\em density}
\begin{eqnarray}
\bF^{\rm pin} (\bX,\bz,t)= 
-\rho_0 \grad_\perp V(\bX + \bv t + \bu(\bX,\bz,t),\bz),
\label{intro.f.pin}
\end{eqnarray}
which implicitly depends on the displacement $\bu$.  The pinning
potential $V$ and the thermal noise $\bzeta$ are assumed to be
Gaussian distributed with correlators (\ref{corr.V}) and
\begin{eqnarray}
\langle \zeta_\alpha (\bX,\bz,t)  \zeta_\beta (\bN,\bN,0) \rangle = 2
\eta_0 T \rho_0 \ \delta_{\bX,\bN} \ \delta_{\alpha \beta} \delta(\bz)
\delta(t)  
\end{eqnarray}
in analogy to equation (\ref{corr.zeta}).

\subsection{Qualitative aspects}

On the basis of this equation of motion we can become more specific in
what sense disorder is `dynamically averaged out' in the limit of
large drift velocities.  In this limit the pinning force $\bF^{\rm
  pin} (\bX,\bz,t)$ acting on a fixed vortex element at position
$(\bX,\bz)$ in the comoving frame changes rapidly as a function of
time.  Since the VLL has finite response times on finite length
scales, the displacements induced by the pinning force decrease with
increasing $\bv$ and the pinning potential is effectively averaged
(`washed') out.  Therefore the effect of disorder vanishes in the
limit of very large driving force.

To substantiate this relation it is instructive to consider a single
vortex line $\bX$ moving with strictly constant velocity (i.e., $\bx =
\bX+ \bv t$), neglecting its response to pinning.  In this
approximation the vortex line experiences a time dependent force
$\bF^{\rm pin}(\bX,\bz,t)= -\rho_0 \grad_\perp V(\bX+ \bv t,\bz)$.  To
some extent the effect of this force can be compared to an additional
thermal noise with a certain `shaking temperature' $\Tsh$.  We choose
the $\bx \equiv (x,y)$ coordinates such that $\bv=(v,0)$ points into
the direction of the first basis vector $\be_x$.  Then
\begin{mathletters}
\mlabel{intro.Tsh}
\begin{eqnarray}
\Tsh_\alpha &\equiv& \frac 1{2 \eta_0 \rho_0} 
\int dt d^{d-2}z \ \overline{F^{\rm   pin}_\alpha(\bX,\bz,t) 
F^{\rm pin}_\alpha(\bX,\bN,0)} 
\nonumber \\
&=& - \frac {\rho_0}{2 \eta_0  v} \int dx_1  \  
\partial_\alpha^2 \Delta(x_1 \be_x)  ,
\\
\Tsh_x &=& 0  ,
\\
\Tsh_y & \simeq & \frac {\rho_0 \Do} {2 \eta_0 v \xi^3} ,
\label{intro.Tsh.y}
\end{eqnarray}
\end{mathletters}
where $\alpha=x,y$ is not to be summed over implicitly.  These
equations are strictly analogous to the equations (\ref{T.sh.mani})
for manifolds.
  
The concept of the `shaking temperature' was introduced by
\tcite{Koshelev+94}, who considered in particular the case $d=2$ and
introduced the `incoherent' shaking temperature $\Tsh \equiv \frac 12
\sum_\alpha \Tsh_\alpha$.  For qualitative purposes they considered
the system driven through disorder as being subject to a total {\em
  effective} temperature $\Teff (v) = T + \Tsh (v)$.  Then the vortex
lattice can be expected to melt at a velocity-dependent temperature
\begin{eqnarray}
\Tm(v) = \Tm- \Tsh (v),
\end{eqnarray}
where $\Tm \equiv \Tm(v=\infty)$ is the melting temperature of the
pure system.  Equivalently, the inverted function $\vm(T)$ defines a
melting velocity above which the vortices freeze into a solid
(`dynamic freezing').  $\Tsh$ vanishes for increasing $v$ and hence
both $\vm$ and the corresponding driving force $\Fm=F(\vm)$ increase
with increasing $T$ for fixed pinning strength.  From the velocity
dependence of $\Tsh_y$ one expects
\begin{eqnarray}
\Tm - \Tm(v) \propto \frac 1v \propto \frac 1F
\label{shift.dyn.melt}
\end{eqnarray}
at large driving forces.  This relation is consistent with
experimental observations \cite{Bhattacharya+93,Hellerqvist+96} and
numerical simulations \cite{Koshelev+94}.

A quantitative analysis of the pinning effects requires taking into
account the interaction between the vortex lines.  For this purpose
\tcite{Koshelev+94} introduced a `coherent' shaking temperature, which
they found to scale like $\Tsh_{\rm coh}(v) \approx ({v_{\rm rel}} /
{v}) \Tsh(v)$ with a characteristic velocity scale $v_{\rm rel}$,
taking into account only the transverse response of the VLL in $d=2$.
However, the question of what physical properties of the system can be
described by such an effective temperature is quite subtle.  This is
evident from equation (\ref{intro.Tsh}), which shows that the shaking
effect is {\em anisotropic} and that one should distinguish the
direction parallel to the driving force from perpendicular directions.

A more careful treatment of disorder will be presented below to
evaluate the effects of disorder on a solid VLL in an elastic
approximation (section \ref{sec.solid.drive}) and to characterize the
structure of the driven lattice, before one can analyze the stability
of the VLL with respect to the proliferation of topological defects,
from what one can locate the melting transition (section
\ref{sec.dyn.melt}).

\subsection{Moving lattice}
\label{sec.solid.drive}

To characterize the structure of the vortex lattice driven in
disorder, we start from the assumption of an elastic and topologically
ordered phase, as exists in the absence of disorder and for low
temperatures in $d \geq 2$.  The first step is to treat pinning
perturbatively in the high-velocity regime. This analysis is most
convenient in Fourier representation, which reads for the displacement
\begin{eqnarray}
\bu(\br,t) = \int_\omega \int_\bq e^{i(\bq \cdot \br - \omega t)} 
\bu(\bq,\omega),
\end{eqnarray}
where $\int_\omega= \int d \omega/2\pi$ and $\int_\bq= \int d^d
q/(2\pi)^d$ is restricted to the first Brillouin zone. The dynamic
response function $G$ of the pure system is ($p=L,T$)
\begin{mathletters} 
\label{response}
\begin{eqnarray}
G_{\alpha\beta}(\bq,\omega)&=&
\sum_p G^p({\bf q},\omega)  P^p_{\alpha\beta} ({\bf q}) ,
\\
G^p({\bf q},\omega) &=&  [ -i \eta_0 \omega
+ c_p q_\perp^2 + c_{44} q_\parallel^2 ]^{-1} .
\end{eqnarray}
\end{mathletters}
To zeroth order in $\bu$ the pinning force acting on the VLL in the
comoving frame has the correlation
\begin{mathletters} 
\mlabel{intro.Xi}
\begin{eqnarray}
\Xi_{\alpha \beta}(\bX,\bz,t) &\equiv&
\overline{F^{\rm pin}_\alpha (\bX,\bz,t) F^{\rm pin}_\beta (\bN,\bN,0)}
\nonumber \\
&=& - \rho_0^2 \partial_\alpha \partial_\beta \Delta(\bX + \bv t)\delta(\bz),
\\
\Xi_{\alpha \beta}(\bq,\omega) 
&=& \rho_0^2 \sum_\bQ k_\alpha k_\beta \Delta(\bk) \delta(\omega + \bv \cdot
\bk)
\nonumber \\
&=& \rho_0^2 \sum_\bQ \Dii_{\alpha \beta}(\bk) \delta(\omega + \bv \cdot \bk),
\end{eqnarray}
\end{mathletters}
where $\bQ$ is a reciprocal lattice vector (RLV) and $\Dii_{\alpha
  \beta}(\bk) \equiv k_\alpha k_\beta \Delta(\bk) $.  The wave vector
$\bk \equiv \bQ+ \bq$ covers the whole Fourier space {\em without}
restriction to the first Brillouin zone (in contrast to $\bq$).

In the following we assume that the vortex lattice is oriented with
one principal axis parallel to the average velocity.
\tcite{Schmid+73} argued that the vortex lattice orients itself in
this direction because this is the direction of minimum entropy
production and of least power dissipation. One can arrive at the same
conclusion from a stability analysis \cite{Mullers+95}.  In this case
there are RLVs $\bQ$ which are perpendicular to the velocity $\bv$ and
which play an important role for the dynamics.  An inspection of the
correlator (\ref{intro.Xi}) for $\omega=0$ and $\bq=\bN$ shows that
the disorder correlator evaluated at these RLVs will determine the
lattice distortions on large length and time scales.  On these scales
the system can be described well by approximating
\begin{eqnarray}
\Xi_{\alpha \beta}(\bq,\omega) \approx \Xi_{\alpha \beta} 
\delta(\omega + \bv \cdot \bq)
\end{eqnarray}
with 
\begin{eqnarray}
\Xi_{\alpha \beta} =\rho_0^2 \sum_{\bQ (\perp \bv)} 
\Dii_{\alpha \beta}(\bQ) \approx
\tDii \delta_{\alpha y} \delta_{\beta y}
\end{eqnarray}
and $\tDii$ as given in equation (\ref{tDii.nochmal}).

From the pinning force correlator (\ref{intro.Xi}) the
zero-temperature displacement correlations follow via
\begin{eqnarray}
W_{\alpha \beta}(\br, t)
&\equiv&
\overline{\langle [u_\alpha(\br,t)-u_\alpha(\bN,0)] 
[u_\beta(\br,t)-u_\beta(\bN,0)]\rangle}
\nonumber \\
&=& 2 \int_\omega \int_\bq  
(1 - e^{i(\bq \cdot \br - \omega t)}) 
 G_{\alpha \gamma}(\bq,\omega) \Xi_{\gamma
  \delta}(\bq,\omega) G_{\delta \beta} (-\bq,-\omega).
\quad \quad
\end{eqnarray}
Within this lowest-order perturbative approach one finds explicitly on
large scales $|\br| \to \infty$ that the fluctuations of $u_x$ are
finite (at $T=0$), whereas $u_y$ is rough in dimensions $d \leq 3$
\cite{Balents+95:to,Giamarchi+96,Balents+97}:
\begin{mathletters}
\mlabel{W.pert.drive}
\begin{eqnarray}
W_{xx}(\br, t=0) &\sim& \frac {2  \rho_0 \tD(\bN)}{\eta_0^2
  v^2}\left(\frac{ 2 \pi}{\xi} \right)^{d-2},
\label{Wxx.drive}
\\
W_{yy}(\br, t=0) &\sim&  \frac {a^{3-d} \Xi_{yy}}{\eta_0 v c_{\rm eff}} 
\ {\cal W} \left(\rho_0 \max\left\{\frac{c_{11}|x|} {\eta_0 v},
y^2,\frac{c_{11}}{c_{44}} \bz^2  \right\}\right),
\label{Wyy.drive}
\end{eqnarray}
\end{mathletters}
with an effective elastic constant $c_{\rm eff}=c_{11}$ in $d=2$ and
$c_{\rm eff}=(c_{11} c_{44})^{1/2}$ in $d=3$ (in general dimension,
$c_{\rm eff}=c_{11}^{2-d/2} c_{44}^{d/2-1}$) and a scaling function
with ${\cal W}(0)=0$ and  
\begin{eqnarray}
{\cal W}(s)&\sim& \left\{
\begin{array}{ll}
\textrm{const.}, & d>3,
\\
\ln(s), & d=3,
\\
s^{(3-d)/2}, &d<3,
\end{array}
\right.
\end{eqnarray}
for $s \to \infty$.

In comparison to the equilibrium case, the driven vortex lattice is
roughened by disorder only in dimensions $d \leq 3$.  Thus, in (the
somewhat academic) dimensions $3<d\leq 4$, the roughness disappears
due to the drive, and disorder is washed out substantially by driving
the VLL.  It is further remarkable that the roughness of the driven
lattice (as found in lowest order perturbation theory) arises from
{\em compression} modes. In contrast to this, in equilibrium even an
incompressible lattice is roughened.  Although the shaking temperature
(\ref{intro.Tsh}) suggests larger fluctuations of $u_y$ than of $u_x$
in agreement with equation (\ref{W.pert.drive}), the roughness of the
lattice {\em cannot} be described by adding the shaking temperature
(\ref{intro.Tsh}) to the physical temperature of the system.  As long
as this temperature is finite, one would expect the lattice to be flat
in $d>2$, which is inconsistent with the roughness found above in
$2<d\leq 3$.

In analogy to the equilibrium case we may define a dynamic Larkin
length, beyond which the perturbative approach breaks down, from
$W(\br, t=0) \equiv W_{xx}(\br, t=0) +W_{yy}(\br, t=0) = \xi^2$, to
which the $y$ component of the displacements gives the dominant
contribution according to equation (\ref{W.pert.drive}).  Because of
the anisotropy of $W_{yy}(\br, t=0)$, the Larkin length strongly
depends on the orientation of $\br$:
\begin{mathletters}
\begin{eqnarray}
\label{Larkin.drive}
L_\xi^{(y)} &\sim&  
\left\{
\begin{array}{ll}
a \exp \left( \frac{\eta_0 v \xi^2 
\sqrt{c_{11} c_{44}}}{\Xi_{yy}} \right) , & d=3,
\\
a \left(\frac{\eta_0 v \xi^2 c_{\rm eff}}{\Xi_{yy} a^{3-d}}
\right)^{1/(3-d)} , & d<3,
\end{array}
\right.
\\
L_\xi^{(x)} &\sim& \frac{\eta_0 v}{c_{11}} \left(L_\xi^{(y)} \right)^2,
\\
L_\xi^{(z)} &\sim& \frac{c_{44}}{c_{11}} L_\xi^{(y)}
\end{eqnarray}
\end{mathletters}
\cite{Giamarchi+96,Balents+98:pm,Ledoussal+98}.

The roughness of the driven vortex lattice (more precisely, of $u_y$)
in $d \leq 3$ implies that the perturbation theory breaks down on
large length scales and progress can be made only by means of a
renormalization-group analysis.  This analysis is very involved
\cite{Balents+98:pm,Ledoussal+98,Scheidl+98:pm} and only its main
results will be presented.

One main feature is that the disorder correlator gets renormalized in
a {\em qualitative} way.  While the original pinning-force correlator
$\Dii_{\alpha \beta}(\bk) \equiv k_\alpha k_\beta \Delta(\bk)$
entering equation (\ref{intro.Xi}) is derived from a random {\em
  potential} such that $\left. \Dii_{\alpha \beta}(\bk)
\right|_{\bk=\bN}=0$, under renormalization it develops a random-{\em
  force} character such that $\left. \Dii_{\infty,\alpha \beta}(\bk)
\right|_{\bk=\bN} \neq 0$.  Thereby the large-scale value of the
pinning-force correlator will be increased to values $\Xi_{\infty,yy}
> \tDii$ and, most important, $\Xi_{\infty,xx} > 0$.
\tcite{Balents+97} pointed out that a roughness of the displacement
component $u_x$ will be generated as a consequence of the finiteness
of $\Xi_{\infty,xx}$.  Thus, on large scales the displacement
correlation function is not described by the perturbative result
(\ref{Wxx.drive}) but by equation (\ref{Wyy.drive}) after the
substitutions $c_{11} \to c_{66}$ and $\Xi_{yy} \to \Xi_{\infty,xx}$.
Since $c_{66} \ll c_{11}$ one expects the fluctuations of $u_y$ to be
larger than the fluctuations of $u_x$ on very large scales
\cite{Balents+97,Balents+98:pm}.

\tcite{Giamarchi+96} pointed out that the displacement component $u_y$
actually shows a frozen pattern in the laboratory frame.  This means
that the vortices move along `static channels', see figure
\ref{fig.channels}.  These channels are oriented parallel to the
average velocity without crossing each other, although they are rough
in $d \leq 3$.  

\begin{figure}[H]
\centering
\epsfig{file=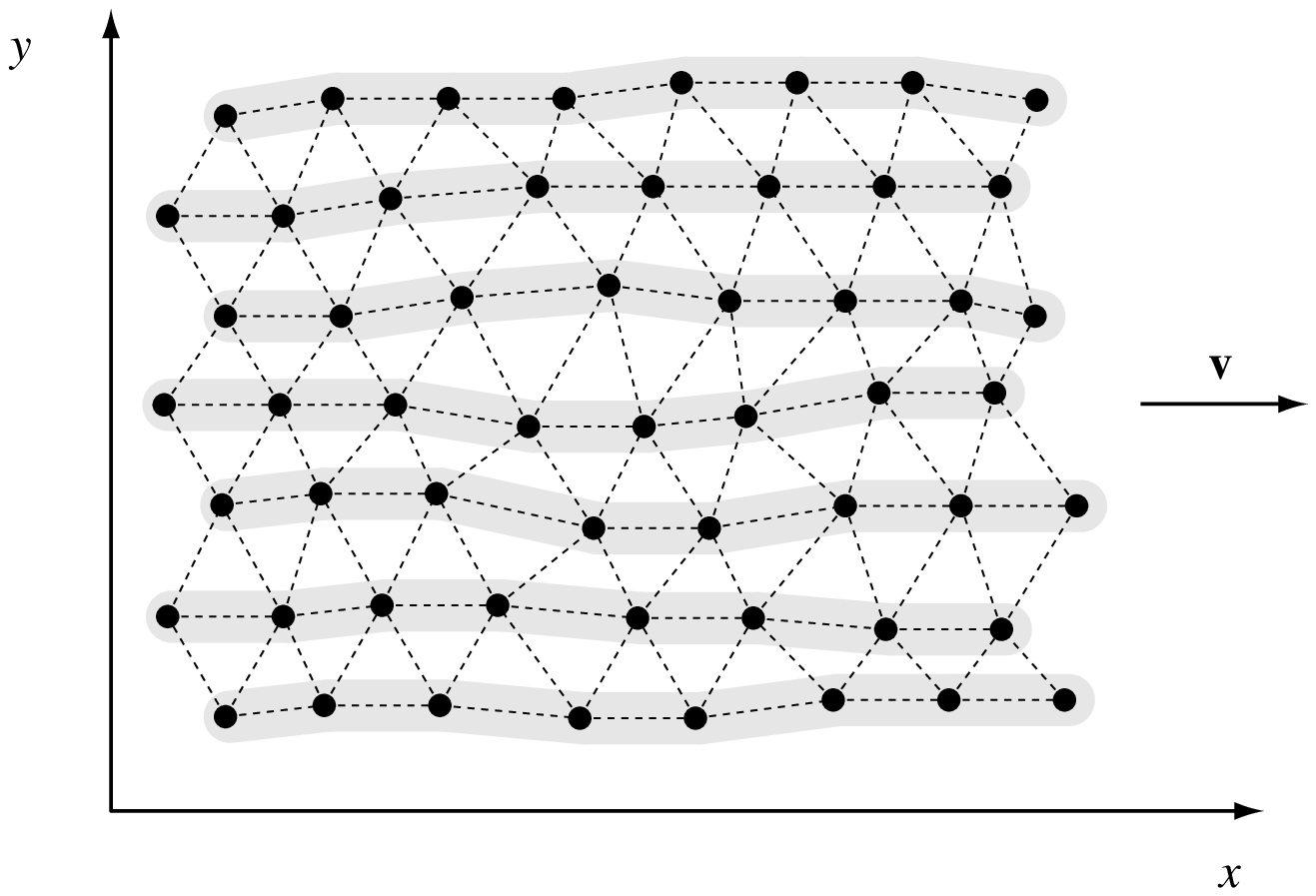,height=6cm}
\fcaption{Structure of the  moving vortex glass
  (intersection at constant $\bz$).  As long as the elastic glass is
  stable (in $d>2$ for sufficiently high drive and low temperature),
  the topological order of the lattice is preserved (dashed lines link
  neighbouring vortices).  Although the moving vortex glass is rough
  (in $d \leq 3$ both displacement components have unbounded
  disorder-induced fluctuations) the vortices flow along channels
  (shaded lines, corresponding to the time-averaged vortex density).}
{fig.channels}
\end{figure}

Thus, in elastic approximation, the driven vortex lattice looks like a
glass with respect to the structure, and hence one may call it `moving
glass' \cite{Giamarchi+96}.  Nevertheless, unlike the Bragg glass the
`moving glass' does not match all criteria of glassiness as specified
in section \ref{sec.mani.sum}.  Since $u_y$ is quenched, one might
expect to find a sub-linear transverse response, $\eta_{yy}(\bv) =
\left.{dF_y(\bv)}/{dv_y}\right|_{v_y=0} = \infty$ at finite
temperature.  This is, however, not the case: a finite transverse
force $F_y$ induces a linear transverse velocity $v_y$
\cite{Ledoussal+98,Scheidl+98:pm}.  Only at zero temperature there is
a finite critical transverse force.  But this is true for any
potential and not indicative for the actual relevance of disorder for
the dynamics.

The renormalization-group analysis reflects characteristic features of
non-equilibrium, including the generation of Kardar-Parisi-Zhang (KPZ)
nonlinearities, and an anisotropic renormalization of the elastic
dispersion of the lattice, of the linear mobility and of the
temperature \cite{Balents+98:pm,Ledoussal+98,Scheidl+98:pm}.  The
detailed discussion of these aspects is beyond the scope of this
article.

\subsection{Moving smectic and dynamic melting}
\label{sec.dyn.melt}

In the previous section we discussed the moving vortex system in the
elastic approximation and found that is has long-range order in $d>3$,
quasi-long-range order in $d=3$ and only short-range order in $d<3$.
In order to examine whether this elastic system is actually stable
with respect to the generation of free dislocations, it is in
principle necessary to examine their dynamic generation process, since
in the non-equilibrium situation the lattice stability can no longer
be examined using energetic criteria.

Nevertheless, if we naively carry over our findings from equilibrium
to non-equilibrium, we might expect the elastic system (the moving
lattice, which is topologically ordered and where vortices move
coherently) in $d \geq 3$ to be stable since it has (quasi-)
long-range order, whereas the short-range order in $d<3$ should imply
the instability.  One can arrive at the same conclusion on the basis
of a dynamical scaling analysis which was performed by
\tcite{Balents+95:to} for CDW systems.

\tcite{Balents+97} argued that because of the dominance of
fluctuations of $u_x$ over fluctuations of $u_y$ on largest scales,
the instability should be generated by dislocations with Burgers
vectors parallel to $\bv$.  Due to the presence of these dislocations
the channels would be decoupled dynamically, i.e. vortices in
different channels could move with locally different velocities.  An
analogous decoupling of charge-density waves in a layered model was
found by \tcite{Vinokur+97} from a variational calculation.

Despite the {\em dynamic} decoupling it is possible that a {\em
  static} channel structure still persists.  This phase is called
`moving smectic' \cite{Balents+97} or `moving transverse glass'
\cite{Ledoussal+98}.  The vortex density can still be modulated with
quasi-long-range order in $d=3$.  In $d<3$ it is possible that even
this channel structure is destroyed, since the power-law roughness of
$u_y$ may also induce dislocations with Burgers vectors having a
component {\em perpendicular} to $\bv$.  Then the vortex system would
be essentially a vortex liquid, which still has a certain anisotropy
since the driving force breaks the rotation symmetry in the $(x,y)$
plane.  Although the moving lattice can exist at sufficiently large
drift velocities in $d=3$, the effective strength of disorder becomes
larger as the velocity is reduced and therefore the moving lattice can
decay first into a moving smectic and then into a moving liquid.  The
schematic phase diagrams
\cite{Balents+98:pm,Ledoussal+98,Scheidl+98:pm} for $d=2$ and $d=3$
are illustrated in figure \ref{fig.dyn.phadi}.

\begin{figure}
\centering
\epsfig{file=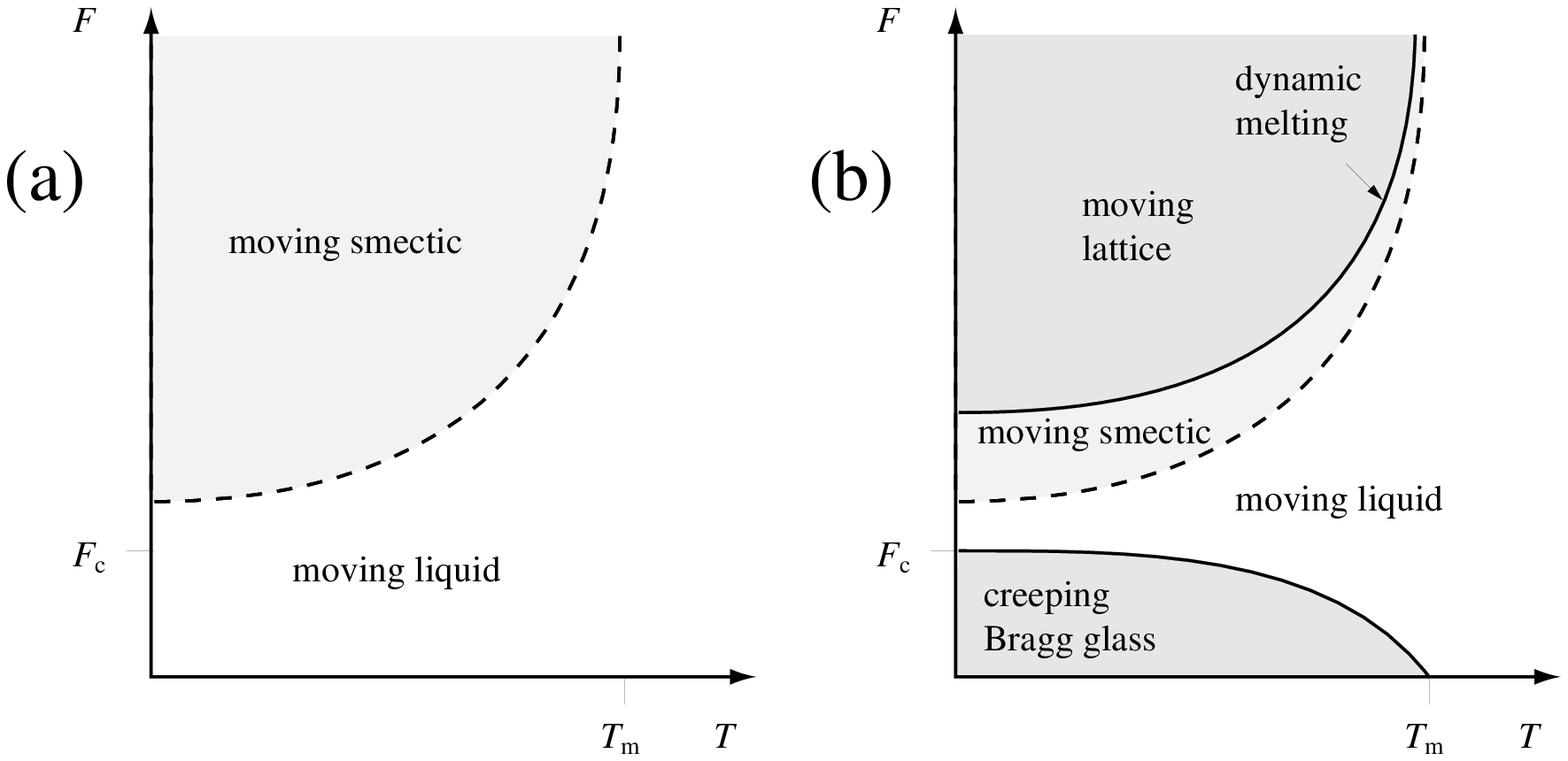,height=6cm}
\fcaption{Schematic illustration of the dynamic phase diagram for
  vortex systems driven in disorder. (a) In $d=2$ dislocations are
  present in the vortex system for zero drive as well as for large
  drive.  Thus at small drive the vortices form a liquid, for which is
  pinned only at $T=0$ and small drive $F<F_{\rm c}$.  At large drive
  the moving vortices can start to follow static channels and to
  become a moving smectic.  (b) In $d=3$ vortices form a Bragg glass
  below melting ($T < \Tm$) and below depinning ($F < F_{\rm c}$). For
  $F>0$ and $T>0$ this lattice will creep until the stresses in the
  lattice become very large such that the lattice breaks into a moving
  liquid. At $T=0$ this destruction of the solid presumably occurs
  very close to depinning. At larger drive, the moving liquid develops
  transverse order as in $d=2$.  Unlike in $d=2$, the moving smectic
  can dynamically freeze into a moving lattice at even larger drive.
  In both dimensions the physics at intermediate drive ($F \simeq
  F_{\rm c}$) lacks a precise theoretical description because in this
  non-equilibrium regime disorder is effectively strong.}
{fig.dyn.phadi}
\end{figure}

A theoretical analysis of the large-scale properties of the driven
vortex system is hampered by the anisotropy of the system, the
relevance of disorder and the dynamic non-linearities such as KPZ
terms that govern the vortex dynamics on large scales.  For these
reasons a rigorous analysis has not been achieved so far.  The
conspiracy of these influences may actually lead to further phase
transitions, such as first-order roughening transitions, which were
found by \tcite{Chen+96} for charge-density-wave systems in $d=1,2$.

There is experimental evidence not only for dynamic melting, but also
the structure of the moving lattice and moving smectic have been
characterized.  The transition was detected by resistive measurements
\cite{Bhattacharya+93,Hellerqvist+96} and the structure of the dynamic
phases was observed by dynamic decoration techniques
\cite{Marchevsky+97,Pardo+98,Troyanovskii+98p}.  Numerical evidence
for dynamic melting was found in $d=2$
\cite{Koshelev+94,Ryu+96,Faleski+96} and also $d=3$
\cite{Dominguez+97}.

In $d=2$ melting can show up only as a crossover which, however, may
be very sharp for weak disorder and restricted system sizes.
\tcite{Spencer+97} presented numerical evidence for the absence of
true topological order even at large velocity in $d=2$.  The
instability of the driven two-dimensional system was demonstrated
analytically by \tcite{Aranson+98:asv} who explicitly examined the
dynamics of single dislocations and their interaction within a driven
vortex lattice.  They were able to show that already the presence of
the KPZ terms leads to a screening of the dislocation interaction on a
{\em finite} length scale (which, however, increases exponentially
with the drift velocity).  Consequently, dislocation--anti-dislocation
pairs will unbind under the additional influence of thermal
fluctuations and even more due to the shaking effect of disorder.

In the remainder of this section we briefly readdress the melting
transition and sketch how it can be captured by the conceptually
simplest approach, a dynamical Lindemann criterion.  Such a
phenomenological approach can be useful in order to locate a
transition semiquantitatively, but it certainly cannot give insight
into large-scale properties.  The Lindemann criterion will not be used
for the total fluctuations of the vortex displacement, but for the
relative displacement of neighbouring vortices.  In this modified form
it has proved to be successful in the static case even for systems
which have no true long-range order, see equation
(\ref{bragg.stab.li}).  In order to extract information about the
anisotropic nature of the driven state it is instructive to look at
the relative displacement fluctuations
\begin{eqnarray}
  w^2(\bb)&=& \overline{\langle [\bu(\bb)-\bu(\bN)]^2 \rangle}
\\
  &=& w^2_x b_x^2/b^2 + w^2_y b_y^2/b^2
\end{eqnarray}
as a function of the {\em orientation} of the vector $\bb$ between the
ideal position of the neighbours. \tcite{Scheidl+98:li} evaluated the
Lindemann criterion $w^2(\bb) \leq \cli^2 a^2$ for the stability of
the lattice and determined the velocity dependence of the melting
transition.  In this context we will not reproduce the detailed
results.  But it is interesting to point out that pinning results in
contributions
\begin{mathletters}
\begin{eqnarray}
  w_x^2 &\propto& \frac 1{v^2},
\\
  w_y^2 &\propto& \frac 1{v}
\end{eqnarray}
\end{mathletters} 
for the bond fluctuations, which imply that for large drift velocities
nearest neighbours separated in $y$ direction have larger bond
fluctuations than nearest neighbours in $x$ direction.  Hence this
phenomenological criterion indicates that the moving lattice (provided
it exists at large drive) will decay into decoupled channels.  This
conclusion from the Lindemann criterion, which examines displacement
fluctuations on {\em short} scales, is in agreement with the analysis
of fluctuations on {\em large} scales described above.  Concerning the
resulting shift of the dynamic melting temperature it agrees with the
finding (\ref{shift.dyn.melt}) from the incoherent shaking temperature
(\ref{intro.Tsh.y}).  Thus the different approaches, the shaking
temperature as measure of the strength of the pinning force, the
Lindemann criterion as a measure of the displacement fluctuations on
small scales and the renormalization-group results for the
displacement fluctuations on large scales give in combination a
consistent picture of the physics of driven vortices.

  \setcounter{equation}{0}
  \section{Summary}
\label{sec.conclusions}

In this article we have reviewed the influence of {\em weak} pinning
by point-like impurities on the vortex-line lattice in type-II
superconductors.  In particular, we have addressed the question to
what extent these superconductors display glass-like properties and by
which order parameter or correlation functions these properties can be
identified.

Hereby it is important to distinguish two concepts of order: (i) order
in the position of the vortex lines, which allows for a breaking of
the continuous translational symmetry that is reflected by a spatial
modulation of the magnetic induction; and (ii) the order of the
superconducting condensate wave function, which is related to the
breaking of the $U(1)$ symmetry and which manifests itself in phase
coherence (ODLRO).

Pinning destroys the positional long-range order of the VLL in $d \leq
4$. However, for weak pinning (and for sufficiently low temperature,
which is always assumed) quasi-long-range positional order persists in
$2< d \leq 4$.  This is true also in superconducting films ($d=2$) in
a parallel field (where the exponent of the algebraic decay depends only
logarithmically on the scale), but not in a perpendicular field, where
dislocations induce short-ranged positional order.

In $d>2$ the vortex lattice behaves as an elastic medium for weak
disorder, i.e., it is topologically ordered since there are no free
dislocations.  Then the vortex system can be called an {\em elastic
  vortex glass}.  (A superconducting film in a parallel field can also
be considered as an elastic glass since dislocations are excluded for
geometrical reasons).  In elastic vortex glasses vortices are
collectively pinned such that vortex motion can take place only due to
thermal activation over arbitrarily high barriers.  Such barriers are
necessary for a vanishing linear resistivity; otherwise the
superconductor in the mixed phase would actually be an Ohmic
conductor.

In the mixed phase vortex fluctuations are decisive not only for the
degree of positional order in the vortex system but also for the phase
coherence.  In pure systems thermal fluctuation destroy ODLRO in $d
\leq 4$ dimensions (if screening is taken into account), including
bulk superconductors and superconducting films in a perpendicular
field.  Preliminary calculations to lowest order in $\epsilon$ for
$d=4-\epsilon$ dimensions suggest that this conclusion remains true
also for systems with weak disorder.  A phase-coherent vortex glass,
where the condensate wave function is quenched, can exist only in
unphysical dimensions $d>4$.  However, higher order terms in
$\epsilon$ may still change this result. This applies in particular to
systems with strong disorder (e.g. gauge-glass models without
screening) that substantially suppresses thermal fluctuations and may
permit the persistence of phase coherence in $d \leq 4$.

When the elastic vortex glass breaks down due to the proliferation of
dislocations, it may still possess a weaker degree of order that might
disappear only at higher temperatures and/or stronger disorder.  Since
the positional order of the elastic vortex glass in $2<d\leq 4$
resembles that of the pure crystal in $d=2$, it is possible that the
elastic vortex glass breaks into a hexatic vortex liquid with
bond-orientational order \cite{Chudnovsky89,Toner91:oo} before it
decays into an isotropic vortex liquid.  A more detailed investigation
of this possibility as well as of further more exotic disordered
phases is left for future studies.

\section*{Acknowledgements}
\addcontentsline{toc}{section}{\numberline{}Acknowledgements}

During the last decade we had enjoyable and valuable discussions with
numerous colleagues.  We wish to express our thanks to all of them.
We are particularly grateful to
S. Bogner, 
G. Eilenberger,
T. Emig,
T. Giamarchi,
D. A. Huse,
J. Kierfeld, 
M. L\"assig,
P. Le Doussal,
R. Ikeda,
M. A. Moore,
L. Radzihovsky,
H. Rieger, and
V. M. Vinokur.
In view of the width of the subject and the desired limitation of
length of this article it was impossible to include all contributions
to the field.  We apologize to the authors of these contribution that
were paid less tribute than they deserve.

  \setcounter{equation}{0}
  \appendix
\section{Pinning of periodic media}
\label{sec.peri}

\renewcommand{\theequation}{\Alph{section}.\arabic{equation}}

In this appendix we present the transformation of the pinning energy
for vortex line lattices, which leads to an effective {\em periodic}
pinning potential.  In our notation $\br=(\bx,\bz)$ denotes a point of
space, with the $D$-dimensional component $\bz$ component along the
vortex lines (in general dimension, a manifold) and the
$N$-dimensional orthogonal component $\bx$ of the displacement.  The
dimension of the embedding space of the VLL then is $d=N+D$.  In the
undistorted lattice, the vortex lines are located at positions $\bX$.
The distorted lattice can be described by the density (lines per
$N$-dimensional volume)
\begin{equation}
\rho_\bu(\br)=\sum_{\bX'}  \delta(\bx-{\bX'}-\bu(\bX',\bz)).
\end{equation}
The average density is denoted by $\rho_0$ and it is related to the
average vortex spacing $a$ through $\rho_0 = a^{-N}$. In
particular, $\rho_0=B/\Phi_0$ for dimension $N=2$.

The pinning energy reads
\begin{eqnarray}
{\cal H}_{\rm pin} &=& \int d^Dz \sum_\bX V(\bX + \bu(\bX,\bz),\bz)=
\int d^dr \ \rho_\bu (\br) V(\br) 
\end{eqnarray}
We always assume disorder to be Gaussian distributed with
correlations
\begin{equation}
\overline{V(\bx,\bz) V(\bx',\bz')}
= \Delta(\bx-\bx') \delta(\bz-\bz').
\end{equation}

Using the Poisson-summation formula \cite{Nattermann+91,Giamarchi+95},
we may rewrite the density
\begin{eqnarray}
\rho_\bu({\bf r})&=&\sum_{\bX'} \int d^N x' \ 
\delta(\bx-\bx' - \bu(\bx',\bz)) \ \delta(\bX'-\bx') 
\nonumber \\
&=& \int d^N x' \  \delta(\bx-\bx'-\bu(\bx',\bz)) \
\rho_0 \sum_\bQ e^{i \bQ \cdot \bx'}
\nonumber \\
&=& 
{\rm det}^{-1}[\delta_{\alpha \beta} + \partial_\alpha u_\beta(\bx,\bz)] 
\rho_0 \sum_\bQ e^{i \bQ \cdot [\bx-\tbu({\bf r})]} 
\nonumber \\
&\approx& 
[1- \partial_\alpha u_\alpha(\bx,\bz)] 
\rho_0 \sum_\bQ e^{i \bQ \cdot [\bx-\tbu({\bf r})]} 
\nonumber \\
&\approx& - \rho_0 \partial_\alpha u_\alpha(\bx,\bz) + 
\rho_0 \sum_{\bQ} \cos\{\bQ \cdot [\bx-\tbu({\bf r})]\} ,
\label{app.manip.rho}
\end{eqnarray}
where $\bQ$ are reciprocal lattice vectors (RLV) and $\partial_\alpha
= \partial / {\partial x_\alpha}$.  The displacement tells us how
vortices are shifted form the ``ideal'' position $\bx'$ to the
``actual'' position $\bx$. In the beginning the displacement is
considered as a function of the ``ideal'' position,
$\bx-\bx'=\bu(\bx',\bz)$.  During these manipulation we have
expressed the displacement as a function of the {\em actual}
position, $\bx-\bx'=\tbu(\bx,\bz)$.

The random potential $V$ couples to the divergence of the displacement
in the density, Eq. (\ref{app.manip.rho}), as an effective {\em
  random-compression} force.  This term is of particular importance in
dimensions $d \leq 2$.

The remaining contributions $\bQ \neq \bN$ to Eq.
(\ref{app.manip.rho}) represent an {\em effective periodic pinning
  potential} since it is invariant under shifts $\tbu (\bx,\bz) \to
\tbu (\bx,\bz) + \bX$ with an arbitrary lattice vector $\bX$.
However, since under this shift $\bx$ is to be held fixed, the shift
means $\bx' \to \bx'-\bX$. Thus it does not mean a translation of the
vortex lattice in the laboratory frame but it means as {\em relabeling}
of the vortices, or a shift of the ideal reference positions $\bx'$.

In practice it is more convenient to start with the exact pinning
energy for a VLL in the replicated system:
\begin{eqnarray}
{\cal H}_{{\rm pin},n} &=& - \frac 1{2T} \sum_{a,b}
\int d^Dz \sum_{\bX,\bX'} \Delta(\bX+\bu^a(\bX,\bz)-\bX'-\bu^b(\bX',\bz))
\nonumber \\
&=& 
 - \frac 1{2T} \sum_{a,b}
\int d^Dz \sum_{\bX,\bX'} \int_\bk \hD(\bk) \
e^{i\bk \cdot[\bX+\bu^a(\bX,\bz)-\bX'-\bu^b(\bX',\bz)]},
\label{app.H.pin.rep}
\end{eqnarray}
where $\int_\bk = \int d^dk/(2\pi)^d$ and the Fourier transform of the
correlator is
\begin{eqnarray}
\hD(\bk) \equiv \int d^Nx \ e^{-i \bk \cdot \bx} \Delta(\bx).  
\end{eqnarray}
The energy (\ref{app.H.pin.rep}) can then be transformed in the
following way: the wave vector $\bk=\bQ+\bq$ is split into a
reciprocal lattice vector $\bQ$ and a vector $\bq$ within the first
Brillouin zone and $\int_\bk = \sum_\bQ \int_\bq$, accordingly.  The
contribution $\bQ=\bN$ represents the pinning energy for a continuous
elastic medium.  This contribution contains a random compression force
(the terms of order $q^2$ in this contribution) which we will not
discuss further here.  We focus on the contributions $\bQ \neq \bN$
which encode the periodicity of the VLL.  For a short disorder
correlation length $\xi \ll a$ one may approximate $\hD(\bQ+\bq)
\approx \hD(\bQ)$ and then perform the the integration over $\bq$,
after which only the contributions $\bX=\bX'$ survive since
$|\bu^a(\bX,\bz)-\bu^b(\bX',\bz)| \ll |\bX-\bX'|$ for a roughness
exponent $\zeta<1$.  After these approximations one ends up with
\begin{eqnarray}
{\cal H}_{{\rm pin},n} &\approx&
- \frac 1{2T} \sum_{a,b}
\int d^Dz \sum_{\bX} \rho_0 \sum_\bQ \Delta(\bQ)
e^{i \bQ \cdot [\bu^a(\bX,\bz)-\bu^b(\bX,\bz)] }
\nonumber 
\\
&=&- \frac 1{2T} \sum_{a,b}
\int d^Dz \sum_{\bX} a^N \tD(\bu^a(\bX,\bz)-\bu^b(\bX,\bz))
\nonumber 
\\
&\approx&- \frac 1{2 T} \sum_{a,b}
\int d^dr \ \tD(\bu^a(\br)-\bu^b(\br))
\label{app.manip.H.n.pin}
\end{eqnarray}
with the effective periodic correlator 
\begin{eqnarray}
\tD(\bu)&=& \rho_0^2\sum_\bQ \Delta(\bQ) e^{i \bQ \cdot \bu}
= \rho_0 \sum_\bX \Delta(\bX + \bu) ,
\label{intro.tDelta}
\end{eqnarray}
which is related to the effective potential via
\begin{eqnarray}
\overline{\tV(\bu,\br) \tV(\bu',\br')} \approx\tD(\bu-\bu') \delta(\br-\br').
\end{eqnarray}
This means that the periodic lattice effectively behaves like an
elastic manifold of internal dimension $D_{\rm eff}=D+N$ in a
effective embedding space $d_{\rm eff}=D+2N$ subject to a periodic
pinning potential.

The advantage of the manipulations (\ref{app.manip.H.n.pin}) as
compared to (\ref{app.manip.rho}) is that we end up with a periodic
correlator for the displacement $\bu$ as a function of the ideal
reference position (which is the actual degree of freedom of the
vortex lattice) rather than for $\tbu$ as a function of the actual
vortex position.

\newpage
\section{List of recurrent symbols}

\begin{center}
\begin{supertabular} {l|l|l}
  symbol & meaning & definition
  \\
  \hline\hline $a$ & effective vortex spacing & $a^2 \equiv \Phi_0/B$
  \\
  $\atr$ & vortex spacing in triangular lattice& $\atr^2 \equiv
  (2/\sqrt 3)a^2$
  \\
  ${\bf A}$ & magnetic vector potential &
  \\
  ${\bf B}$ & magnetic induction &
  \\
  $\co$ & velocity of light &
  \\
  $\cg$, $c_1$, $c_2$, \dots & numerical constants &
  \\
  $c_{11}$ & compression modulus &
  \\
  $c_{44}$ & tilt modulus &
  \\
  $c_{66}$ & shear modulus &
  \\
  $\cli$ & Lindemann number & (\ref{intro.Linde})
  \\
  $C_{\rm VG}$ & phase-coherent vortex-glass correlation &
  (\ref{C_VG})
  \\
  $d$ & (total) spatial dimension & page \pageref{intro.d}
  \\
  $D$ & (internal) spatial dimension & page \pageref{intro.D}
  \\
  ${\cal E}$ & energy &
  \\
  $\bF$ & driving force & (\ref{intro.eqmo})
  \\
  ${\cal F}$ & free energy &
  \\
  $G$ & displacement response function & (\ref{intro.G})
  \\
  $\Gi$ & Ginzburg number & (\ref{intro.Gi})
  \\
  ${\bf H}$ & magnetic field &
  \\
  $\Hc$ & thermodynamic critical field & page \pageref{intro.Hc}
  \\
  $\Hci$ & lower critical field & page \pageref{intro.Hci}
  \\
  $\Hcii$ & upper critical field & page \pageref{intro.Hcii}
  \\
  $\Hmi$ & lower melting field & page \pageref{intro.Hmi}
  \\
  $\Hmii$ & upper melting field & (\ref{intro.Hmii})
  \\
  ${\cal H}$ & Hamiltonian &
  \\
  $j$ & current density &
  \\
  $J$ & effective stiffness &(\ref{intro.J})
  \\
  ${\bf k}$ & wave vector, without restriction & $\bk=\bQ+\bq$
  \\
  $L$ & system size &
  \\
  $L_a$ & crossover length & (\ref{intro.L_a})
  \\
  $\Ls$ & screening length & (\ref{intro.L_s})
  \\
  $L_\xi$ & Larkin length & (\ref{intro.L_xi}), (\ref{def.L.R_xi})
  \\
  $n$ & number of replicas &
  \\
  $N$ & number of displacement components & page \pageref{intro.N}
  \\
  $\bP^{L/T}$ & longitudinal/transverse projector & (\ref{intro.P})
  \\
  ${\bf q}$ & wave vector, restricted to first Brillouin zone &
  \\
  ${\bf Q}$ & reciprocal lattice vector (RLV) &
  \\
  $\cQ$ & thermodynamic quantity & (\ref{scal.aniso})
  \\
  $\Qtr$ & length of first RLV in triangular lattice & $\Qtr^2 \equiv
  16 \pi^2/3\atr^2$
  \\
  $R_a$ & crossover length & (\ref{intro.R_a})
  \\
  $R_t$ & rescaled length & (\ref{intro.r_t})
  \\
  $R_\xi$ & Larkin length & (\ref{def.L.R_xi})
  \\
  $s$ & film thickness &
  \\
  $S$ & translational order parameter correlation & (\ref{eq.D4}),
  (\ref{intro.Psi_k})
  \\
  $S_{\rm PG}$ & positional glass correlation & (\ref{eq.D4})
  \\
  $\hS$ & structure factor & (\ref{intro.S})
  \\
  ${\cal S}$ & entropy &
  \\
  $\Tc$ & superconducting transition temperature &
  \\
  $\Tco$ & mean-field transition temperature & page
  \pageref{intro.Tco}
  \\
  $\Tg$ & glass transition temperature & (\ref{intro.T_G.1+1}),
  (\ref{T_G.2D})
  \\
  $\Th$ & hexatic/isotropic liquid transition temperature &
  (\ref{T_H})
  \\
  $\Tm$ & melting temperature & (\ref{T_M.2D})
  \\
  $T_{\rm KT}$ & Kosterlitz-Thouless transition temperature &
  (\ref{intro.T_KT})
  \\
  $\bu$ & vortex displacement &
  \\
  $U$ & interaction potential &
  \\
  $\bv$ & velocity &
  \\
  $V$ & pinning potential &
  \\
  $W$ & variance of displacement difference & (\ref{intro.W}),
  (\ref{intro.W.mat})
  \\
  $x$ & component of $\bx$ parallel to $\bv$ &
  \\
  $\bx$ & vector component perpendicular to $\bB$&
  \\
  $y$ & component of $\bx$ perpendicular to $\bv$ &
  \\
  $\bz$ & vector component parallel to $\bB$&
  \\
  $z$ & dynamical exponent & (\ref{intro.z})
  \\
  $z_l$, $z_t$ & rescaled lengths & (\ref{def.z_t})
  \\
  $\gamma$ & ratio of elastic constants & $\gamma \equiv
  c_{66}/c_{11}$, (\ref{intro.gamma})
  \\
  $\gE$ & Euler's constant & $\gE=0.577 \cdots$
  \\
  $\Gamma$ & elastic dispersion relation & (\ref{intro.dispers})
  \\
  $\delta$ & anisotropy parameter & (\ref{def.delta(gamma)})
  \\
  $\Delta$ & disorder correlator & (\ref{intro.Delta})
  \\
  $\Do$ & integral of disorder correlator &(\ref{intro.Delta_0})
  \\
  $\Dii$ & moment of disorder correlator & (\ref{intro.Delta2})
  \\
  $\tD$ & periodic disorder correlator & (\ref{intro.tDelta})
  \\
  $\tDii$ & moment of periodic disorder correlator & page
  \pageref{intro.tDii}, (\ref{tDii.nochmal})
  \\
  $\epsilon$ & anisotropy & page \pageref{intro.eps.aniso}
  \\
  $\epsilon$ & dimension & $\epsilon=4-D$
  \\
  $\eo$ & energy scale & (\ref{intro.eps0})
  \\
  $\st$ & stiffness of manifold & (\ref{H.el.mani})
  \\
  $\zeta$ & roughness exponent & (\ref{intro.W})
  \\
  ${\bbox \zeta}$ & thermal noise & (\ref{intro.eqmo})
  \\
  $\zeta_{\rm F}$ & Flory roughness exponent & (\ref{zeta.F})
  \\
  $\zeta_{\rm rf}$ & random force roughness exponent & (\ref{zeta.rf})
  \\
  $\zeta_{\rm rfi}$ & random field roughness exponent &
  (\ref{zeta.rfi})
  \\
  $\zeta_{\rm rm}$ & random manifold roughness exponent &
  (\ref{zeta.rm})
  \\
  $\zeta_{\rm th}$ & thermal roughness exponent & (\ref{zeta.th})
  \\
  $\eta$ & correlation exponent & (\ref{intro.eta})
  \\
  $\eta_0$ & unrenormalized friction coefficient & (\ref{intro.eqmo})
  \\
  $\theta$ & scaling exponent of energy fluctuations &
  (\ref{intro.theta})
  \\
  $\kappa$ & Ginzburg-Landau parameter & $\kappa \equiv \lambda/\xi$
  \\
  $\lambda$ & magnetic penetration length & (\ref{intro.lambda})
  \\
  $\Lambda_T$ & thermal length & (\ref{intro.Lambda_T})
  \\
  $\mu$ & random tilt & (\ref{intro.mu})
  \\
  $\mu$ & creep exponent & (\ref{def.mu.creep})
  \\
  $\mu_{\rm mag}$ & magnetic permeability & (\ref{intro.mu_mag})
  \\
  $\xi$ & coherence length & (\ref{intro.xi})
  \\
  $\xi$ & disorder correlation length &
  \\
  $\Xi$ & pinning force correlator & (\ref{intro.Xi})
  \\
  $\rho$ & vortex density & (\ref{intro.rho})
  \\
  $\rho_0$ & average vortex density & $\rho_0=B/\Phi_0$
  \\
  $\rho_{\rm n}$ & normal state resistivity & (\ref{eta.BS})
  \\
  $\sigma$ & random tilt strength & (\ref{intro.sigma})
  \\
  $\tau$ & relative temperature distance & $\tau \equiv 1-T/\Tg$,
  (\ref{intro.tau})
  \\
  $\Phi_0$ & flux quantum & $\Phi_0 \equiv h\co/2e$
  \\
  $\psi$ & barrier scaling exponent & (\ref{scale.barrier})
  \\
  $\psi_\bk$ & translational order parameter & (\ref{intro.psi_k})
  \\
\end{supertabular}
\end{center}
\bigskip

Note that functions and their Fourier transforms have the same name
although they are different functions.  This ambiguity is removed by
the argument of the function, which is either a length (such as $\bx$,
$\bz$, $\br$ etc.) or a wave vector (such as $\bk$, $\bq$, $\bQ$
etc.).  The only exceptions are $\Delta$ and $\tD$. Their common
Fourier transform is denoted by $\hD$.


  \addcontentsline{toc}{section}{\numberline{}References}
  \bibliographystyle{bibstyle}
  \bibliography{library}



\end{document}